\documentclass[trackchanges, twocolumn]{aastex701}

\usepackage{textcomp}
\usepackage{gensymb}

\newcommand{\commentred}[1]{{\color{black}#1}}

\newcommand{\oiii}{[O\,{\sc iii}]}

\newcommand{\nii}{[N\,{\sc ii}]}
\newcommand{\sii}{[S\,{\sc ii}]}

\newcommand{\feii}{[Fe\,{\sc ii}]}

\received{2026 January 26}
\revised{2026 October 15}
\accepted{2026 October 26}
\submitjournal{ApJ}
\shorttitle{JWST Census of Faint SMGs in the HUDF}
\shortauthors{Kiyota et al.}
\graphicspath{{./}{figures/}}
\usepackage{amsmath}

\begin{document}

\title{
JWST Spectroscopic Census of ALMA Faint Submillimeter Galaxies\\
in the Hubble Ultra Deep Field
}

\author[orcid=0009-0004-4332-9225]{Tomokazu Kiyota}
\affiliation{Astronomical Science Program, Graduate Institute for Advanced Studies, SOKENDAI, 2-21-1 Osawa, Mitaka, Tokyo 181-8588, Japan}
\affiliation{National Astronomical Observatory of Japan, 2-21-1 Osawa, Mitaka, Tokyo, 181-8588, Japan}
\email[show]{tomokazu.kiyota@grad.nao.ac.jp}

\author[orcid=0000-0002-1049-6658]{Masami Ouchi}
\affiliation{National Astronomical Observatory of Japan, 2-21-1 Osawa, Mitaka, Tokyo, 181-8588, Japan}
\affiliation{Institute for Cosmic Ray Research, The University of Tokyo, 5-1-5 Kashiwanoha, Kashiwa, Chiba 277-8582, Japan}
\affiliation{Astronomical Science Program, Graduate Institute for Advanced Studies, SOKENDAI, 2-21-1 Osawa, Mitaka, Tokyo 181-8588, Japan}
\affiliation{Kavli Institute for the Physics and Mathematics of the Universe (WPI), University of Tokyo, Kashiwa, Chiba 277-8583, Japan}
\email{ouchims@icrr.u-tokyo.ac.jp}

\author[orcid=0000-0002-2364-0823]{Daisuke Iono}
\affiliation{National Astronomical Observatory of Japan, 2-21-1 Osawa, Mitaka, Tokyo, 181-8588, Japan}
\affiliation{Astronomical Science Program, Graduate Institute for Advanced Studies, SOKENDAI, 2-21-1 Osawa, Mitaka, Tokyo 181-8588, Japan}
\email{d.iono@nao.ac.jp}

\author[orcid=0000-0001-7201-5066]{Seiji Fujimoto}
\affiliation{David A. Dunlap Department of Astronomy and Astrophysics, University of Toronto, 50 St. George Street, Toronto, Ontario, M5S 3H4, Canada}
\affiliation{Dunlap Institute for Astronomy and Astrophysics, 50 St. George Street, Toronto, Ontario, M5S 3H4, Canada}
\email{seiji.fujimoto@utoronto.ca}

\author[orcid=0000-0002-4052-2394]{Kotaro Kohno}
\affiliation{Institute of Astronomy, Graduate School of Science, The University of Tokyo, 2-21-1 Osawa, Mitaka, Tokyo 181-0015, Japan}
\affiliation{Research Center for the Early Universe, Graduate School of Science, The University of Tokyo, 7-3-1 Hongo, Bunkyo-ku, Tokyo 113-0033, Japan}
\email{kkohno@ioa.s.u-tokyo.ac.jp}

\author[orcid=0000-0001-7821-6715]{Yoshihiro Ueda}
\affiliation{Department of Astronomy, Kyoto University, Kyoto 606-8502, Japan}
\email{ueda@kusastro.kyoto-u.ac.jp}

\author[orcid=0000-0003-2965-5070]{Kimihiko Nakajima}
\affiliation{Institute of Liberal Arts and Science, Kanazawa University, Kakuma-machi, Kanazawa, Ishikawa, 920-1192, Japan} 
\affiliation{National Astronomical Observatory of Japan, 2-21-1 Osawa, Mitaka, Tokyo, 181-8588, Japan} 
\email{knakajima@staff.kanazawa-u.ac.jp}

\author[orcid=0000-0003-4321-0975]{Moka Nishigaki}
\affiliation{Astronomical Science Program, Graduate Institute for Advanced Studies, SOKENDAI, 2-21-1 Osawa, Mitaka, Tokyo 181-8588, Japan}
\affiliation{National Astronomical Observatory of Japan, 2-21-1 Osawa, Mitaka, Tokyo, 181-8588, Japan}
\email{moka.nishigaki@grad.nao.ac.jp}

\author[orcid=0000-0002-1319-3433, sname=]{Hidenobu Yajima}
\affiliation{Center for Computational Sciences, University of Tsukuba, Ten-nodai, 1-1-1 Tsukuba, Ibaraki 305-8577, Japan} 
\email{yajima@ccs.tsukuba.ac.jp}

\begin{abstract}

%249 words
We present a JWST/NIRSpec rest-frame optical spectroscopic census of ALMA 1-mm continuum sources in the Hubble Ultra Deep Field (UDF) identified by the deep ALMA UDF and ASPECS programs. Our sample is composed of the ALMA flux-limited ($S_{1\,\mathrm{mm}}\gtrsim 0.1\,\mathrm{mJy}$) sources observed with medium-resolution NIRSpec spectroscopy from JADES and SMILES, 16 faint submillimeter galaxies (SMGs) at spectroscopic redshifts of $z\sim 1$--$4$. These SMGs show bright longer-wavelength optical lines (H$\alpha$, \nii$\lambda\lambda6548,658\commentred{4}$, and \sii$\lambda\lambda6717,6731$) and faint shorter-wavelength optical lines (H$\beta$ and \oiii$\lambda\lambda4959,5007$) with a large nebular attenuation, $E(B-V)\sim0.3$--$1.8$. We test the SMGs using BPT diagnostics and Chandra X-ray fluxes, and find that most SMGs are classified as AGNs; the AGN fraction is $\sim80\%$ for the SMGs at $M_*>10^{10.5} M_\odot$. 
We find only one SMG ($<10\%$) with a broad Balmer line, indicating that the SMGs are predominantly obscured AGNs. With the optical lines, we estimate the metallicities of the SMGs to be moderately high, $\sim0.4$--$2 Z_\odot$, exceeding the model-predicted dust-growth critical metallicity ($\sim0.1$--$0.2Z_\odot$), which naturally explains the dusty nature of the SMGs. Interestingly, the SMGs fall in the mass-metallicity relation and the star-formation main sequence, showing no significant differences from other high-$z$ galaxies. Similarly, we find electron densities of $n_e\sim10^2$--$10^3\,\mathrm{cm}^{-3}$ for the SMGs that are comparable with other high-$z$ galaxies. 
Together with the high SMG fraction ($\sim 100$\%) at the massive end ($M_*>10^{10.5} M_\odot$), these results indicate that the SMGs are mostly not special, but typical massive star-forming galaxies at high redshift.

\end{abstract}


\section{Introduction} 
\label{sec:intro}

Over the past few decades, large (sub)millimeter continuum surveys have been carried out with single-dish telescopes and interferometers such as the Submillimetre Common-User Bolometer Array (SCUBA; \citealt{holland99}) \commentred{on the James Clerk Maxwell Telescope}, the South Pole Telescope (SPT; \citealt{carlstrom11}), \commentred{the Large APEX BOlometer CAmera (LABOCA; \citealt{siringo09}) on the Atacama Pathfinder Experiment 12 m telescope (APEX)} and, more recently, the Atacama Large Millimeter/submillimeter Array (ALMA; \citealt{wootten09}). 
These surveys detect rest-frame far-infrared (FIR) dust emission from high-redshift ($z\gtrsim1$) galaxies and provide constraints on their dust-obscured star formation and cold interstellar medium (ISM; e.g., \citealt{blain93, smail97, blain02, fujimoto16, fujimoto17, dunlop17, hatsukade18, aravena19, aravena20, bouwens20, dudzevivciute20, fujimoto24, fujimoto25}). 

The bright dust-continuum sources, commonly referred to as submillimeter galaxies (SMGs), contain large dust and molecular gas reservoirs and sustain extreme star-formation rates of $10^2$--$10^3~M_\odot~\mathrm{yr^{-1}}$ (\citealt{casey14, hodge20} for reviews). 
Their large stellar and gas masses, short gas-depletion timescales, and number densities suggest that high-redshift SMGs are plausible progenitors of the quiescent and massive elliptical galaxies observed at lower redshifts (e.g., \citealt{hopkins08, toft14, gomez-guijarro18, dudzevivciute20}). 
In this context, high-redshift dusty galaxies, including SMGs, are key laboratories for understanding the evolutionary pathway of massive galaxies. 

ALMA has further revealed this population down to fainter dust-continuum sources, and extensive observations \commentred{have been} conducted with multi-wavelength facilities. 
One remarkable example is the Great Observatories Origins Deep Survey (GOODS; \citealt{giavalisco04}) south (GOODS-S) field, especially in the Hubble Ultra Deep Field (HUDF; \citealt{beckwith06}).  In this field, deep ALMA dust continuum (e.g., \citealt{walter16, dunlop17, hatsukade18, franco18, aravena20, gonzalezlopez20, gomez-guigarro22-goods-alma}), CO lines (e.g., \citealt{aravena19, boogaard19, decarli19}), Chandra X-ray Observatory (Chandra; e.g., \citealt{weisskopf02}) data (e.g., \citealt{giacconi02, lehmer05, wang13, luo17, ueda18}), and the Hubble Space Telescope (HST) optical to near-infrared (NIR) imaging (e.g., \citealt{grogin11, illingworth13}) exist. 
These data provide multi-wavelength spectral energy distributions (SEDs) from the ultraviolet (UV) to the FIR, as well as key FIR emission lines, \commentred{enabling us to probe stellar populations, molecular gas, and dust content.}
% which reveal stellar populations, molecular gas, and dust content. 
% However, these surveys and bright SMGs observations lack the overall rest-frame optical emission lines (e.g., \citealt{casey17}). 
\commentred{However, despite these extensive multi-wavelength data, systematic rest-frame optical emission-line diagnostics remain unavailable for much of the dusty galaxy population, particularly for the fainter dust-continuum sources.} (see also \citealt{casey17}). 

Rest-frame optical emission lines provide powerful diagnostics of gas-phase metallicity, ionization state (e.g., \citealt{curti20, sanders21, nakajima23, nishigaki25}), electron density (e.g., \citealt{topping25}), and the presence of active galactic nuclei (AGN; e.g., \citealt{kokorev23}), but such measurements have been challenging for SMGs: 
strong dust attenuation renders their rest-UV/optical continua and lines faint, and these diagnostics are redshifted into the NIR, where even 8--10~m class ground-based telescopes had limited sensitivity and wavelength coverage. 
Thus, even basic quantities of dust-continuum sources remain poorly constrained (see also, e.g., \citealt{swinbank04, casey17, shapley20}). 

The James Webb Space Telescope (JWST; \citealt{gardner23}) has dramatically changed this situation. 
Its NIR sensitivity and spectral coverage enable us to investigate rest-frame optical nebular lines in high-redshift dusty galaxies. 
Early JWST/Near Infrared Spectrograph (NIRSpec; \citealt{jakobsen22}) observations have started to probe the ionized gas conditions and central powering sources \commentred{of} bright SMGs (e.g., \citealt{arribas24, jones24, ubler24-GN20, jones25, parlanti24, parlanti25, cooper25} and see also JWST imaging studies, e.g., \citealt{cheng23, rujopakarn23, boogaard24, gillman24, bodansky25, hodge25, ikeda25, mckay25, liu26}). 
% However, 
% %existing samples remain 
% \commentred{many previous spectroscopic studies have focused on individually targeted bright SMGs}
% % heterogeneous 
% and limited in size. A systematic census of rest-frame optical spectra for ALMA flux-limited galaxies spanning a broad range of redshifts and IR luminosities is still lacking. 
However, 
\commentred{many previous JWST spectroscopic studies have focused on individually targeted bright SMGs, and their samples have been limited in size.} 
A systematic census of rest-frame optical spectra for galaxies drawn from an ALMA flux-limited sample, spanning a broad range of redshifts and IR luminosities, is still lacking.

This paper presents \commentred{the} first census of rest-frame optical to NIR JWST/NIRSpec spectra for 16 ALMA flux-limited dust continuum sources in the GOODS-S/HUDF field at $z_\mathrm{spec}=1$--$4$ with IR luminosity spanning $\log (L_\mathrm{IR}/L_\odot)\sim11$--$13$. 
We integrate these datasets to examine their rest-frame optical spectral features, ISM conditions, and origins of the dust continuum \commentred{emission}. 

The structure of this paper is as follows. 
Section~\ref{sec:data} describes our sample and datasets. 
Section~\ref{sec:analysis} shows our analysis of the spectroscopic and photometric data, including emission-line flux measurements and SED fitting. 
Section~\ref{sec:results} presents the results and the physical properties of the dust-continuum sources. 
In Section~\ref{sec:discussion-SMGs}, we discuss the implications of these results. 
Section~\ref{sec:conclusions} summarizes our main findings. 
Throughout this paper, we assume a \citet{chabrier03} initial mass function and a flat $\Lambda$CDM cosmology with $H_0 = 67.7~\mathrm{km~s^{-1}~Mpc^{-1}}$, $\Omega_m = 0.31$, and $\Omega_\Lambda = 0.69$ (\citealt{planck20}). 
We adopt a solar metallicity of $12 + \log(\mathrm{O/H}) = 8.69$ \citep{asplund09}. 
All magnitudes are in the AB system \citep{oke83}.

\section{Sample and Data} 
\label{sec:data} 

In this section, we describe the galaxy sample and the multi-wavelength datasets used in this study.
Table~\ref{tab:FIR} summarizes the sample and several key measurements from the literature, including dust-continuum flux densities ($S_\mathrm{1.2\,mm}$ and/or $S_\mathrm{1.3\,mm}$), IR luminosities ($L_\mathrm{IR}$), X-ray luminosities ($L_\mathrm{X}$), and molecular gas masses estimated from CO lines ($M_\mathrm{mol,\,CO}$).

\subsection{Galaxy Sample} 
\label{subsec:sample} 

We compile ALMA flux-limited dust continuum sources in the HUDF. 
We adopt the sources detected in the HUDF survey \citep{dunlop17} and the ALMA Spectroscopic Survey in the Hubble Ultra Deep Field Large Program (ASPECS-LP; \citealt{aravena20}) as the parent sample (38 sources in total). 
The rms depths of the HUDF survey and the ASPECS-LP are $S_\mathrm{1.3\,mm}=30~\mu\mathrm{Jy\,beam^{-1}}$ (\citealt{dunlop17}) and $S_\mathrm{1.2\,mm}=9.3~\mu\mathrm{Jy\,beam^{-1}}$ (\citealt{aravena20}), respectively. 
The detection procedures are summarized in \citet{dunlop17} and \citet{aravena20}. 

\begin{figure*}
% \gridline{\fig{figures/fig_LIR_redshift_v4.pdf}{1.0\linewidth}{}} 
\gridline{\fig{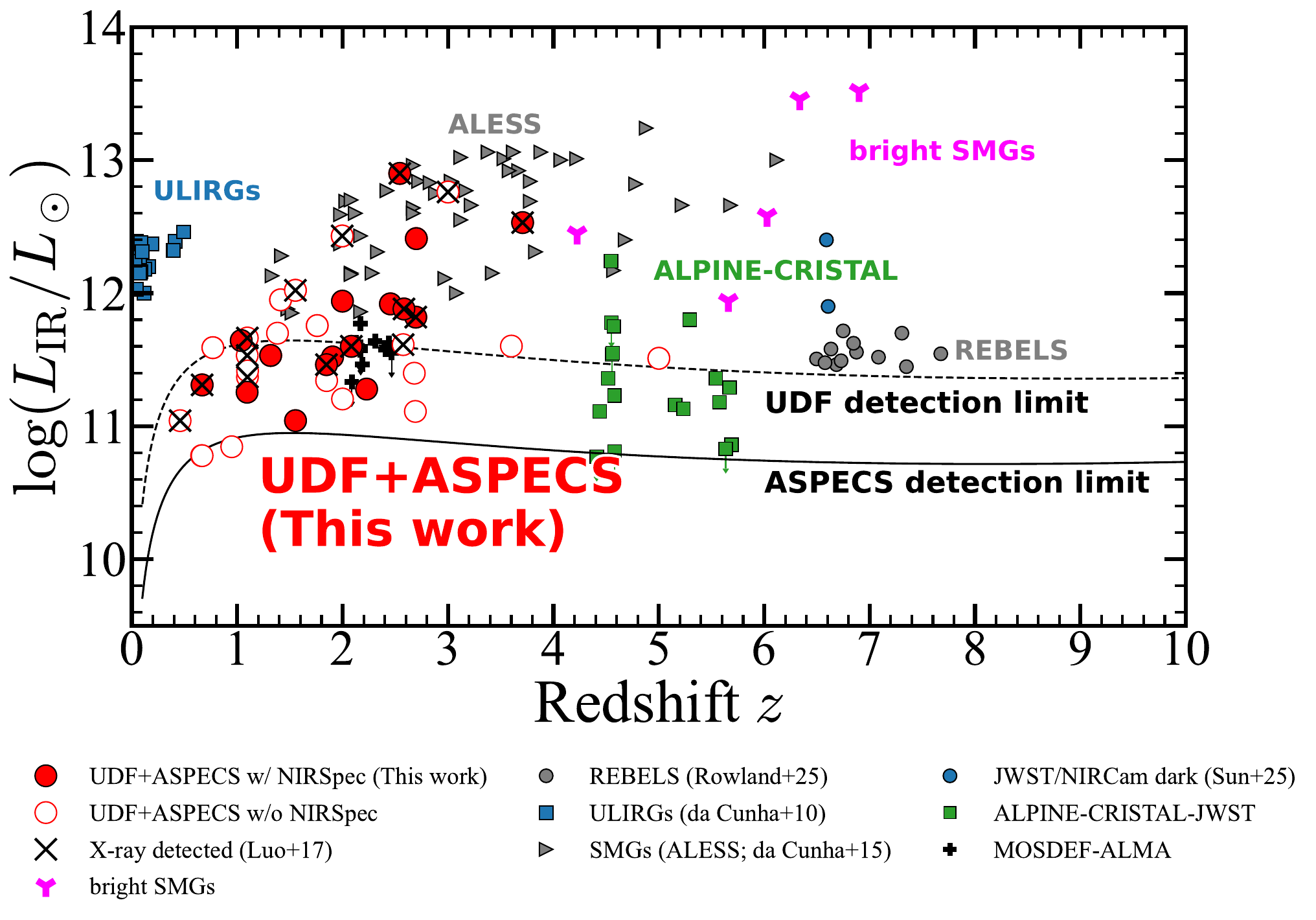}{1.0\linewidth}{}} 
\caption{
Relation between IR luminosity and redshift. 
Dust-continuum sources in HUDF are shown as open red circles (UDF: \citealt{dunlop17}; ASPECS: \citealt{aravena20}). 
Filled red circles indicate ALMA sources with JWST/NIRSpec (UDF+ASPECS sample in this study). 
The black crosses mark X-ray detections \citep{luo17}. 
Comparison samples are local ULIRGs at $z<0.5$ \citep[blue squares;][]{dacunha10}, ALESS \citep[gray triangles;][]{hodge13,dacunha15}, JWST/NIRCam dark galaxies at $z=6.6$ \citep[blue circles;][]{sun25}, 
\commentred{MOSDEF-ALMA \citep[black
plus symbols;][]{sanders23-co,popping23}, ALPINE-CRISTAL observed with JWST \citep[green squares;][]{bethermin20, mitsuhashi24, faisst26}} 
and REBELS \citep[gray circles;][]{bouwens22,rowland25}. 
Some high-redshift bright SMGs are also highlighted as magenta (SPT0311-58: e.g., \citealt{marrone18}, HFLS3: e.g., \citealt{riechers13}, HZ10: e.g., \citealt{capak15}, SPT0418-47: e.g.,  \citealt{cathey24}, \commentred{G09.83808 at $z=6.03$: e.g., \citealt{zavala18}}). 
The black solid and dashed curves show the $L_{\rm IR}$ corresponding to the ASPECS ($S_{\rm 1.2\,mm}>0.03\,{\rm mJy}$; $3.3\sigma$) and HUDF ($S_{\rm 1.3\,mm}>0.12\,{\rm mJy}$; $3.5\sigma$) detection limits, respectively, assuming dust temperature of $T_{\rm dust}=40$ K and emissivity index of $\beta_{\rm IR}=1.5$. 
}
\label{fig:LIR-redshift}
\end{figure*}

From this parent sample, we focus on 16 ALMA sources observed with the JWST/NIRSpec micro-shutter assembly (MSA) (hereafter, the UDF+ASPECS sample). 
Figure~\ref{fig:LIR-redshift} shows the relation between the IR luminosities and redshift of the sample (\commentred{filled} red: UDF+ASPECS sample). 
The IR luminosities in this paper are taken from the literature (\citealt{dunlop17, ueda18, aravena20}) and summarized in Table~\ref{tab:FIR}. 
The resulting UDF+ASPECS sample covers a wide redshift ($z=1$--$4$) and IR luminosity range ($\log(L_\mathrm{IR}/L_\odot)=11$--$13$). 
For comparison, we also show other dusty or main-sequence galaxies, including local ($z<0.5$) ultra luminous infrared galaxies (ULIRGs), the LABOCA Extended Chandra Deep Field South Survey \commentred{with ALMA} (ALESS; \citealt{hodge13,dacunha15}), \commentred{the MOSFIRE Deep Evolution Field (MOSDEF)-ALMA \citep{sanders23-co, popping23}, ALPINE-CRISTAL-JWST sample, which combines observations from the ALMA Large Program to INvestigate C$^{+}$ at Early Times (ALPINE), the [C\,{\sc ii}] Resolved ISM in STar-forming galaxies with ALMA (CRISTAL) survey, and JWST/NIRSpec \citep{bethermin20, mitsuhashi24, faisst26}}, the Reionization Era Bright Emission Line Survey (REBELS; \citealt{bouwens22}), JWST/NIRCam dark galaxies \citep{sun25}, and other bright SMGs (SPT0311-58: e.g., \citealt{marrone18}, HFLS3: e.g., \citealt{riechers13}, HZ10: e.g., \citealt{capak15}, SPT0418-47: e.g.,  \citealt{cathey24}, \commentred{G09.83808: e.g.,} \citealt{zavala18}). 
\commentred{Although these samples are not all redshift-matched to UDF+ASPECS, we include them because some of them provide published measurements of metallicity with JWST/NIRSpec (ALPINE-CRISTAL-JWST, REBELS-IFU), dust, and gas properties that are useful for interpreting the ISM properties of the UDF+ASPECS sources in comparison with other dusty star-forming galaxies.}

\begin{figure}
    \centering
    \includegraphics[width=1.0\linewidth]{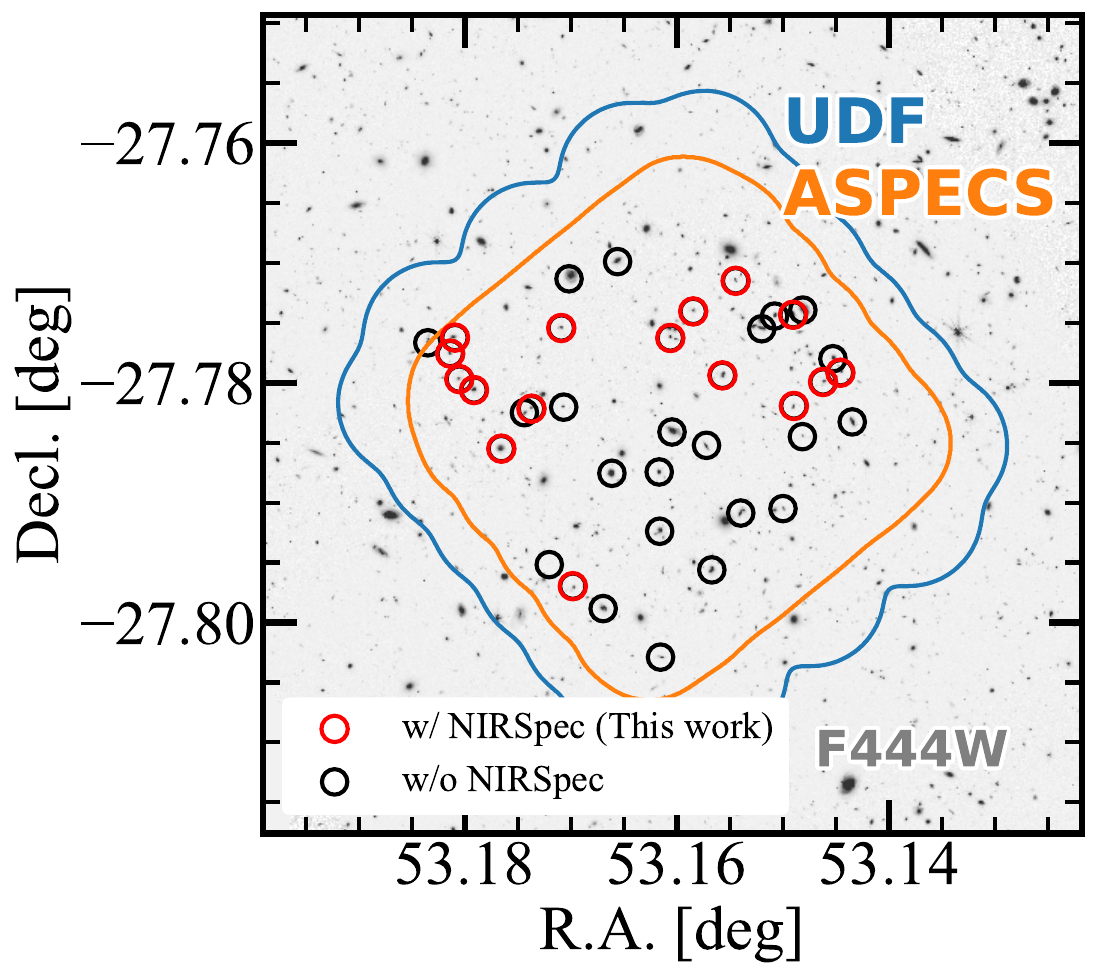}
    \caption{
    UDF \citep{dunlop17} and ASPECS \citep{aravena20} sky regions and the dust continuum source positions overlaid on the JWST/NIRCam F444W image. 
    The blue and orange regions represent the survey areas of UDF and ASPECS, respectively.
    ALMA dust-continuum sources reported in the UDF and the ASPECS are shown as black open circles. The red open circles indicate the UDF or ASPECS sources with JWST/NIRSpec (UDF+ASPECS sample in this study). 
    }
    \label{fig:sky-footprint}
\end{figure}

\begin{figure}
    \centering
    % \includegraphics[width=1.0\linewidth]{figures/fig_MUV_Sv_v1.pdf}
    % \plotone{figures/fig_MUV_Sv_v1.pdf}
    % \plotone{figures/fig_MUV_Sv_v2.pdf}
    \plotone{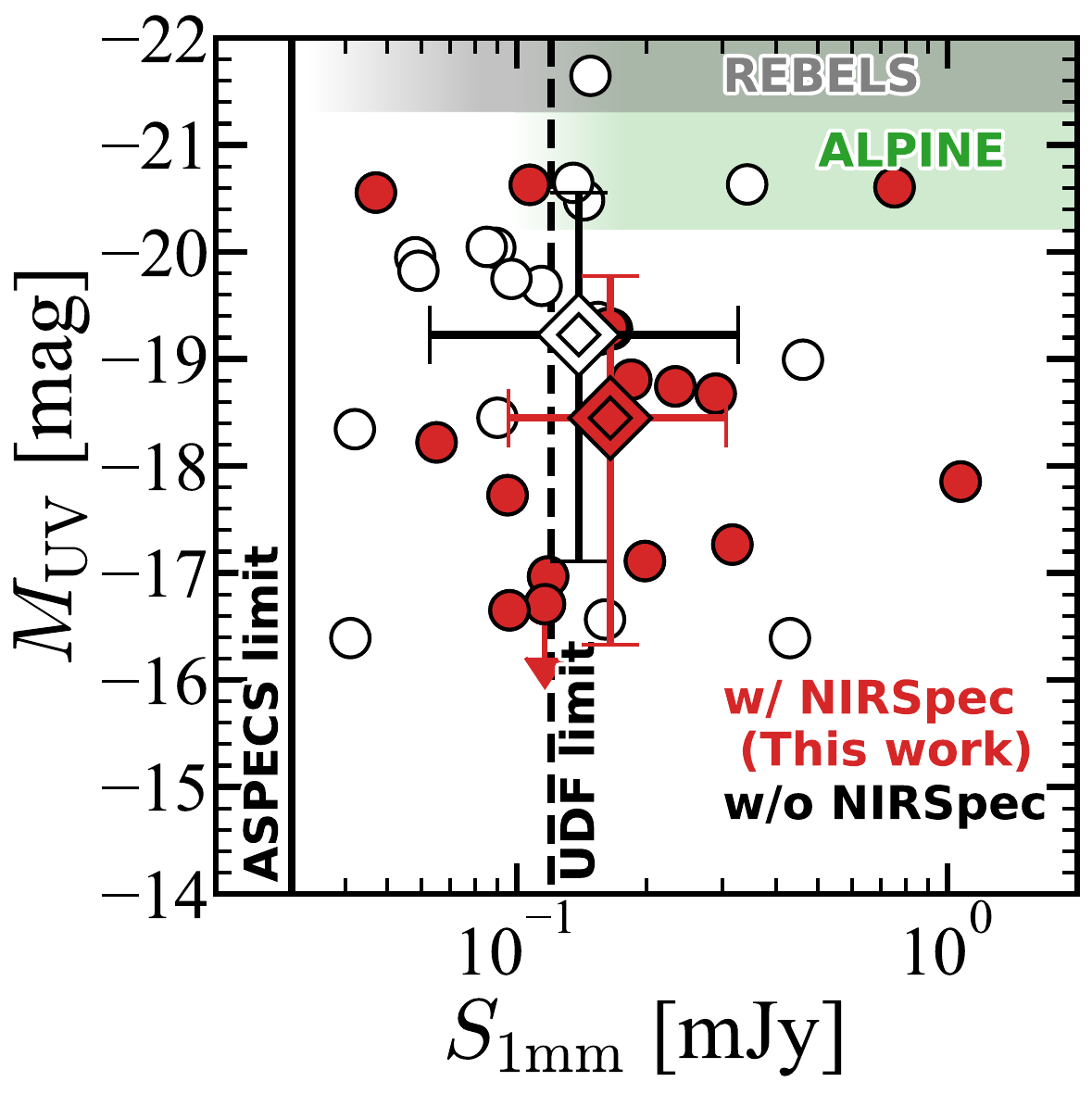}
    \caption{
    Relation between UV absolute magnitude ($M_\mathrm{UV}$) and 1\,mm flux density ($S_\mathrm{1\,mm}$).
    Dust-continuum sources reported in the UDF survey \citep{dunlop17} and the ASPECS survey \citep{aravena20} are shown as open circles (parent sample). 
    Filled red circles indicate the sources with JWST/NIRSpec in the parent sample (UDF+ASPECS sample). 
    If sources are detected in both ASPECS (1.2 mm) and UDF (1.3 mm), we plot the 1-mm flux density reported in the ASPECS \citep{aravena20}. 
    Sources with non-detections in the rest-frame UV are shown as $3\sigma$ lower limits on $M_\mathrm{UV}$ as arrows. 
    The black and red double diamonds show the medians for the parent sample and the UDF+ASPECS sample, respectively. 
    The associated error bars indicate the 16th–84th percentile ranges.
    The black solid and dotted lines show the ASPECS ($S_{\rm 1.2\,mm}>0.03~{\rm mJy}$; $3.3\sigma$) and UDF ($S_{\rm 1.3\,mm}>0.12~{\rm mJy}$; $3.5\sigma$) detection limits, respectively. 
    \commentred{The shaded regions illustrate the UV selection limits of REBELS (gray; $M_{\mathrm{UV}}<-21.3$) and ALPINE (green; $M_{\mathrm{UV}}<-20.2$), together with representative continuum sensitivities of the parent surveys \citep{inami22, faisst20}. The fading spans approximate point-source flux thresholds. Flux densities and sensitivity thresholds are shown at their native observing wavelengths, without conversion to a common $1\,\mathrm{mm}$ wavelength.}
    }
    \label{fig:MUV-S}
\end{figure}

Figure~\ref{fig:sky-footprint} shows the sky distribution of the dust-continuum sources. 
Figure~\ref{fig:MUV-S} presents the relation between the UV absolute magnitude ($M_\mathrm{UV}$) and the 1\,mm flux density ($S_\mathrm{1\,mm}$). 
$M_\mathrm{UV}$ is calculated by following \citet{xu25} using the photometry described in Section~\ref{subsec:uv-optical-phot}. 
To obtain the magnitude, we use the photometry of the filter whose central wavelength
is closest to the rest-frame wavelength of 1500\,\AA\ in the HST+JWST photometric catalog used in this study (see Section~\ref{subsec:uv-optical}). 
In Figure~\ref{fig:MUV-S}, we compare the full dust-continuum sample reported by \citet{dunlop17} and \citet{aravena20} with the subset observed by NIRSpec (UDF+ASPECS sample). 
We find no clear trend between $M_\mathrm{UV}$ and $S_\mathrm{1\,mm}$ within the UDF+ASPECS sample. 
\commentred{The UDF+ASPECS sample probes a region of the $M_\mathrm{UV}$--$S_\mathrm{1\,mm}$ parameter space complementary to those probed by the REBELS survey (gray shaded region in Figure~\ref{fig:MUV-S}) and the ALPINE (green shaded region). Whereas REBELS and ALPINE primarily target UV-bright galaxies, the UDF+ASPECS sample is drawn from blind, flux-limited millimeter-continuum surveys and includes UV-faint galaxies as well.}
Sections~\ref{subsec:uv-optical} and \ref{subsec:x-ray} describe the details of the data and cross-matching procedures.

%%%% LIR vs. redshift %%%%%%%
%%%%%%%%%%%%%%%%%%%%%%%%%

%%%% fig: FoV %%%%%%%%%%%
%%%%%%%%%%%%%%%%%%%%%%%%%

%%% fig: M_UV vs. S_1mm %%
%%%%%%%%%%%%%%%%%%%%%%%%%%%

\subsection{UV to Infrared Data} 
\label{subsec:uv-optical} 

\subsubsection{Spectroscopy} 
\label{subsec:uv-optical-spec} 

For the spectroscopic data of the UDF+ASPECS sample, we use publicly available NIRSpec-MSA medium-resolution grating data ($R\sim1000$) from the JWST Advanced Deep Extragalactic Survey (JADES; \citealt{eisenstein23a, rieke23-jadesDR1, deugenio25, eisenstein25, curtis-Lake25, scholtz25}) data release (DR) 4 \citep{curtis-Lake25, scholtz25} and the Systematic Mid-Infrared Instrument Legacy Extragalactic Survey (SMILES; \citealt{alberts24, rieke24, zhu25}) DR2 \citep{zhu25}. 

The JADES DR4 provides the NIRSpec-MSA spectra for 5190 sources in the GOODS field. The observations were conducted with a low-dispersion prism ($R\sim100$), three medium-resolution gratings ($R\sim1000$), and, for a subset, the higher-resolution G395H grating ($R\sim2700$). From JADES DR4, we use medium-resolution grating data covering $1$--$5~\mu\mathrm{m}$. The SMILES DR2 provides NIRSpec-MSA medium-resolution grating spectra for 166 galaxies at $0<z<7.5$ in the HUDF. The observations and reduction processes are described in \citet{curtis-Lake25, scholtz25} and \citet{zhu25} for JADES and SMILES, respectively. We use the publicly released reduced spectra in our analysis. 

We cross-match NIRSpec-MSA spectra to ALMA sources using a $0\farcs5$ radius. 
The matched source IDs are listed in Table~\ref{tab:FIR}. 
When both JADES and SMILES spectra are available, we use the spectrum that covers more of the key diagnostic lines (H$\beta$, \oiii, H$\alpha$, \nii, and \sii). 
Table~\ref{tab:emission-lines} includes a flag indicating whether the adopted spectrum is from JADES or SMILES. 
We note that UDF10 shows a $0\farcs54$ offset between the JWST counterpart and the dust-continuum peak (see Figure~\ref{fig:snapshots2}). 
We treat these as the same physical source and proceed with our analysis under this assumption, as \citet{dunlop17}. 
UDF12 has spectra in JADES DR4, but these spectra lack emission lines or continuum, and there are no counterparts at the slit position. 
We exclude UDF12 from our final sample. 
UDF12 is originally reported to have a photometric redshift of $z_\mathrm{phot}=5.0$ \citep{dunlop17}.

\subsubsection{Photometry} 
\label{subsec:uv-optical-phot} 

For the rest-frame UV–optical photometry of the UDF+ASPECS sample, we use the public JADES DR2 photometric catalog (hereafter, the JADES photometric catalog; \citealt{eisenstein25}). 
The JADES photometric catalog provides the HST and JWST Near Infrared Camera (NIRCam; \citealt{rieke23}) imaging in the GOODS-S/HUDF and incorporates the NIRCam medium-band data from the JWST Extragalactic Medium-band Survey (JEMS; \citealt{williams23}). 
We use the eight HST broadbands (Advanced Camera for Surveys F435W, F606W, F775W, F814W; Wide Field Camera 3 (WFC3) F105W, F125W, F140W, F160W) and 13 NIRCam broad and medium bands, from F090W to F444W (F090W, F115W, F182M, F200W, F210M, F277W, F335M, F356W, F410M, F430M, F444W, F460M, F480M). 
The details of the image reduction and catalog construction are described in \citet{rieke23-jadesDR1} and \citet{eisenstein25}. 

To extend our wavelength coverage into the mid-infrared, we complement the JADES photometric catalog together with the SMILES DR1 photometric catalog, which provides JWST/Mid-Infrared Instrument (MIRI; \citealt{bouchet15, rieke15, wright23}) broadband imaging in the GOODS-S (see \citealt{alberts24} for details). 
From this catalog, we use the eight MIRI broadbands from F560W to F2550W (F560W, F770W, F1000W, F1280W, F1500W, F1800W, F2100W, F2550W). 

For both catalogs, we use the 2.5$\times$-scaled Kron-aperture fluxes measured on raw-resolution images that are not PSF-matched. 
We adopt the associated random-aperture uncertainties provided in the catalogs as flux uncertainties. 
We do not correct the photometry for Galactic extinction. 
The expected extinction in the GOODS-S/HUDF is very small (typically $E(B-V) \lesssim 0.01$; e.g., \citealt{schlafly11}, see also \citealt{iani24} for the correction factors of the NIRCam and MIRI bands). 
Its impact on our results is negligible. 

We also utilize Herschel \citep{pilbratt10} photometry from the FourStar Galaxy Evolution Survey (ZFOURGE) catalog \citep{straatman16}. 
We use the Herschel/Photodetector Array Camera and Spectrometer (PACS; \citealt{poglitsch10}) green (100~$\mu\mathrm{m}$) and red (160~$\mu\mathrm{m}$) band data to constrain the shorter-wavelength dust continuum. 

We associate these photometric catalogs with the ALMA dust-continuum sources by position. 
Starting from the ALMA continuum peak coordinates reported by \citet{dunlop17} and \citet{aravena20}, we search for the NIRCam and MIRI counterparts (Herschel counterparts) within a radius of $0\farcs5$ ($1\farcs0$). 
All sources with NIRSpec data are cross-matched with NIRCam and MIRI photometry. 
The matched catalog IDs are listed in Table~\ref{tab:FIR}. 
These data are used for the SED fitting described in Section~\ref{subsec:SED-fitting}.

\subsection{X-ray Data} 
\label{subsec:x-ray}

In the GOODS-S field, ultra-deep Chandra observations have been accumulated over the past two decades (e.g., \citealt{giacconi02, lehmer05, xue11, luo17}).
For the UDF+ASPECS sample, we make use of the Chandra Deep Field-South (CDF-S) 7~Ms source catalog presented by \citet{luo17} (see also \citealt{ueda18}). 
We use these X-ray detections to identify AGNs among our dust-continuum sources (Section~\ref{subsec:AGN-SMGs}). 
We cross-match the 16 ALMA sources with JWST/NIRSpec spectra (UDF+ASPECS sample) to the CDF-S 7~Ms catalog within $1\arcsec$, yielding seven matches. 
The CDF-S 7~Ms source catalog \citep{luo17} provides absorption-corrected $0.5$--$7~\mathrm{keV}$ X-ray luminosities. 
We summarize the corresponding source IDs (CID) and X-ray luminosities in Table~\ref{tab:FIR}.

\begin{deluxetable*}{lccccccccccccc}
    \tablecaption{Far-infrared and X-Ray Properties of the UDF+ASPECS Sources. \label{tab:FIR}}
    \tablewidth{0pt}
    \tablehead{
    \colhead{ID} & \colhead{ID} & \colhead{ID} & \colhead{ID} & \colhead{ID} & \colhead{CID} & \colhead{R.A.} & \colhead{Decl.} & \colhead{Redshift} & \colhead{$S_\mathrm{1.3\,mm}$} & \colhead{$S_\mathrm{1.2\,mm}$} & \colhead{$\log L_\mathrm{IR}$} & \colhead{$\log L_\mathrm{X}$} & \colhead{$M_\mathrm{mol,\,CO}$}  \\
    \colhead{(UDF)} & \colhead{(ASPECS)} & \colhead{(NIRSpec)} & \colhead{(NIRCam)} & \colhead{(MIRI)} & \colhead{} & \colhead{(J2000)} & \colhead{(J2000)} & \colhead{} & \colhead{($\mu$Jy)} & \colhead{($\mu$Jy)} & \colhead{($L_\odot$)} & \colhead{(erg s$^{-1}$)} & \colhead{($10^{10}\,M_\odot$)} 
    } 
    \colnumbers
    \startdata
    UDF2 & C06 & \nodata & 208820 & 1257 & \nodata & $53.18137$ & $-27.77757$ & 2.697 & 996$\pm$87 & 1071$\pm$47 & 12.41$^{+0.10}_{-0.10}$$^a$ & \nodata & 11.0$\pm$1.3 \\
    UDF3 & C01 & 14279 & 209117 & 1271 & 718 & $53.16062$ & $-27.77627$ & 2.543 & 863$\pm$84 & 752$\pm$24 & 12.90$^{+0.04}_{-0.05}$$^a$ & 42.7 & 13.3$\pm$0.5 \\
    UDF4 & C04 & 209357 & 209357 & 1284 & \nodata & $53.17090$ & $-27.77544$ & 2.453 & 303$\pm$46 & 316$\pm$12 & 11.92$^{+0.13}_{-0.16}$$^a$ & \nodata & 5.0$\pm$0.5 \\
    UDF7 & C11 & \nodata & 208030 & 1231 & 797 & $53.18051$ & $-27.77970$ & 2.694 & 231$\pm$48 & 289$\pm$21 & 11.82$^{+0.20}_{-0.19}$$^a$ & 42.6 & \nodata \\
    UDF9 & \nodata & 209108 & 209108 & 1272 & 799 & $53.18092$ & $-27.77624$ & 0.668 & 198$\pm$39 & \nodata & 11.31$^{+0.48}_{-0.48}$$^b$ & 40.9 & \nodata \\
    UDF10 & \nodata & 50039680 & 203108 & 1014 & 756 & $53.16981$ & $-27.79697$ & 2.084 & 184$\pm$46 & \nodata & 11.60$^{+0.22}_{-0.22}$$^b$ & 42.4 & \nodata \\
    UDF13 & C07 & 51236 & 208000 & 1225 & 655 & $53.14622$ & $-27.77994$ & 2.583 & 174$\pm$45 & 233$\pm$12 & 11.88$^{+0.05}_{-0.07}$$^a$ & 42.4 & \nodata \\
    UDF15 & \nodata & 10007444 & 124908 & 1200 & \nodata & $53.14897$ & $-27.78194$ & 1.904 & 166$\pm$46 & \nodata & 11.52$^{+0.31}_{-0.31}$$^b$ & \nodata & \nodata \\
    UDF16 & C15 & \nodata & 206183 & 1156 & \nodata & $53.17655$ & $-27.78550$ & 1.317 & 155$\pm$44 & 118$\pm$13 & 11.53$^{+0.05}_{-0.05}$$^a$ & \nodata & 4.7$\pm$0.4 \\
    \nodata & C08 & 209777 & 209777 & 1305 & 715 & $53.15846$ & $-27.77406$ & 3.711 & \nodata & 163$\pm$10 & 12.53$^{+0.17}_{-0.05}$$^a$ & 43.7 & \nodata \\
    \nodata & C13 & \nodata & 207739 & 1773 & \nodata & $53.17912$ & $-27.78061$ & 1.038 & \nodata & 116$\pm$16 & 11.64$^{+0.04}_{-0.05}$$^a$ & \nodata & 3.7$\pm$0.4 \\
    \nodata & C14a & \nodata & 207227 & 1196 & \nodata & $53.17371$ & $-27.78217$ & 1.997 & \nodata & 96$\pm$10 & 11.94$^{+0.08}_{-0.14}$$^a$ & \nodata & \nodata \\
    \nodata & C18 & 208134 & 208134 & 1236 & 703 & $53.15571$ & $-27.77939$ & 1.846 & \nodata & 107$\pm$16 & 11.46$^{+0.16}_{-0.12}$$^a$ & 42.2 & \nodata \\
    \nodata & C20 & \nodata & 209480 & 1301 & \nodata & $53.14904$ & $-27.77433$ & 1.094 & \nodata & 95$\pm$16 & 11.26$^{+0.07}_{-0.05}$$^a$ & \nodata & \nodata \\
    \nodata & C26 & 208267 & 208267 & 1237 & \nodata & $53.14458$ & $-27.77917$ & 1.552 & \nodata & 65$\pm$15 & 11.04$^{+0.13}_{-0.09}$$^a$ & \nodata & \nodata \\
    \nodata & C31 & 10008071 & 210730 & 1345 & \nodata & $53.15446$ & $-27.77150$ & 2.227 & \nodata & 47$\pm$12 & 11.28$^{+0.49}_{-0.05}$$^a$ & \nodata & \nodata \\
    \enddata
    \tablecomments{
    Far-infrared properties. 
    (1) UDF source ID \citep{dunlop17}; 
    (2) ASPECS ID \citep{aravena20}; 
    (3) JADES NIRSpec ID in the JADES DR4 spectroscopic catalog \citep{curtis-Lake25, scholtz25}; 
    (4) NIRCam ID in JADES DR2 photometric catalog \citep{eisenstein25}; 
    (5) MIRI ID in SMILES DR1 catalog \citep{zhu25}; 
    (6) Chandra source ID in the CDF-S 7~Ms source catalog \citep{luo17}; 
    (7) and (8) ALMA source position (J2000) based on the dust continuum detection \citep{dunlop17, aravena20}; 
    (9) redshift measured from JWST/NIRSpec spectra; 
    (10) total flux densities at 1.3~mm (UDF; \citealt{dunlop17}); 
    (11) total flux densities at 1.2~mm (ASPECS; \citealt{aravena20}); 
    (12) infrared luminosity: $^a$ based on \citet{aravena20}; $^b$ based on \citet{dunlop17, ueda18};  
    (13) absorption-corrected intrinsic 0.5--7.0~keV luminosity \citep{luo17}; 
    (14) \commentred{molecular} gas mass estimated from CO lines \citep{aravena19, aravena20}. \\
    }
\end{deluxetable*}

\begin{deluxetable*}{lccccccccccc}
    \tablecaption{Rest-frame Optical Emission-line Flux Measurements. \label{tab:emission-lines}}
    \tablewidth{0pt}
    \tablehead{
    \colhead{ID} & \colhead{ID} & \colhead{NIRSpec ID} & \colhead{ID} & \colhead{Redshift} & \colhead{H$\beta$} & \colhead{\oiii} & \colhead{H$\alpha$} & \colhead{\nii} & \multicolumn{2}{c}{\sii} & \colhead{Flag} \\
    \cline{10-11}
    \colhead{(UDF)} & \colhead{(ASPECS)} & \colhead{(JADES)} & \colhead{(MIRI)} & \colhead{} & \colhead{} & \colhead{$\lambda$5007} & \colhead{} & \colhead{$\lambda$6584} & \colhead{$\lambda$6717} & \colhead{$\lambda$6731} & \colhead{}
    } 
    \startdata
    UDF2 & C06 & \nodata & 1257 & 2.697 & 10.1$\pm$2.5 & 20.9$\pm$2.4 & 55.0$\pm$1.5 & 30.3$\pm$1.4 & 12.9$\pm$1.3 & 9.2$\pm$1.2 & S \\
    UDF3 & C01 & 14279 & 1271 & 2.543 & 69.0$\pm$4.6 & 100.1$\pm$4.1 & 289.7$\pm$3.6 & 74.3$\pm$2.6 & 34.0$\pm$2.2 & 26.6$\pm$2.1 & J \\
    UDF4 & C04 & 209357 & 1284 & 2.453 & 18.5$\pm$2.7 & 14.9$\pm$2.5 & 117.6$\pm$2.2 & 59.5$\pm$1.8 & 16.8$\pm$1.3 & 17.7$\pm$1.4 & S \\
    UDF7 & C11 & \nodata & 1231 & 2.694 & 23.1$\pm$3.8 & 41.8$\pm$3.4 & 148.4$\pm$3.1 & 78.5$\pm$2.9 & 36.1$\pm$5.4 & 18.7$\pm$2.8 & S \\
    UDF9 & \nodata & 209108 & 1272 & 0.668 & 37.7$\pm$11.9 & $<26.5$ & 317.9$\pm$8.9 & 116.7$\pm$5.2 & 58.7$\pm$4.2 & 54.2$\pm$4.2 & J \\
    UDF10 & \nodata & 50039680 & 1014 & 2.084 & 52.2$\pm$2.7 & 97.5$\pm$3.4 & 419.9$\pm$4.8 & 140.8$\pm$3.3 & 36.6$\pm$2.7 & 33.3$\pm$2.2 & S \\
    UDF13 & C07 & 51236 & 1225 & 2.583 & $<7.4$ & 9.0$\pm$2.2 & 76.3$\pm$1.9 & 80.3$\pm$1.9 & 14.1$\pm$1.5 & 13.0$\pm$1.4 & J \\
    UDF15 & \nodata & 10007444 & 1200 & 1.904 & 17.9$\pm$1.5 & 44.6$\pm$2.0 & 89.7$\pm$1.6 & 13.7$\pm$1.3 & 14.0$\pm$1.2 & 11.5$\pm$1.2 & J \\
    UDF16 & C15 & \nodata & 1156 & 1.317 & $<6.1$ & $<6.1$ & 96.6$\pm$2.7 & 105.4$\pm$3.4 & 12.6$\pm$2.4 & 11.9$\pm$2.4 & S \\
    \nodata & C08\textsuperscript{a} & 209777 & 1305 & 3.711 & 17.3$\pm$2.5 & 88.9$\pm$2.6 & 105.6$\pm$3.6 & 53.0$\pm$3.3 & $<5.6$ & 7.5$\pm$1.7 & J \\
    \nodata & C13 & \nodata & 1773 & 1.038 & 8.8$\pm$2.5 & $<5.9$ & 203.1$\pm$2.5 & 110.4$\pm$2.2 & 28.2$\pm$1.7 & 23.5$\pm$1.6 & S \\
    \nodata & C14a & \nodata & 1196 & 1.997 & $<4.6$ & $<4.8$ & 44.2$\pm$1.6 & 32.1$\pm$1.5 & 11.8$\pm$1.4 & 13.2$\pm$1.4 & S \\
    \nodata & C18 & 208134 & 1236 & 1.846 & 19.5$\pm$2.7 & 17.4$\pm$2.8 & 96.6$\pm$3.1 & 38.3$\pm$2.7 & 15.7$\pm$2.6 & 14.9$\pm$2.4 & J \\
    \nodata & C20 & \nodata & 1301 & 1.094 & $<18.5$ & $<16.6$ & 181.4$\pm$7.1 & 88.6$\pm$6.0 & $<17.0$ & $<17.2$ & S \\
    \nodata & C26 & 208267 & 1237 & 1.552 & 30.8$\pm$2.0 & 21.8$\pm$1.9 & 229.6$\pm$3.5 & 67.8$\pm$3.0 & 35.1$\pm$2.9 & 23.7$\pm$2.6 & J \\
    \nodata & C31 & 10008071 & 1345 & 2.227 & 89.6$\pm$2.3 & 396.9$\pm$3.7 & 359.4$\pm$2.9 & 35.4$\pm$1.6 & 39.8$\pm$1.4 & 30.9$\pm$1.5 & S \\
    \enddata
    \tablecomments{
    Emission-line flux measurements of NIRSpec medium-resolution data. The fluxes are not corrected for dust extinction and are in units of $10^{-19}\,\mathrm{erg\,s^{-1}\,cm^{-2}}$. Errors are the $1\sigma$ uncertainties. Upper limits are $3\sigma$ levels. 
    The flag indicates which dataset is used for the emission-line flux measurements (J: slit-loss-corrected JADES DR4 spectra extracted with a 5-pixel aperture and 3-nod pattern; S: SMILES DR2 spectra without slit-loss correction; see \citealt{zhu25} for details of the SMILES spectra). \\
    \textsuperscript{a}: C08 shows the prominent broad Balmer lines. Only the narrow line fluxes are shown (Figure~\ref{fig:snapshots3}; see also \citealt{juodbalis25}). 
    }
\end{deluxetable*}

\begin{deluxetable*}{lcccccccccccc}
    \tablecaption{Physical Properties of the UDF+ASPECS Sources. \label{tab:properties}}
    \tablewidth{0pt}
    \tablehead{
    \colhead{ID} & \colhead{ID} & \colhead{$\log M_*$} & \colhead{$\log\mathrm{SFR}$} & $E(B-V)$ & \colhead{\sii$6717/6731$} & \colhead{$n_e$(\sii)} & \multicolumn{2}{c}{$12+\log(\mathrm{O/H})$ (N2)} & \colhead{BPT} & \colhead{\commentred{BPT}} &\colhead{X-ray} & \colhead{AGN} \\
    \cline{8-9}
    \colhead{(UDF)} & \colhead{(ASPECS)} & \colhead{$(M_\odot)$} & \colhead{($M_\odot~\mathrm{yr^{-1}}$)} & (mag) & \colhead{}  & \colhead{($\mathrm{cm^{-3}}$)} & \colhead{(S21)} & \colhead{(C20)} & \colhead{\commentred{(K03)}} & \colhead{\commentred{(K01)}} & \colhead{} & \colhead{}
    } 
    \colnumbers
    \startdata
    UDF2 & C06 & 10.89$\pm$0.04 & 2.48$\pm$0.05 & 0.57$\pm$0.21 & 1.41$\pm$0.23 & $28_{-28}^{+339}$ & 9.01 & 8.54 & T & F & F & T \\
    UDF3 & C01 & 9.80$\pm$0.02 & 2.81$\pm$0.02 & 0.35$\pm$0.06 & 1.28$\pm$0.13 & $193_{-168}^{+229}$ & 8.70 & 8.34 & F & F & T & T \\
    UDF4 & C04 & 10.06$\pm$0.24 & 1.97$\pm$0.06 & 0.70$\pm$0.12 & 0.95$\pm$0.11 & $1014_{-354}^{+544}$ & 8.97 & 8.52 & T & F & F & T \\
    UDF7 & C11 & 10.63$\pm$0.12 & 1.37$\pm$0.54 & 0.71$\pm$0.14 & 1.92$\pm$0.40 & $<494$ & 8.99 & 8.53 & T & F & T & T \\
    UDF9 & \nodata & 9.92$\pm$0.08 & 0.43$\pm$0.05 & 0.95$\pm$0.27 & 1.08$\pm$0.11 & $582_{-247}^{+354}$ & 8.83 & 8.44 & F & F & T & T \\
    UDF10 & \nodata & 10.12$\pm$0.09 & 1.20$\pm$0.17 & 0.90$\pm$0.05 & 1.10$\pm$0.11 & $541_{-231}^{+316}$ & 8.80 & 8.41 & T & F & T & T \\
    UDF13 & C07 & 10.77$\pm$0.02 & 1.38$\pm$0.09 & $>1.12$ & 1.08$\pm$0.17 & $584_{-337}^{+567}$ & \nodata & 8.71 & T & T & T & T \\
    UDF15 & \nodata & 10.05$\pm$0.03 & 1.00$\pm$0.02 & 0.50$\pm$0.07 & 1.22$\pm$0.17 & $294_{-239}^{+367}$ & 8.53 & 8.21 & F & F & F & F \\
    UDF16 & C15 & 10.78$\pm$0.02 & 1.36$\pm$0.07 & $>1.48$ & 1.05$\pm$0.29 & $669_{-560}^{+1550}$ & \nodata & 8.72 & T & F & F & T \\
    \nodata & C08\textsuperscript{a} & 10.86$\pm$0.03 & 2.09$\pm$0.02 & 0.67$\pm$0.13 & \nodata & \nodata & 8.96 & 8.52 & T & T & T & T \\
    \nodata & C13 & 10.47$\pm$0.04 & 1.32$\pm$0.04 & 1.80$\pm$0.25 & 1.20$\pm$0.11 & $330_{-175}^{+231}$ & 9.00 & 8.54 & F & F & F & F \\
    \nodata & C14a & 10.13$\pm$0.02 & 1.62$\pm$0.07 & $>1.06$ & 0.89$\pm$0.14 & $1262_{-550}^{+1018}$ & 9.18 & 8.61 & T & F & F & T \\
    \nodata & C18 & 10.51$\pm$0.05 & 1.51$\pm$0.04 & 0.49$\pm$0.12 & 1.05$\pm$0.24 & $669_{-489}^{+1114}$ & 8.86 & 8.46 & F & F & T & T \\
    \nodata & C20 & 10.64$\pm$0.04 & 0.89$\pm$0.05 & $>1.08$ & \nodata & \nodata & 8.95 & 8.51 & F & F & F & F \\
    \nodata & C26 & 10.16$\pm$0.05 & 1.02$\pm$0.13 & 0.84$\pm$0.06 & 1.48$\pm$0.21 & $<604$ & 8.75 & 8.38 & F & F & F & F \\
    \nodata & C31 & 9.80$\pm$0.06 & 1.60$\pm$0.03 & 0.31$\pm$0.02 & 1.29$\pm$0.08 & $188_{-104}^{+123}$ & 8.40 & 8.10 & F & F & F & F \\
    \enddata
    \tablecomments{
    (1)--(2) source IDs;
    (3) stellar mass in units of solar mass from CIGALE SED fitting;
    (4) SFR averaged in recent 10~Myr in units of $M_\odot~\mathrm{yr^{-1}}$ from CIGALE SED fitting;
    (5) Dust extinction $E(B-V)$ derived from H$\alpha$/H$\beta$ ratio (see Section~\ref{subsec:morphology-SMGs}). Lower limits correspond to the $3\sigma$ levels.
    (6) \sii\ doublet ratio that is not corrected for the dust extinction; 
    (7) electron density derived from \sii\ doublet. Upper limits correspond to the $2\sigma$ levels;
    (8) and (9) metallicity derived from the N2 index with the calibrations of \citet{sanders21} (S21) and \citet{caravalho20} (C20), respectively;
    (10) BPT flag (T: above the \citealt{kauffmann03} curve in the BPT diagram, F: below the \citealt{kauffmann03} curve in the BPT diagram); 
    \commentred{(11) BPT flag (T: above the \citealt{kewley01} curve in the BPT diagram, F: below the \citealt{kewley01} curve in the BPT diagram);}
    (12) X-ray flag (T: X-ray detection, F: no X-ray detection; \citealt{luo17}); 
    (13) AGN flag (T: X-ray detection or BPT diagram above \citealt{kauffmann03} curve, F: non-AGN). \\
    \textsuperscript{a}: C08 shows broad components in \oiii\ and the Balmer lines (Figure~\ref{fig:snapshots3}). \commentred{The $E(B-V)$ and metallicity of C08 are based on the narrow-line fluxes.} Because their contribution to the \sii\ doublet may not be reliably separated, we conservatively do not report an electron density. 
    }
\end{deluxetable*}

\section{Analysis} 
\label{sec:analysis} 

\subsection{Emission-line Fitting and Flux Measurements} 
\label{subsec:flux-measurements}

We measure the key optical emission-line fluxes of H$\beta$, \oiii$\lambda\lambda4959, 5007$, H$\alpha$, \nii$\lambda\lambda6548, 6584$, and \sii$\lambda\lambda6717, 6731$ using the JWST/NIRSpec-MSA spectra. 
Emission-line fluxes are derived by least-squares fitting of Gaussian profiles and a linear continuum, implemented with the \texttt{scipy.optimize} package \citep{virtanen20}. 
We use the noise spectrum to weight the fit. 
\commentred{The line flux is defined as the integrated area of the best-fit Gaussian profile.}
In the Gaussian fitting, we generally model each emission line with a single Gaussian profile. 

For each spectrum, we simultaneously fit the main nebular emission lines with Gaussians. 
H$\beta$ and \oiii$\lambda\lambda4959, 5007$ are fit together with a common line width and redshift, and we fix the flux ratio $f_\mathrm{[O\,III]\lambda5007}/f_\mathrm{[O\,III]\lambda4959} = 2.98$ \citep{storey00} to reduce the number of free parameters. 
Likewise, H$\alpha$, \nii$\lambda\lambda6548, 6584$, and \sii$\lambda\lambda6717, 6731$ are fit simultaneously with a common line width and redshift, and fixing the flux ratio $f_\mathrm{[N\,II]\lambda6584}/f_\mathrm{[N\,II]\lambda6548} = 2.94$. 

We constrain the Gaussian width to be larger than the instrumental broadening, assuming a spectral resolution of $R = 1000$ for the medium-resolution NIRSpec-MSA. 
The velocity dispersion is then calculated assuming $\mathrm{\sigma_{int}} = \sqrt{\mathrm{\sigma_{obs}^2 - \sigma_{inst}^2}}$, where $\mathrm{\sigma_{int}}$, $\mathrm{\sigma_{obs}}$, and $\mathrm{\sigma_{inst}}$ are intrinsic, observed, and instrumental dispersions, respectively. 

When the Balmer lines (e.g., H$\alpha$) show clear broad emission, we fit them with the sum of narrow and broad Gaussian components, while tying the redshift and velocity dispersion of the narrow components across all lines. 
Additionally, if broad components are also present in non-Balmer lines, we also fit an alternative model that includes both narrow and broad Gaussian components for all lines. 
To assess whether broad components are warranted, we use the Akaike Information Criterion (AIC; \citealt{akaike74}):
\begin{equation}
    \mathrm{AIC} = -2\ln(L) + 2k,
\end{equation}
where $L$ is the maximum likelihood and $k$ is the number of free parameters. 
We adopt the model including broad components when $\Delta\mathrm{AIC} \equiv \mathrm{AIC_{without~broad}} - \mathrm{AIC_{with~broad}} > 20$ (e.g., \citealt{harikane23}).

Unless otherwise noted, we adopt the best-fit redshift from the H$\alpha$+\nii+\sii\ narrow lines as the source redshift throughout this paper (see Table~\ref{tab:FIR}). 
For sources without H$\alpha$ coverage or detection, we use the redshift of the H$\beta$+\oiii\ fit instead. 

Flux uncertainties are estimated by summing in quadrature the errors of the spectral bins within a wavelength interval of $2 \times \mathrm{FWHM}$ (full width at half-maximum) centered on the peak of the best-fit Gaussian profile. 
All rest-frame wavelengths are quoted on the vacuum wavelength scale (e.g., 5008.240\,\AA\ for \oiii$\lambda5007$), although we follow the long-standing nomenclature based on their air wavelengths. 
We summarize the emission-line measurements in Table~\ref{tab:emission-lines}. 
The flux measurements are broadly consistent with the JADES and SMILES emission-line flux catalog \citep{curtis-Lake25, scholtz25, zhu25}.

\subsection{SED Fitting} 
\label{subsec:SED-fitting}

One strength of the UDF+ASPECS sample is the rich infrared to FIR photometry covered by JWST/MIRI, Herschel, and ALMA. 
These data constrain the dust-obscured star formation. 
We perform SED fitting with the Code Investigating GALaxy Emission (CIGALE version 2025.0; \citealt{burgarella05, noll09, boquien19, yang20, yang22}) to derive stellar masses and star formation rates (SFR) of the UDF+ASPECS sample. 

We use JADES HST+NIRCam photometry and SMILES MIRI photometry (Section~\ref{subsec:uv-optical-phot}). 
We also include the Herschel photometry from the ZFOURGE catalog and the ALMA dust continuum measurements reported in \citet{dunlop17} and \citet{aravena20} (see Section~\ref{subsec:uv-optical-phot}). 
CIGALE computes model SEDs in an energy-conserving manner and derives physical properties as likelihood-weighted means over the full model grid (a Bayesian-like approach). 
The SED-fitting configuration is summarized below, and Appendix~\ref{appendix:sed-model} lists the full set of parameters. 

We adopt a delayed star-formation history (SFH) with an additional recent instantaneous variation in the SFR \citep{ciesla17}. 
Stellar emission is modeled with the simple stellar population (SSP) templates of \citet{bruzual03}. 
Nebular emission is included using templates based on the photoionization models of \citet{inoue11}. 
Dust attenuation is described by a modified \citet{calzetti00} attenuation law allowing a steeper slope in the UV \citep{leitherer02}. 
Dust emission is modeled with the THEMIS dust model \citep{Jones17}. 
The AGN component is included with the SKIRTOR2016 module \citep{stalevski12, stalevski16}. 
In this analysis, three AGN parameters are treated as free: the average edge-on optical depth at 9.7~$\micron$ ($\tau_\mathrm{9.7}$), the inclination angle ($i$), and the AGN fraction ($f_\mathrm{AGN}$). 

We fix the redshift to the JWST/NIRSpec value (Section~\ref{subsec:flux-measurements}; Table~\ref{tab:FIR}). 
For the HST, JWST/NIRCam, and MIRI photometry, we adopt a 5\% uncertainty floor to account for remaining systematic uncertainties (e.g., \citealt{robertson24, weibel24}). 
For bands with a signal-to-noise ratio (S/N) below three, we treat the measurements as non-detections and use the corresponding $3\sigma$ upper limits in the SED fitting. 
\commentred{These upper limits are incorporated in the likelihood calculation but are not imposed as strict hard boundaries. The best-fit SED is not necessarily required to lie below every plotted upper limit when constrained by the full multi-wavelength photometry \citep{boquien19}.}

\begin{figure}
    % \plotone{figures/fig_M_SFR_v4.pdf}
    \plotone{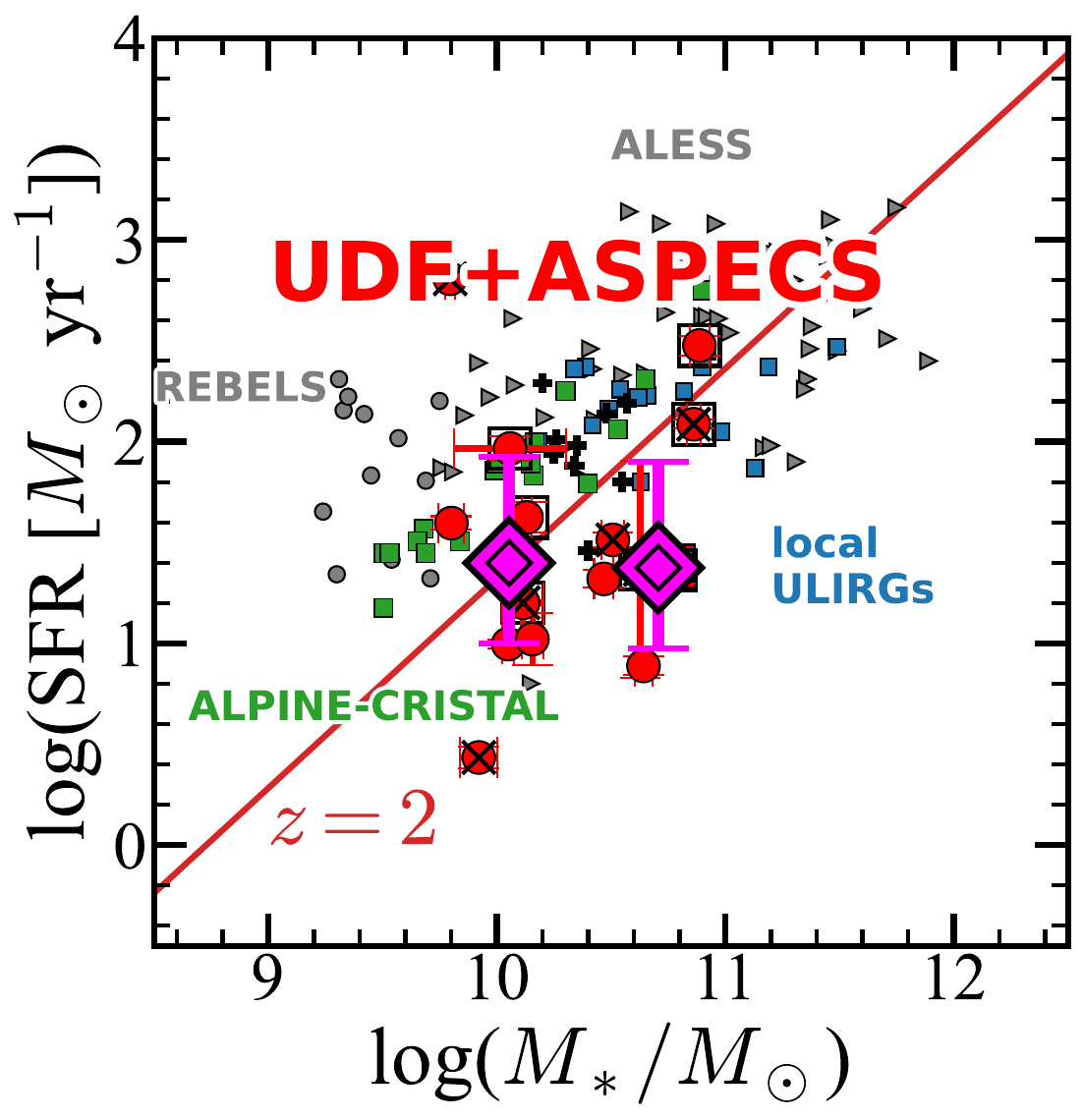}
    \caption{
    Relation between star formation rate (SFR) and stellar mass. 
    The red circles show the UDF+ASPECS sample. 
    The X-ray-detected sources and BPT-AGN sources (Section~\ref{subsec:AGN-SMGs}) are shown as the black crosses and black squares, respectively. 
    The magenta double diamonds with error bars show the median and the 16th–84th percentiles of the UDF+ASPECS sample in two stellar-mass bins ($M_*<10^{10.4}~M_\odot$ and $M_*>10^{10.4}~M_\odot$). 
    The red line shows the star formation main sequence at $z\sim2$ \citep{santini17}. 
    The ALESS \citep[gray triangle;][]{dacunha15}, \commentred{MOSDEF-ALMA \citep[black plus symbol;][]{sanders23-co, popping23}, ALPINE-CRISTAL \citep[green square;][]{bethermin20,mitsuhashi24,faisst26},} and the REBELS \citep[gray circle;][]{rowland25}, and the local ULIRGs \citep[blue squares;][]{dacunha10} are also shown. 
    }
\label{fig:LIR-Mstar}
\end{figure}

Table~\ref{tab:properties} summarizes the stellar masses and SFRs of the UDF+ASPECS sample derived from the SED fitting. 
Figure~\ref{fig:LIR-Mstar} shows the SFR–stellar mass relation, together with literature \commentred{samples (local ULIRGs, ALESS, MOSDEF-ALMA, ALPINE-CRISTAL-JWST and REBELS)}. The UDF+ASPECS sources are mainly on the star-forming main sequence at $z\sim2$ (see also, e.g., \citealt{dunlop17, elbaz18, aravena20})\commentred{, and have smaller stellar masses and SFRs than those of the other dusty galaxies (e.g., ALESS and local ULIRGs). The REBELS sample exhibits comparable SFRs and IR luminosities, but systematically lower stellar masses, reflecting the redshift evolution of the star-forming main sequence. Compared to the MOSDEF-ALMA, which are UV-selected galaxies at similar redshift, they show comparable stellar mass and SFR to the UDF+ASPECS sample.}  
These UDF+ASPECS results (stellar mass and SFR) are broadly consistent with previous estimates obtained without JWST photometry (e.g., \citealt{dunlop17, aravena20}). 

\begin{figure*}
    \centering
    \includegraphics[width=0.49\linewidth]{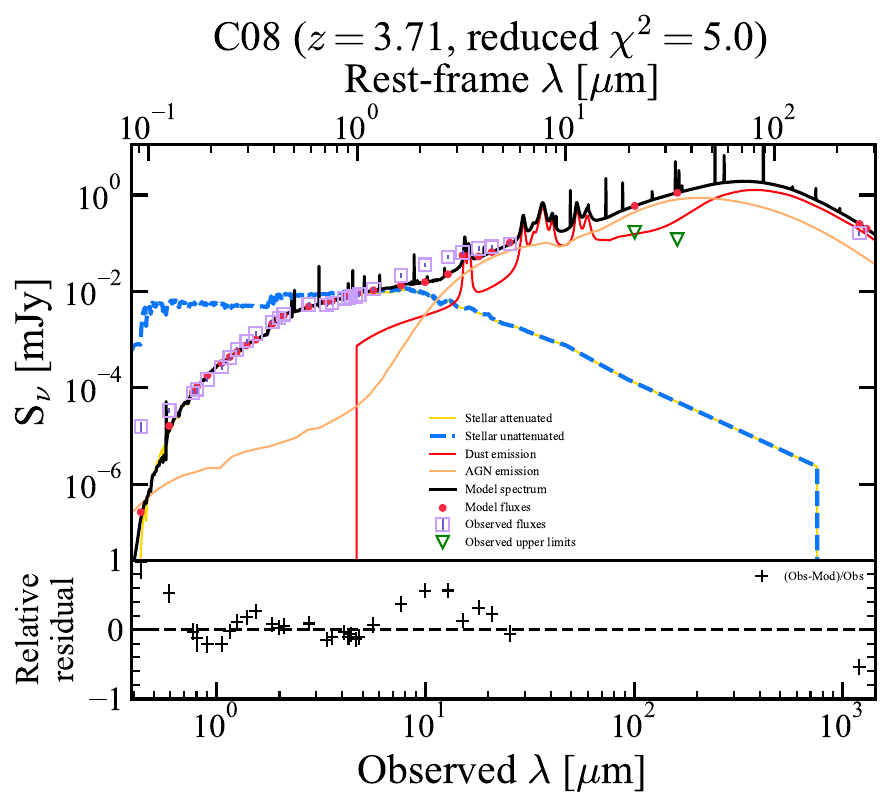}
    \includegraphics[width=0.49\linewidth]{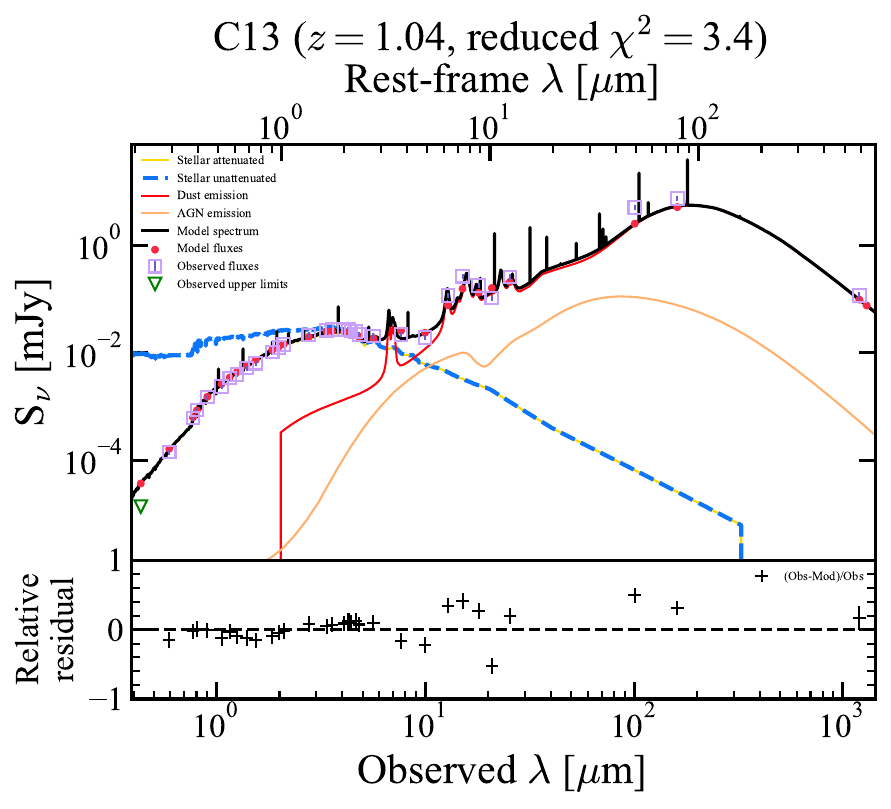}
    \caption{
    Example SED-fitting results (left: C08 as an example of an AGN; right: C13 \commentred{as an example of a source with prominent PAH emission}). 
    Top: The purple squares and green triangles indicate the observed photometry and the $3\sigma$ upper limits, respectively. The black curve shows the total best-fit SED, while the red, orange, blue-dashed, and yellow curves show the dust emission, AGN emission, dust-unattenuated, and dust-attenuated stellar emission, respectively. The reduced $\chi$-square of the best-fit SED is indicated above the panels. 
    Bottom: Relative residuals, $(F_\mathrm{obs}-F_\mathrm{model})/F_\mathrm{obs}$, between the observed photometry and the total best-fit SED. 
    We note that AGN fractions ($f_\mathrm{AGN}$) of C08 and C13 are $0.50\pm0.03$ and $0.01\pm0.02$, respectively. 
    }
    \label{fig:SED-expample}
\end{figure*}

Figure~\ref{fig:SED-expample} shows examples of the SED-fitting results: C08 and C13 as examples of an AGN and prominent polycyclic aromatic hydrocarbon (PAH) emission, respectively. 
We note that C08 shows a relatively large reduced $\chi^2$ ($=5.0$); however, this discrepancy is driven mainly by the mid-infrared photometry, which is likely dominated by AGN emission. The stellar mass is mainly constrained by the rest-frame optical–NIR, and the SFR by the FIR. Thus, these estimates are not strongly affected by the mid-infrared discrepancy.
These SED results, especially PAH emission, are also discussed in Section~\ref{subsec:PAH}. 
\commentred{The SEDs and best-fit models for the remaining sources are shown in Appendix~\ref{appendix:sed-results}.}

%%% fig: SFR vs. Mstar %%%
%%%%%%%%%%%%%%%%%%%%%%%%%%

%%% SED example %%%%%%%%%%
%%%%%%%%%%%%%%%%%%%%%%%%%%

\section{Results} 
\label{sec:results} 

\subsection{Morphology and Spectra} 
\label{subsec:morphology-SMGs}

%%% fig: snapshots and spectra %%%
\begin{figure*}
\gridline{\fig{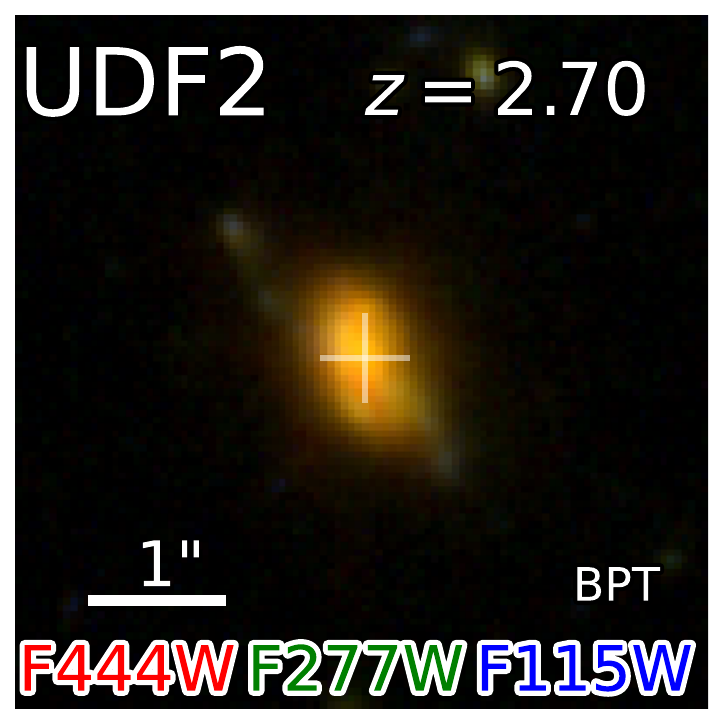}{0.25\textwidth}{}
          \fig{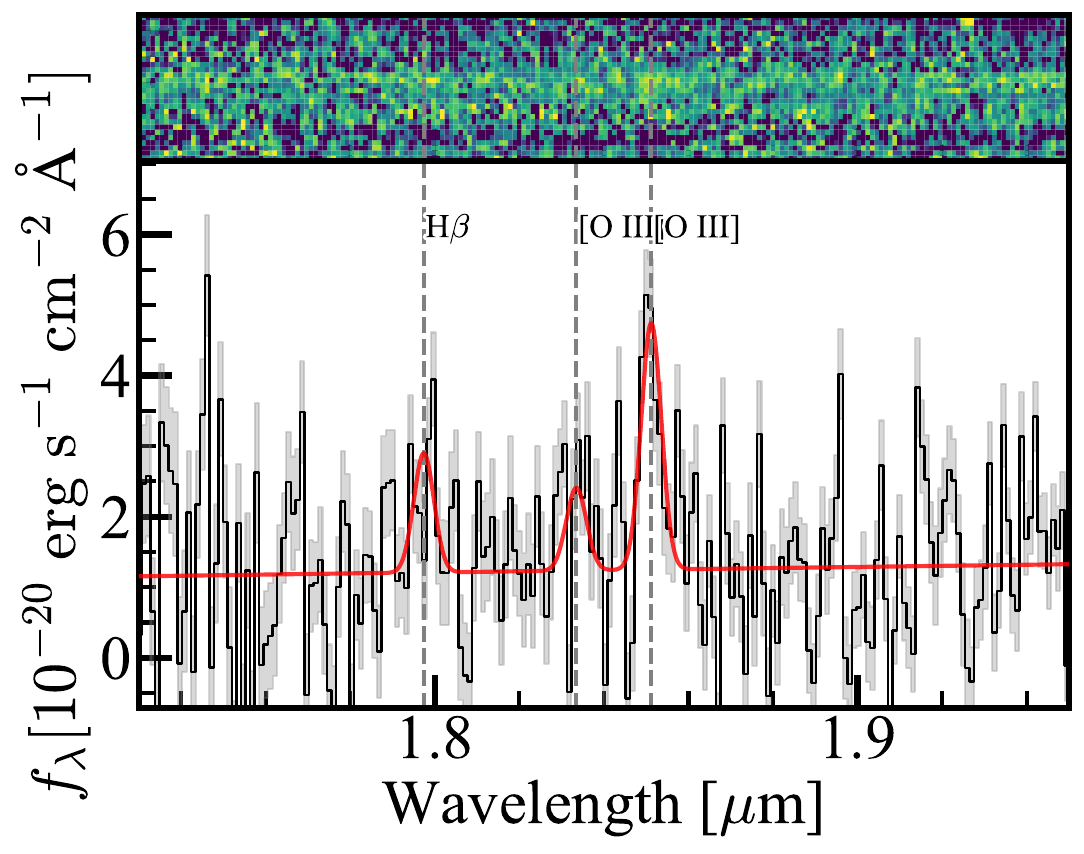}{0.35\textwidth}{}
          \fig{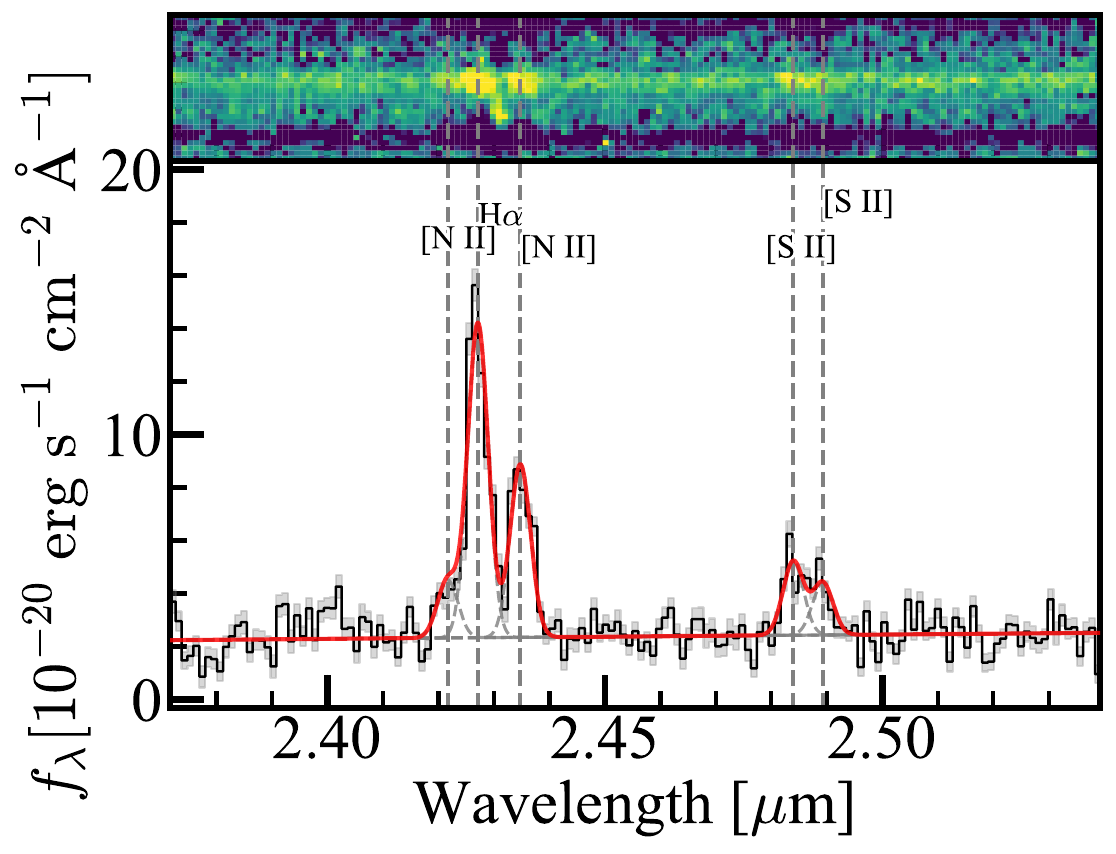}{0.36\textwidth}{}
          } 
\gridline{\fig{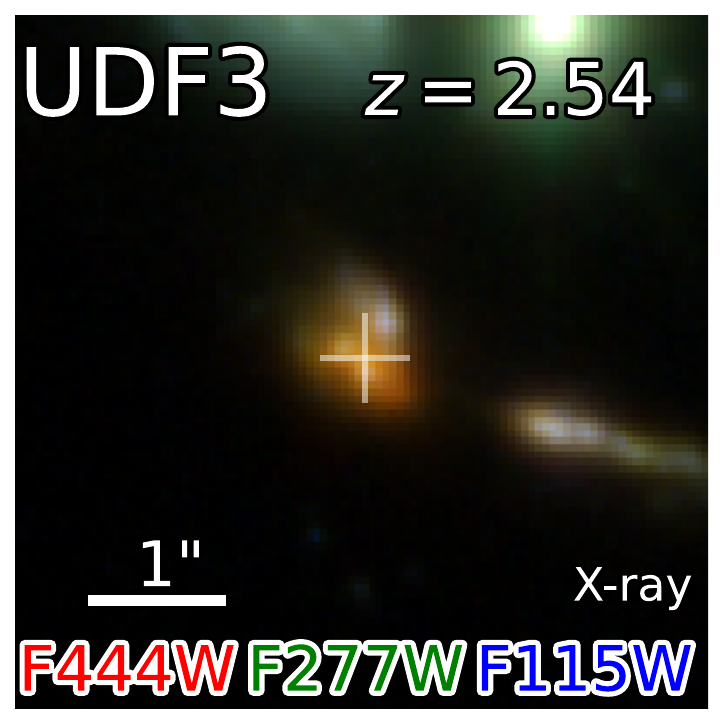}{0.25\textwidth}{}
          \fig{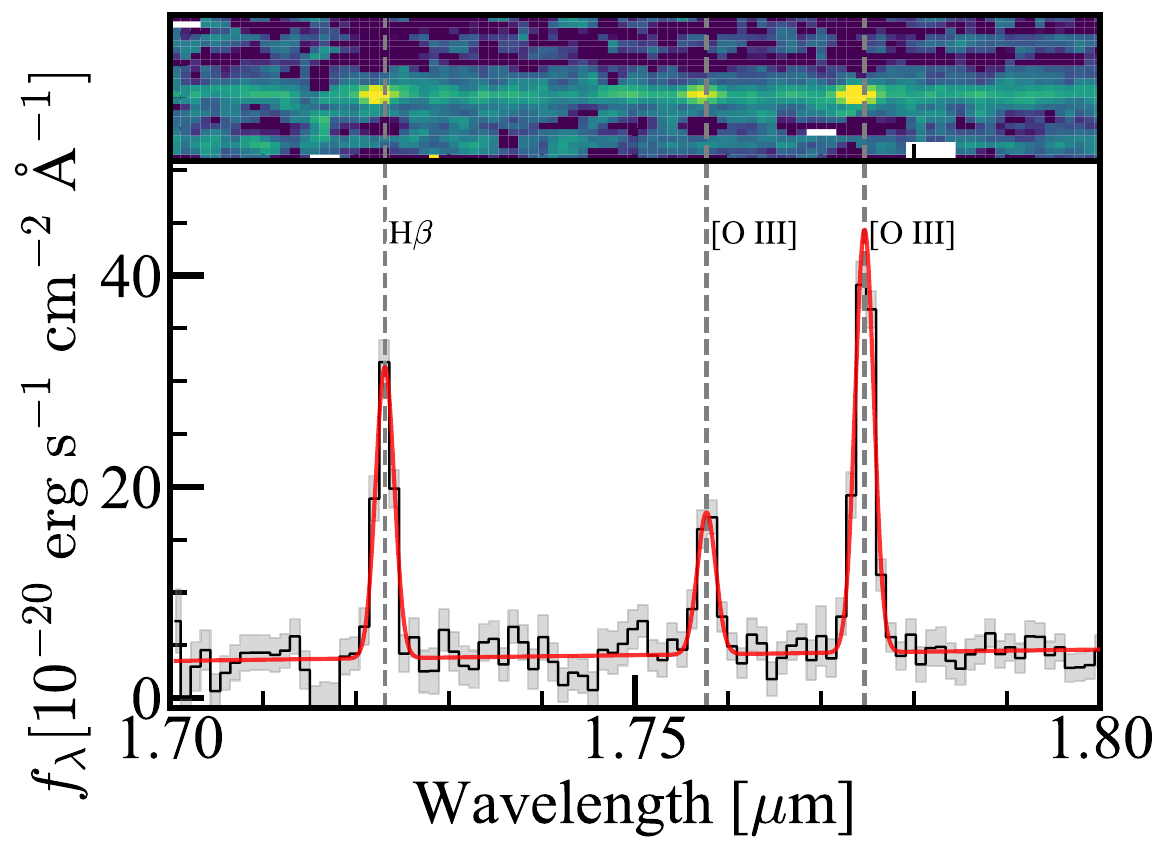}{0.36\textwidth}{}
          \fig{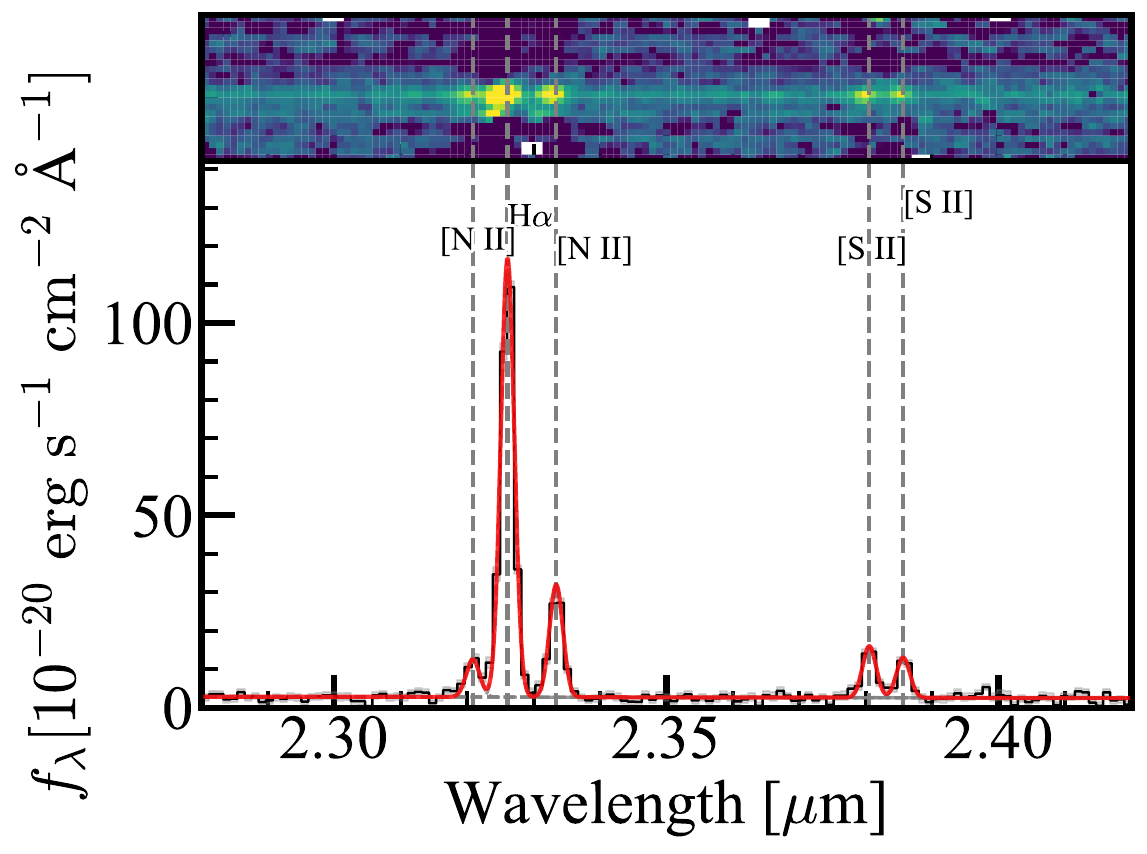}{0.355\textwidth}{}
          } 
\gridline{\fig{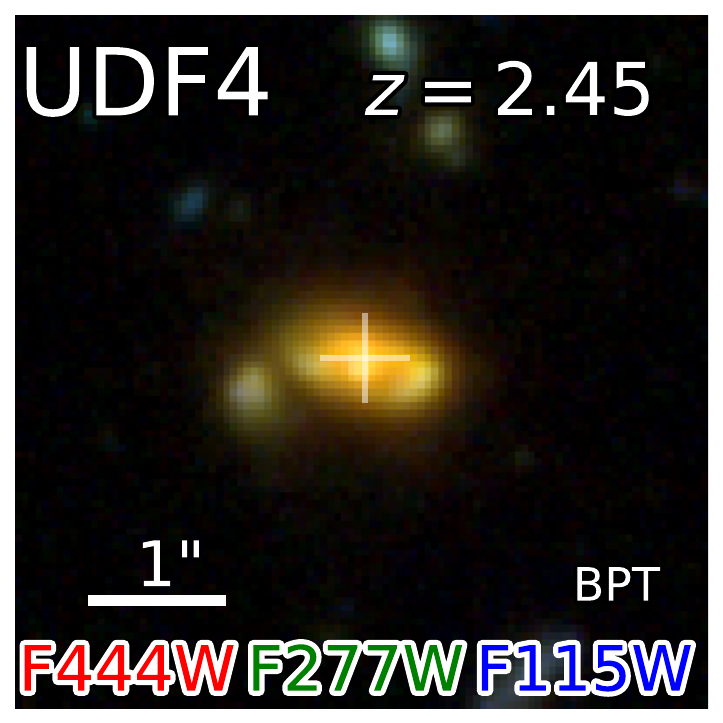}{0.25\textwidth}{}
          \fig{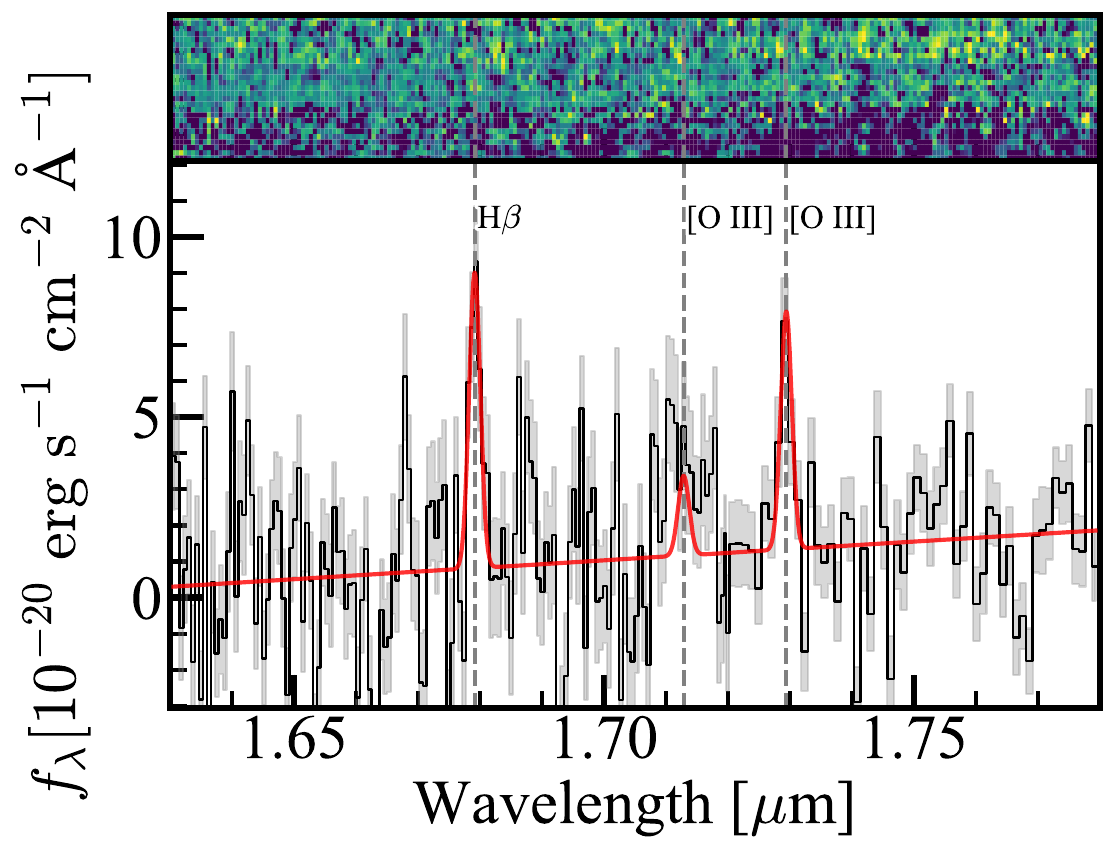}{0.36\textwidth}{}
          \fig{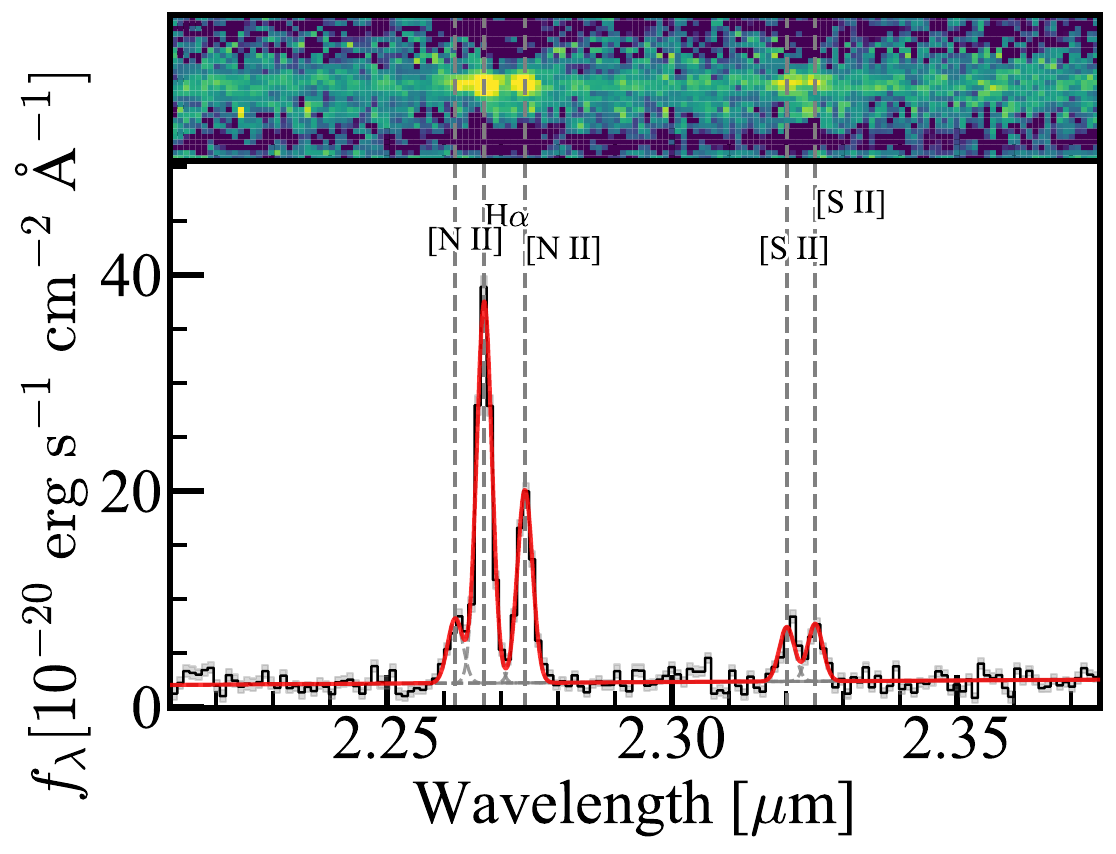}{0.36\textwidth}{}
          } 
\caption{
NIRCam images and NIRSpec spectra of UDF2, UDF3, and UDF4.  
Left: 
pseudocolor images ($5\arcsec\times5\arcsec$; blue: NIRCam F115W, green: NIRCam F277W, red: NIRCam F444W). 
The white cross shows the position of the dust continuum peak \citep{dunlop17, aravena20}. 
The bar in the bottom-left corner shows a $1\arcsec$ scale (8.3~kpc at $z=2.5$). 
The labels ‘X-ray' and/or ‘BPT' \commentred{indicate} an X-ray detection and/or a source classified as an AGN in the BPT diagram using the \citet{kauffmann03} curve.
Middle: 
NIRSpec spectrum around the wavelength of H$\beta$ and \oiii$\lambda\lambda4959, 5007$ emission lines. 
The top and bottom panels show the two-dimensional spectrum and the one-dimensional spectrum, respectively. 
The black solid lines and gray shades show the observed spectrum and associated $1\sigma$ uncertainties, respectively. 
The red solid curves show the best-fit spectrum. 
Right: 
Same as in the middle panel but for H$\alpha$, \nii$\lambda\lambda6548, 6584$, and \sii$\lambda\lambda6717, 6731$ emission lines.
}
\label{fig:snapshots1}
\end{figure*}
%%%%%%%%%%%%%%%%%%%%%%%%%%%%%%%%%%

%%% fig: snapshots and spectra %%%
\begin{figure*}[p]
\centering
% \vspace*{\fill}
\rotatebox{90}{
  \begin{minipage}{\textheight}
  \centering
  \gridline{\fig{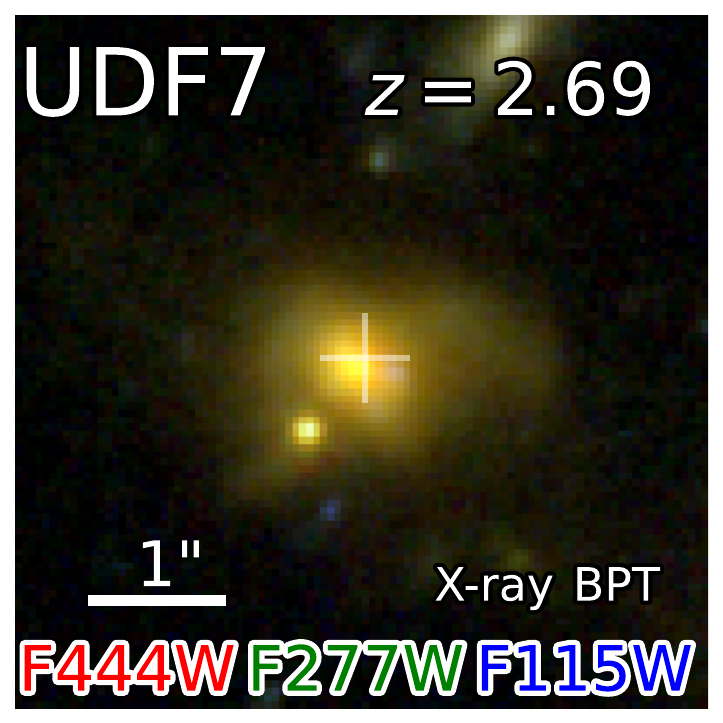}{0.1\textwidth}{}
            \fig{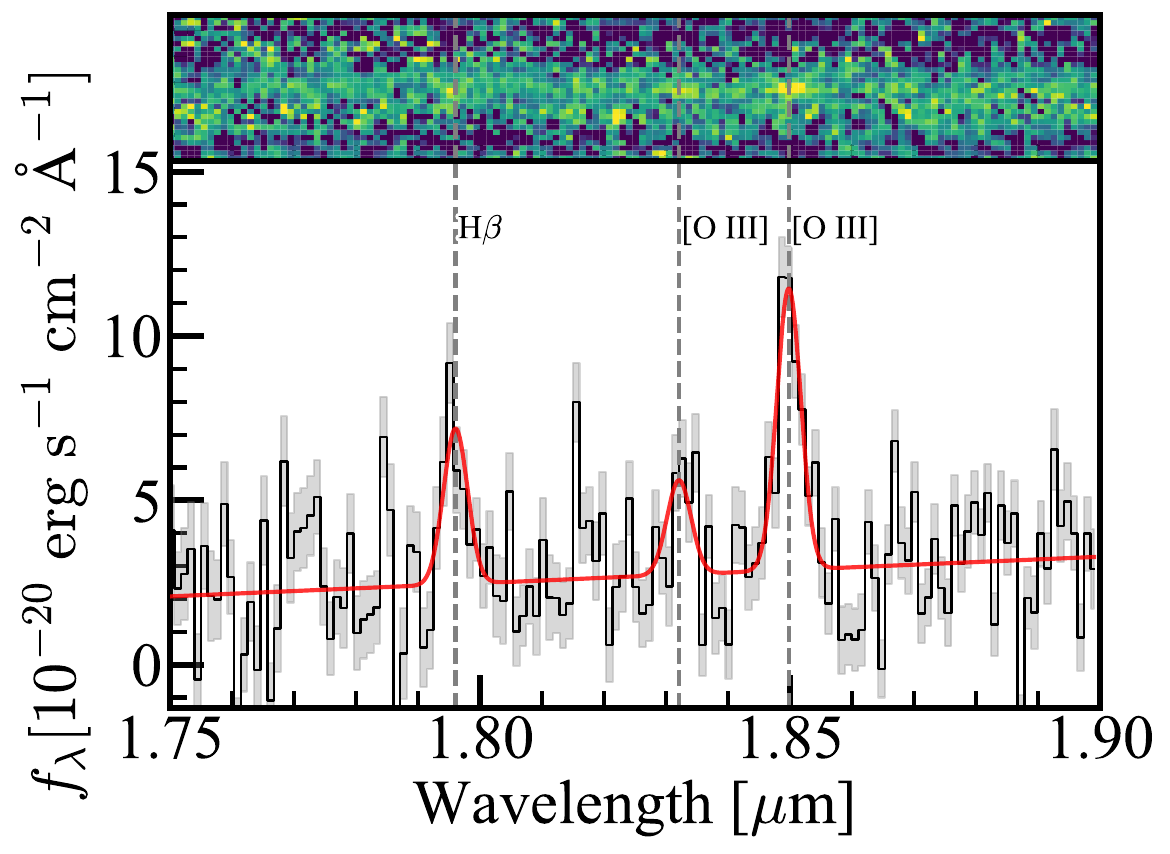}{0.15\textwidth}{}
            \fig{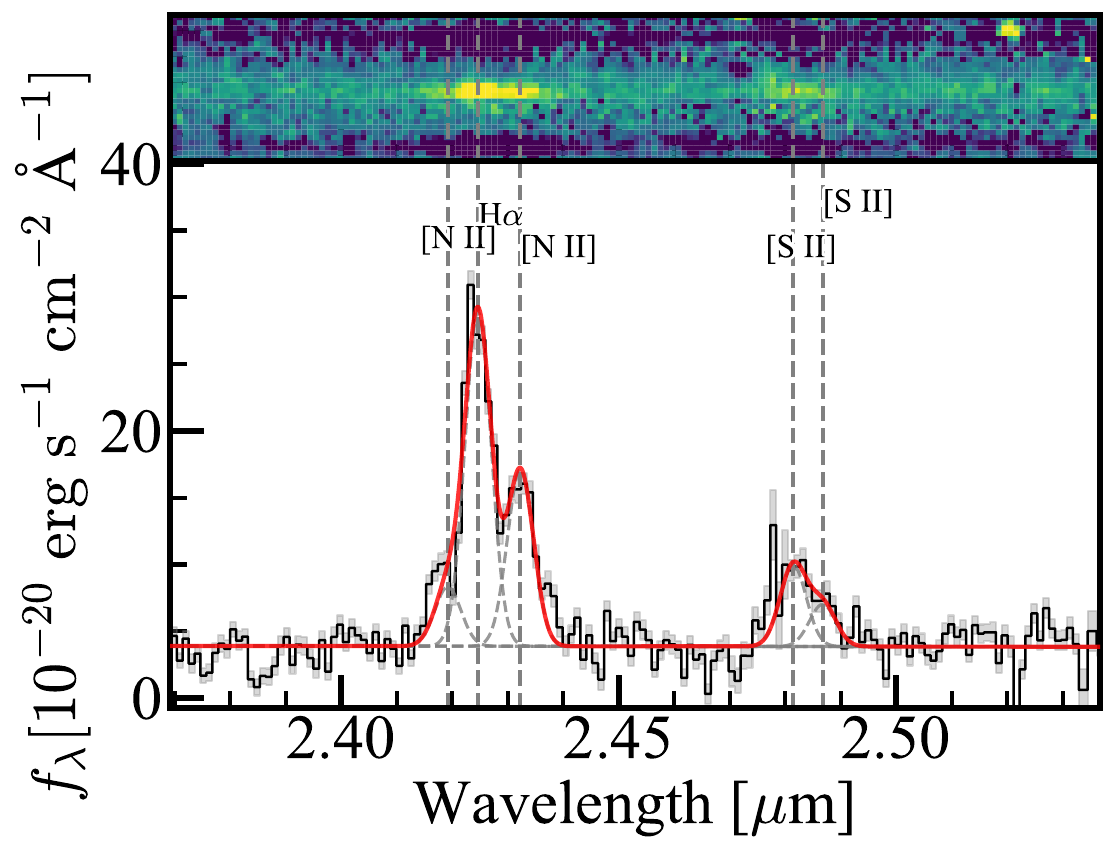}{0.15\textwidth}{}
            \fig{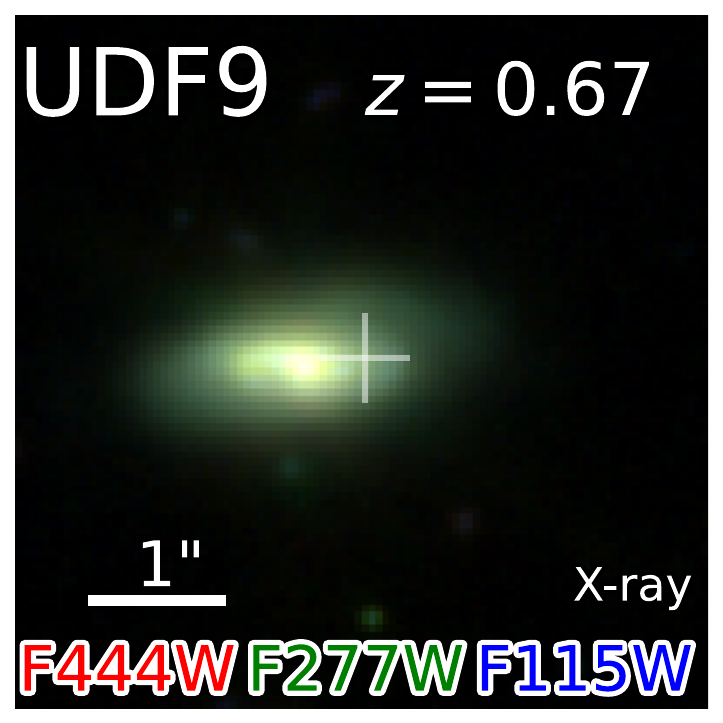}{0.1\textwidth}{}
            \fig{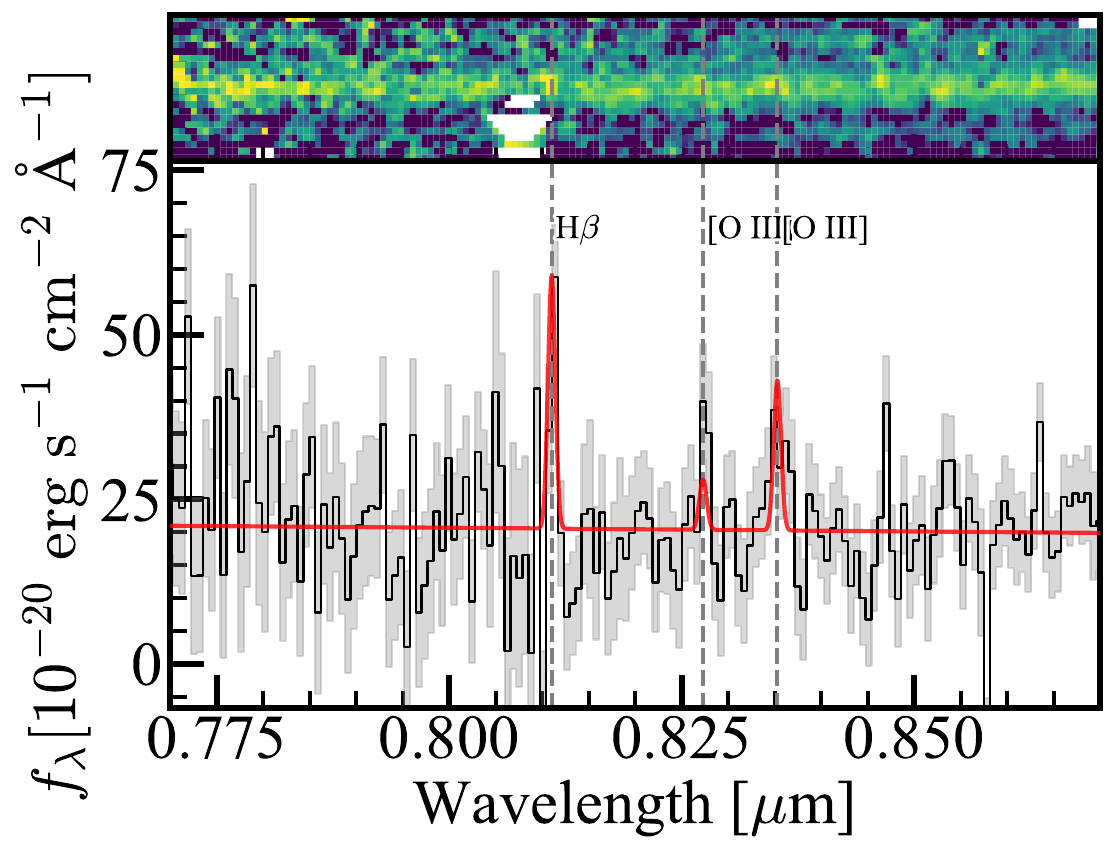}{0.15\textwidth}{}
            \fig{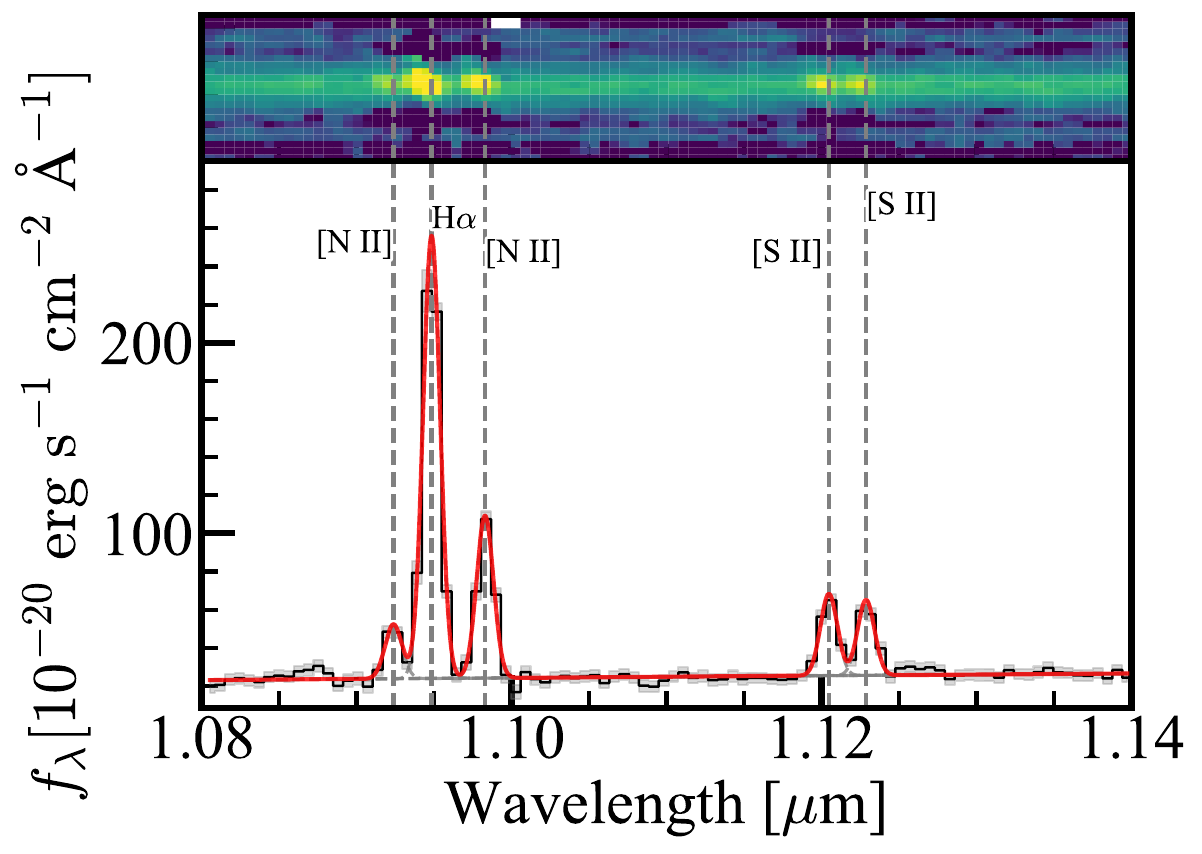}{0.15\textwidth}{}}
  \gridline{\fig{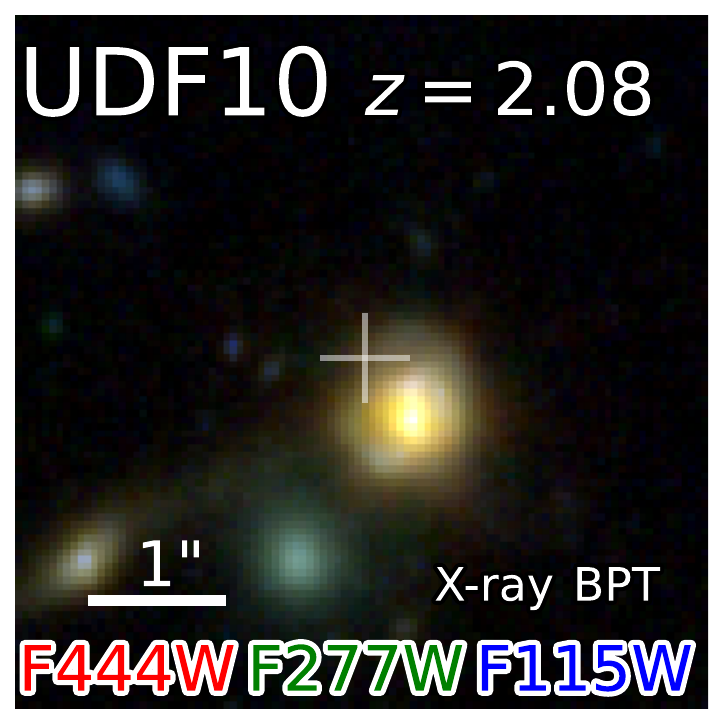}{0.1\textwidth}{}
            \fig{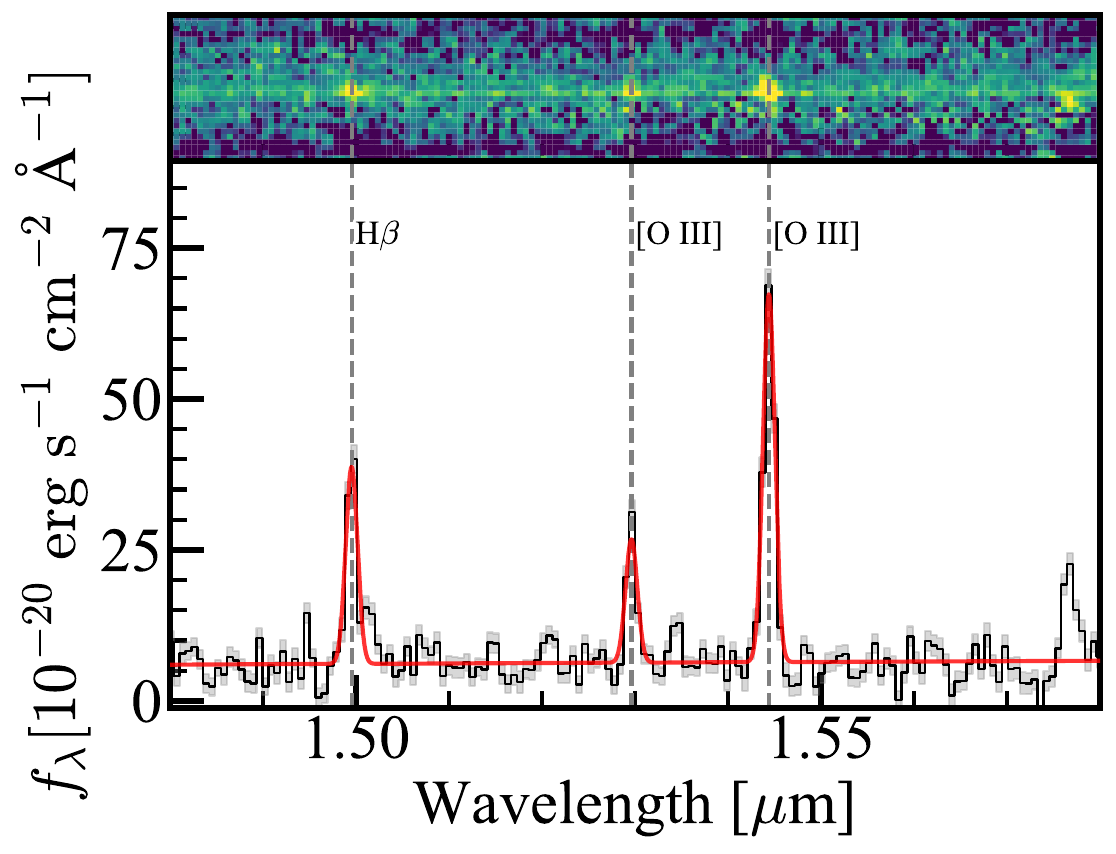}{0.15\textwidth}{}
            \fig{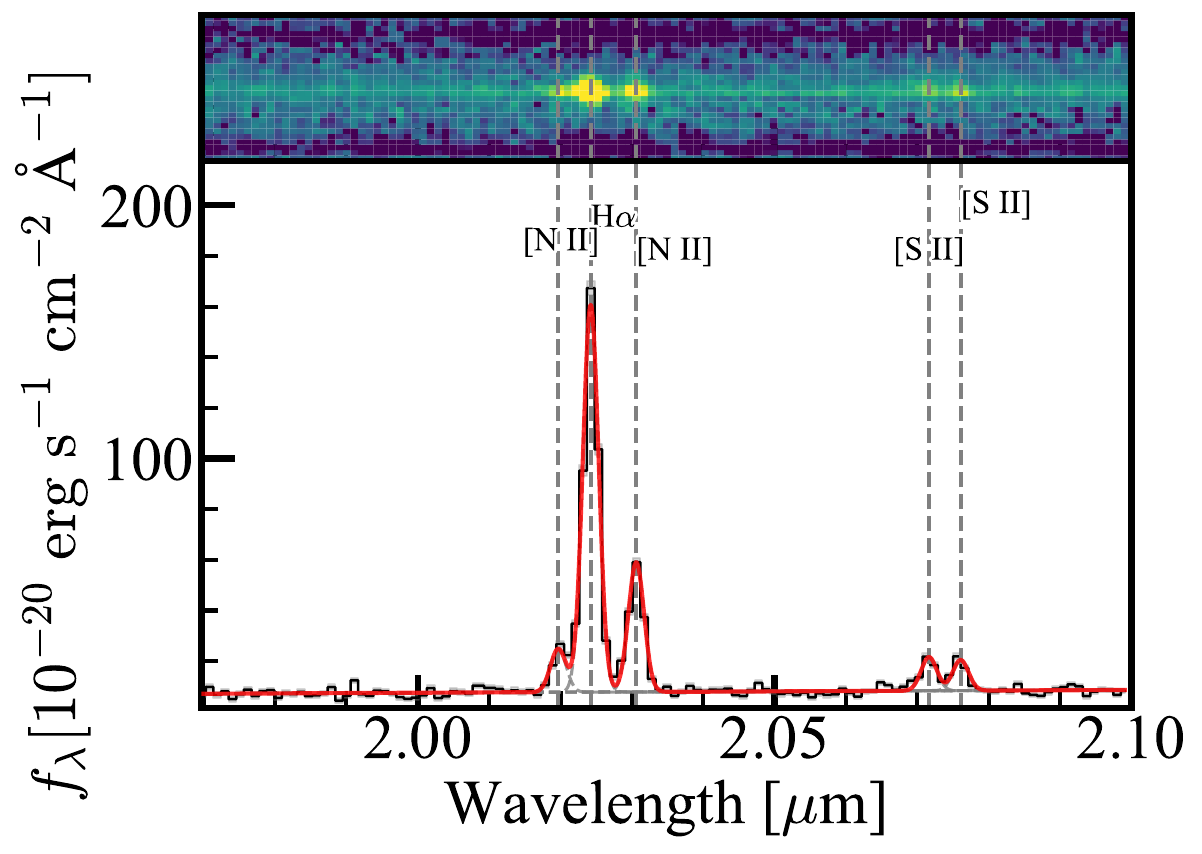}{0.15\textwidth}{}
            \fig{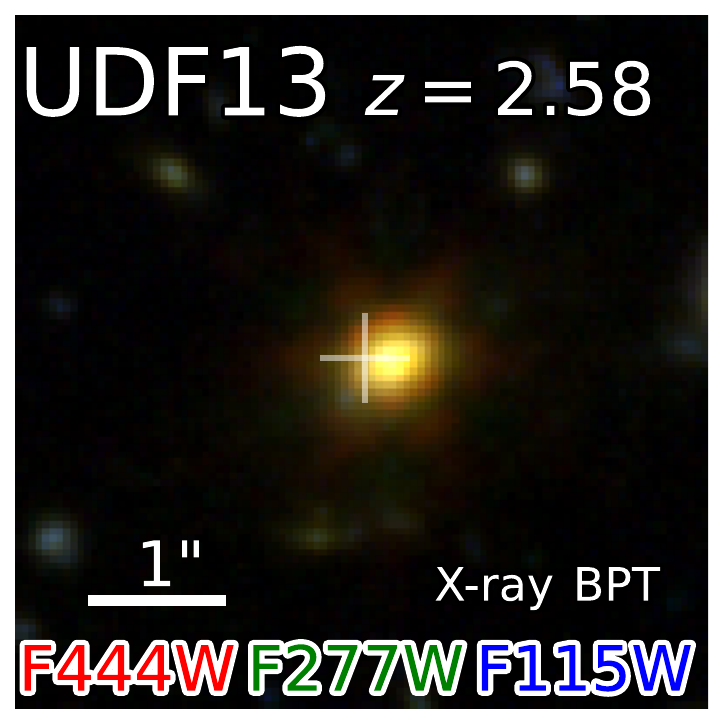}{0.1\textwidth}{}
            \fig{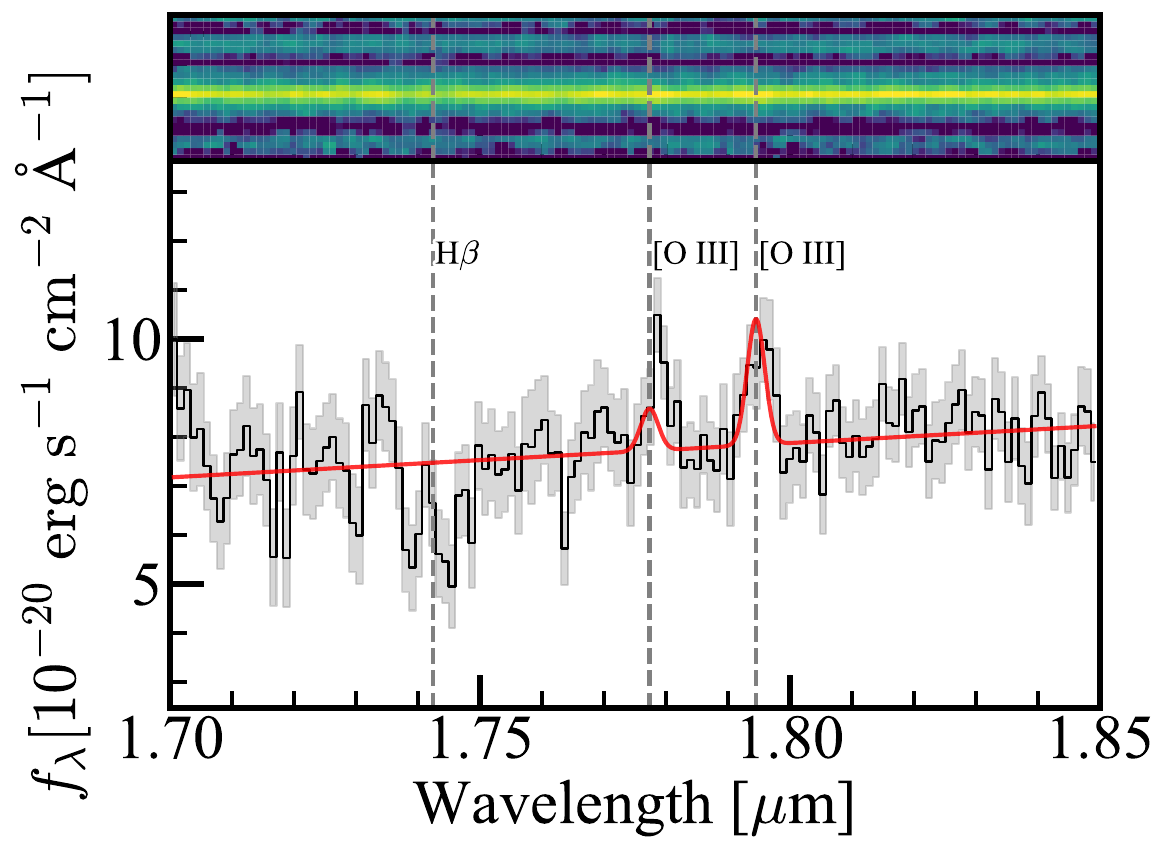}{0.15\textwidth}{}
            \fig{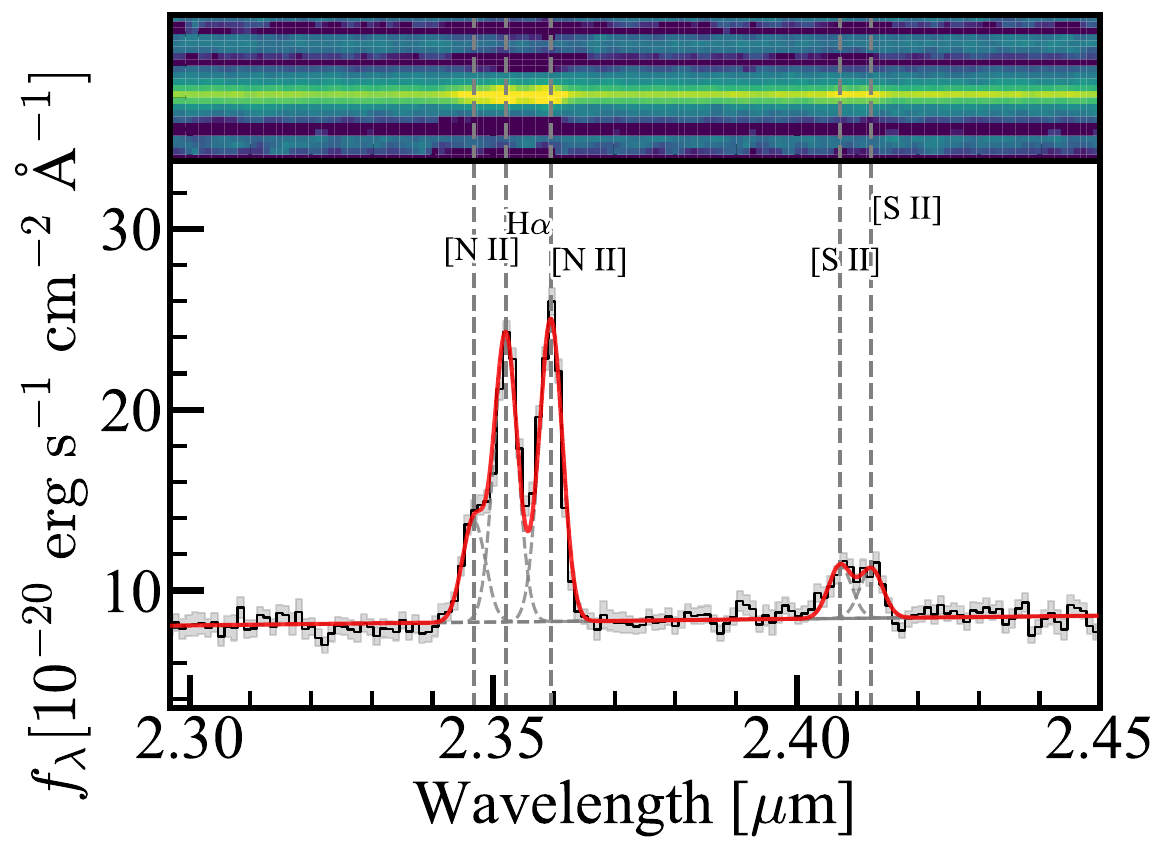}{0.15\textwidth}{}}
  \gridline{\fig{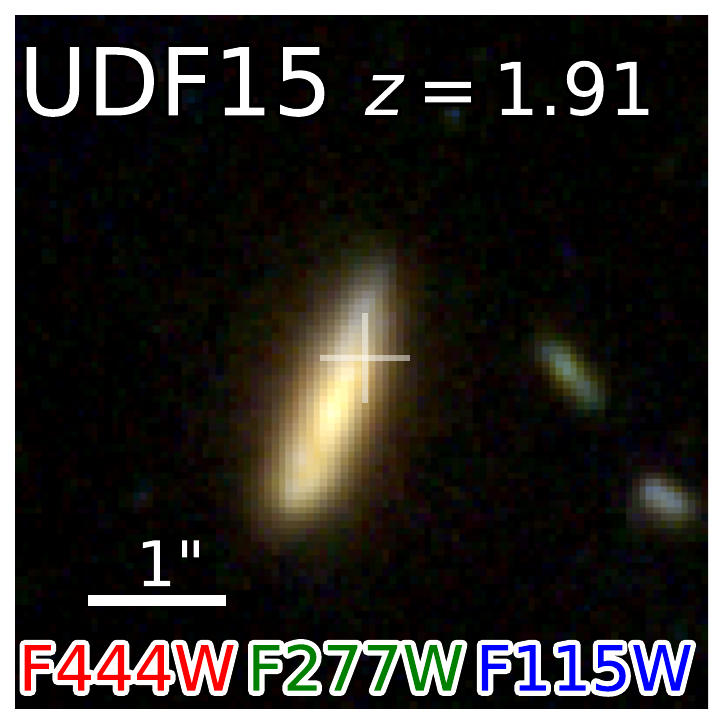}{0.1\textwidth}{}
            \fig{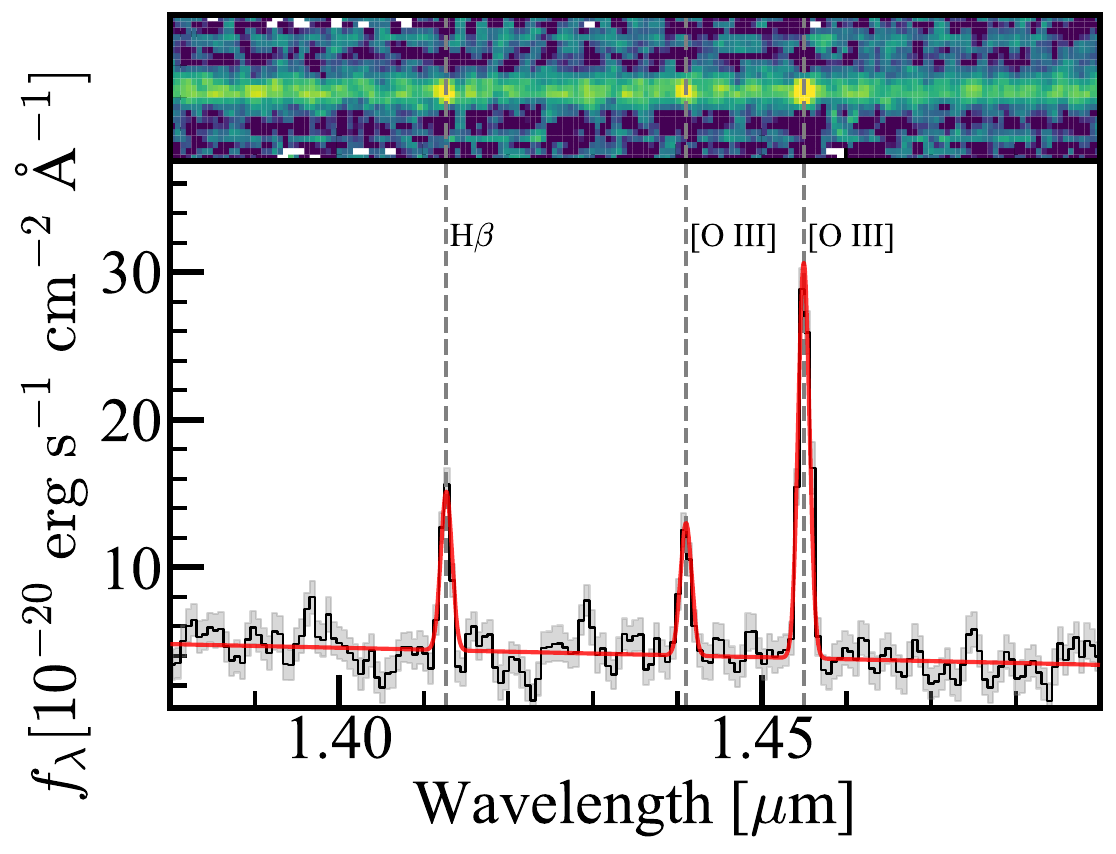}{0.15\textwidth}{}
            \fig{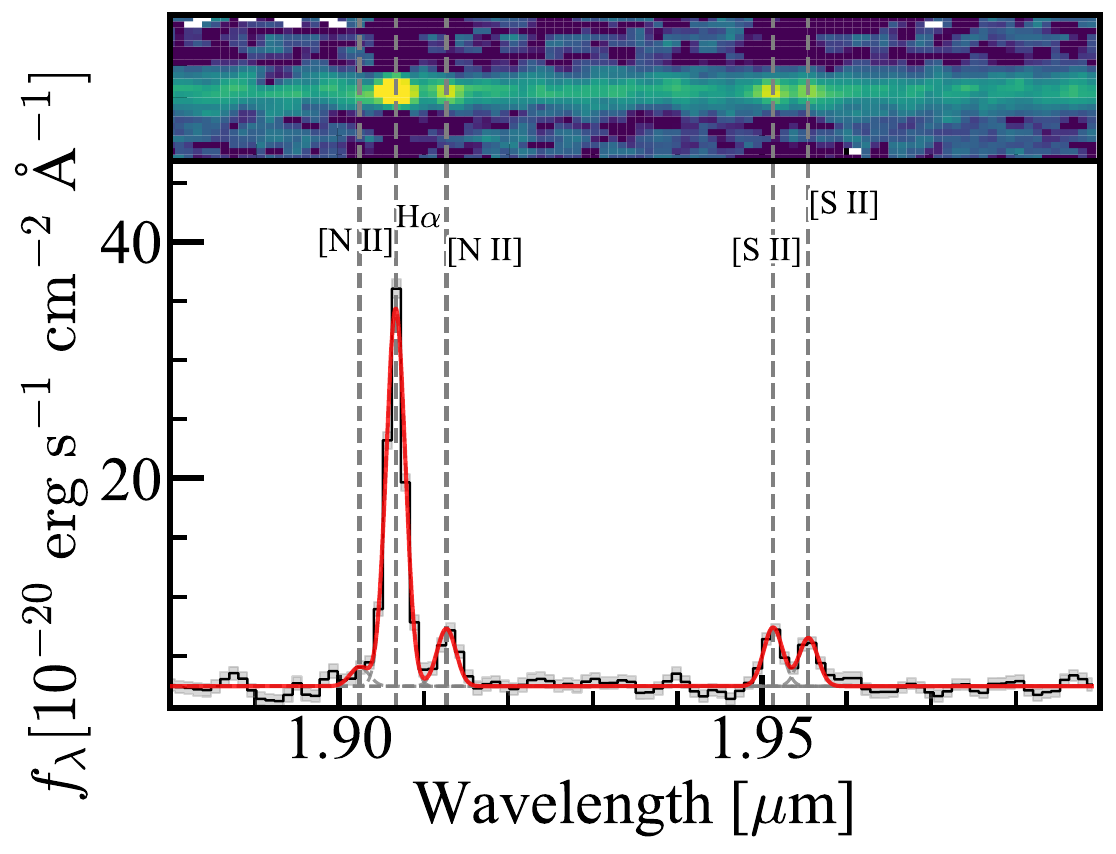}{0.15\textwidth}{}
            \fig{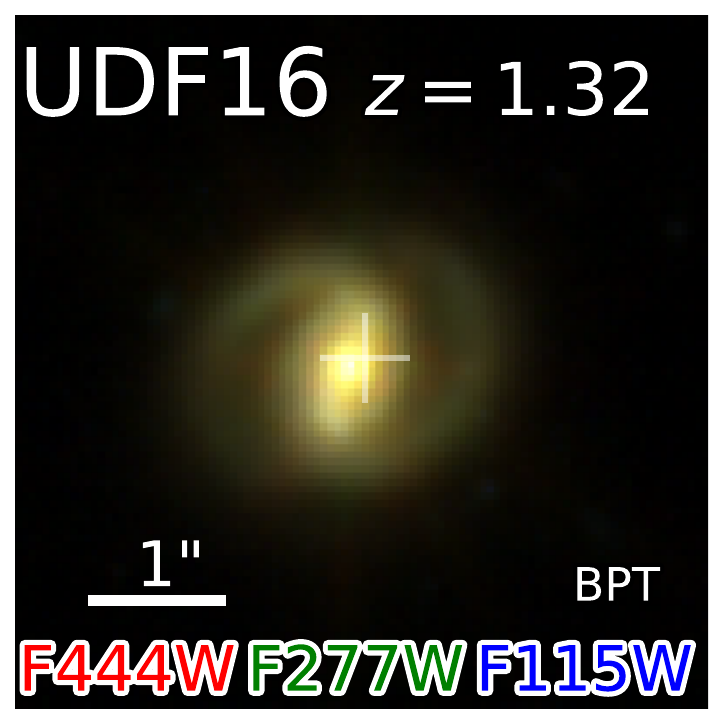}{0.1\textwidth}{}
            \fig{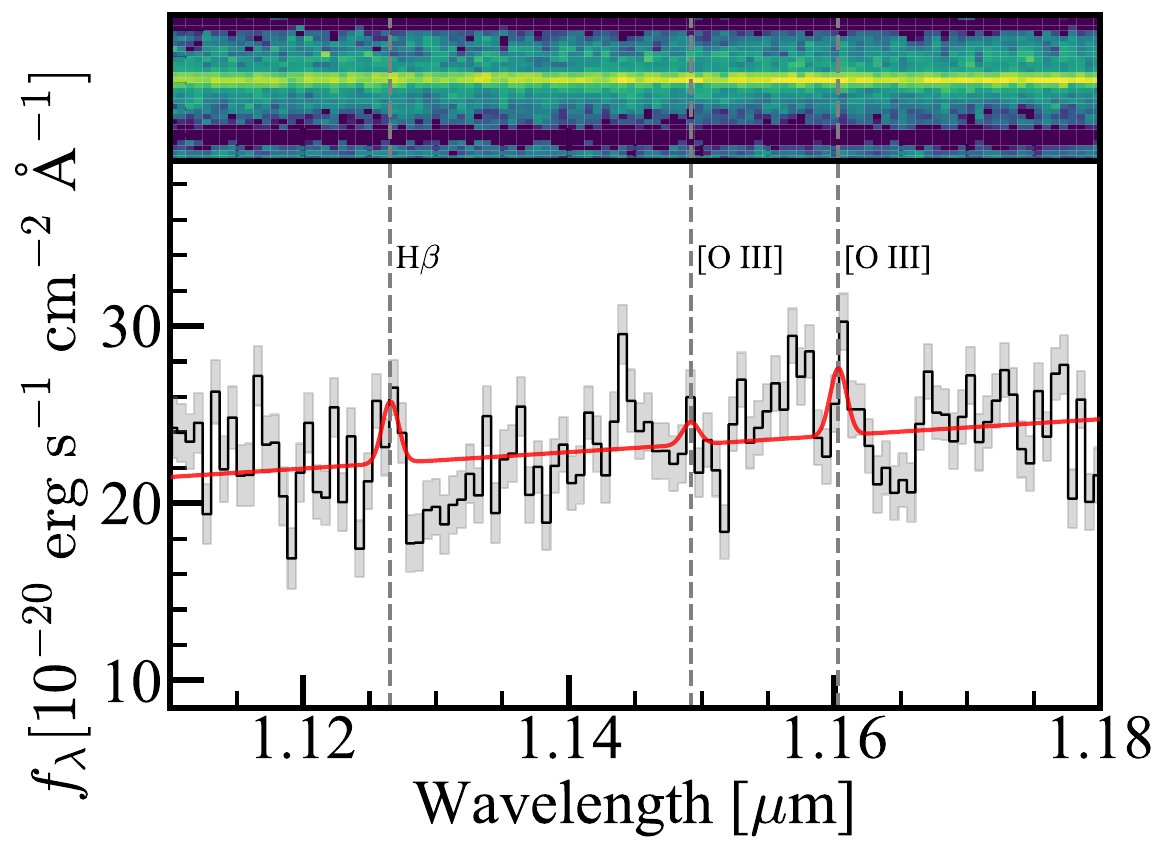}{0.15\textwidth}{}
            \fig{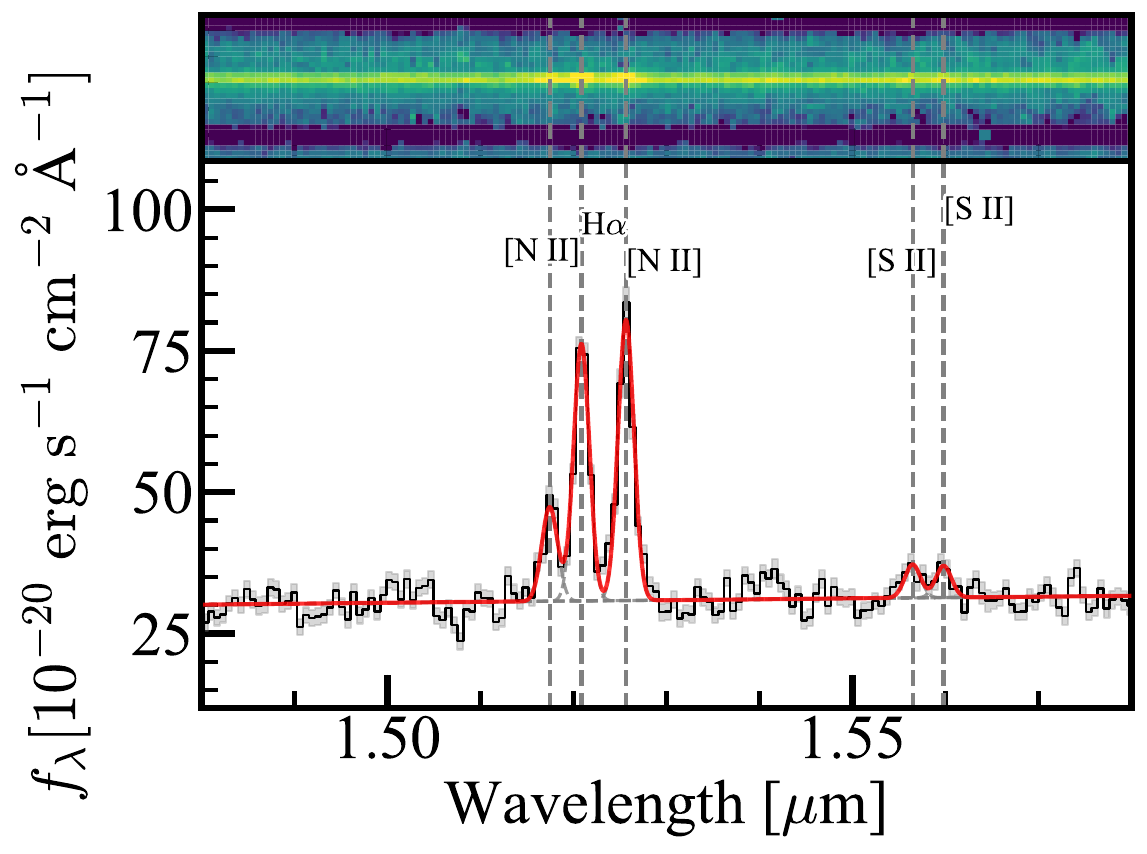}{0.15\textwidth}{}}
  \end{minipage}
}
\caption{Same as in Figure~\ref{fig:snapshots1} but for UDF7--UDF16. \label{fig:snapshots2}}
\end{figure*}
%%%%%%%%%%%%%%%%%%%%%%%%%%%%%%%%%%

%%% fig: snapshots and spectra %%%
\begin{figure*}[p]
\centering
% \vspace*{\fill}
\rotatebox{90}{
  \begin{minipage}{\textheight}
  \centering
  \gridline{\fig{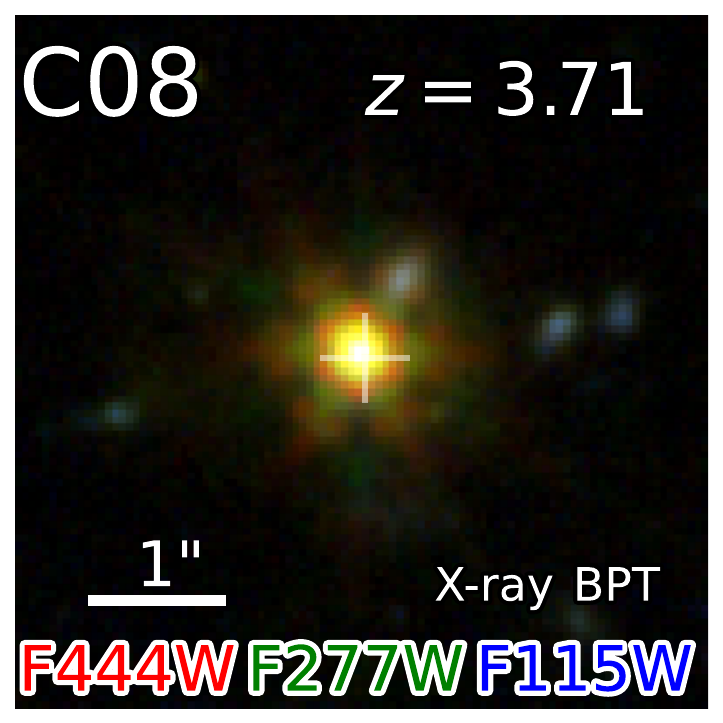}{0.1\textwidth}{}
            \fig{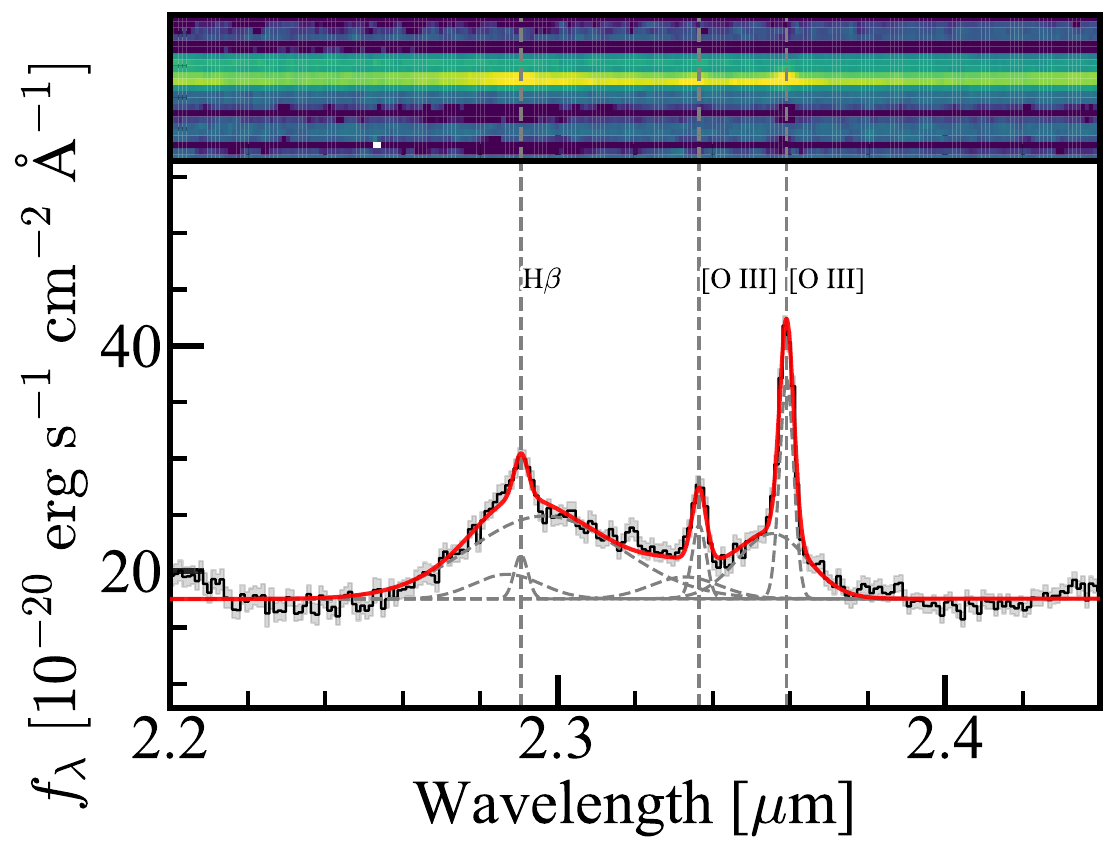}{0.15\textwidth}{}
            \fig{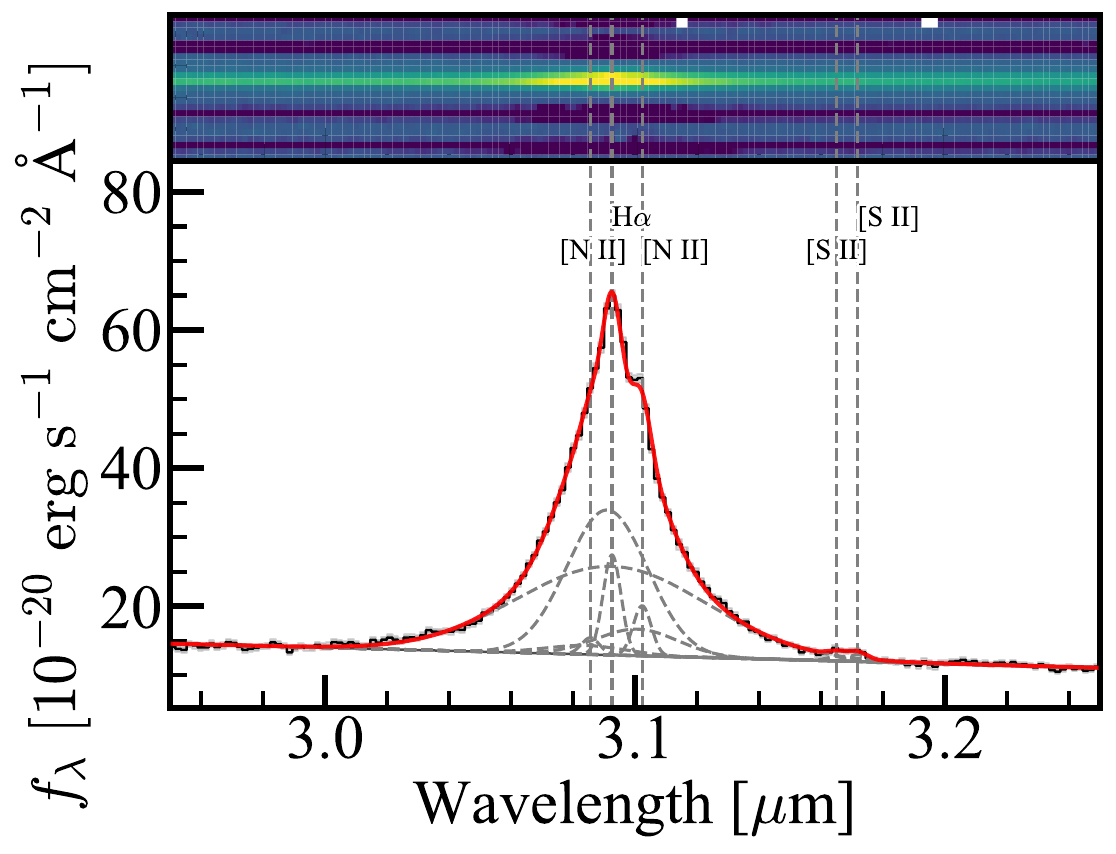}{0.15\textwidth}{}
            \fig{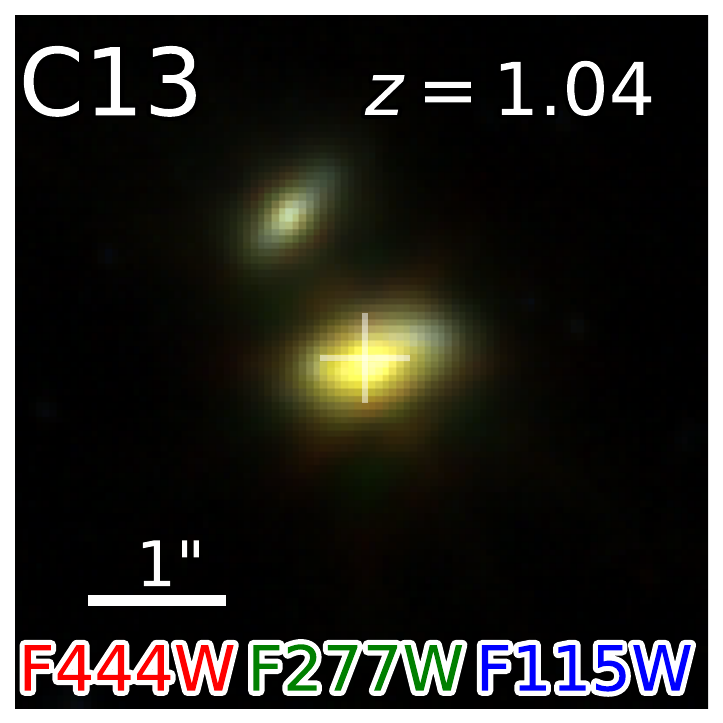}{0.1\textwidth}{}
            \fig{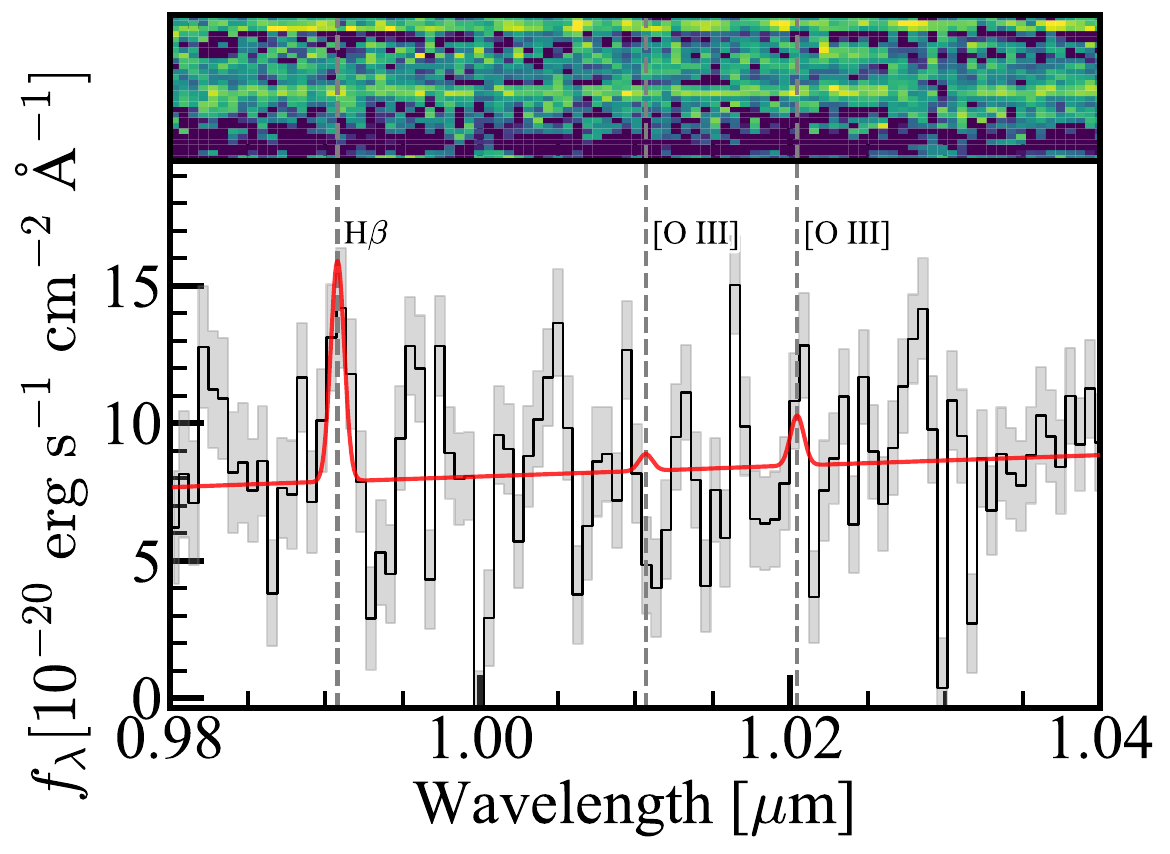}{0.15\textwidth}{}
            \fig{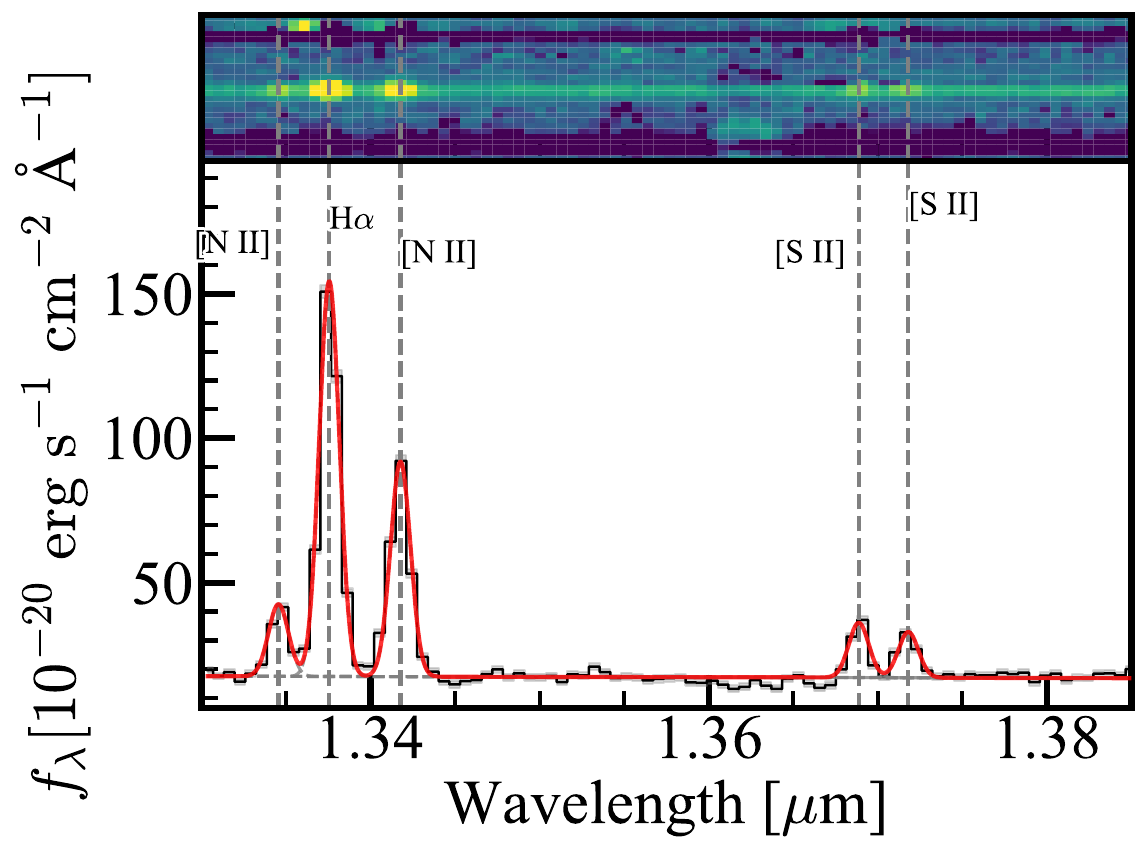}{0.15\textwidth}{}}
  \gridline{\fig{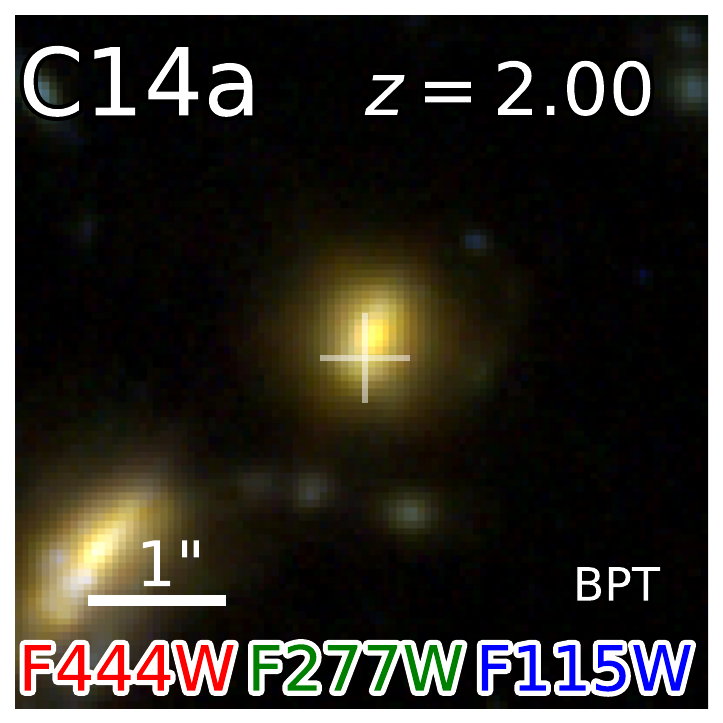}{0.1\textwidth}{}
            \fig{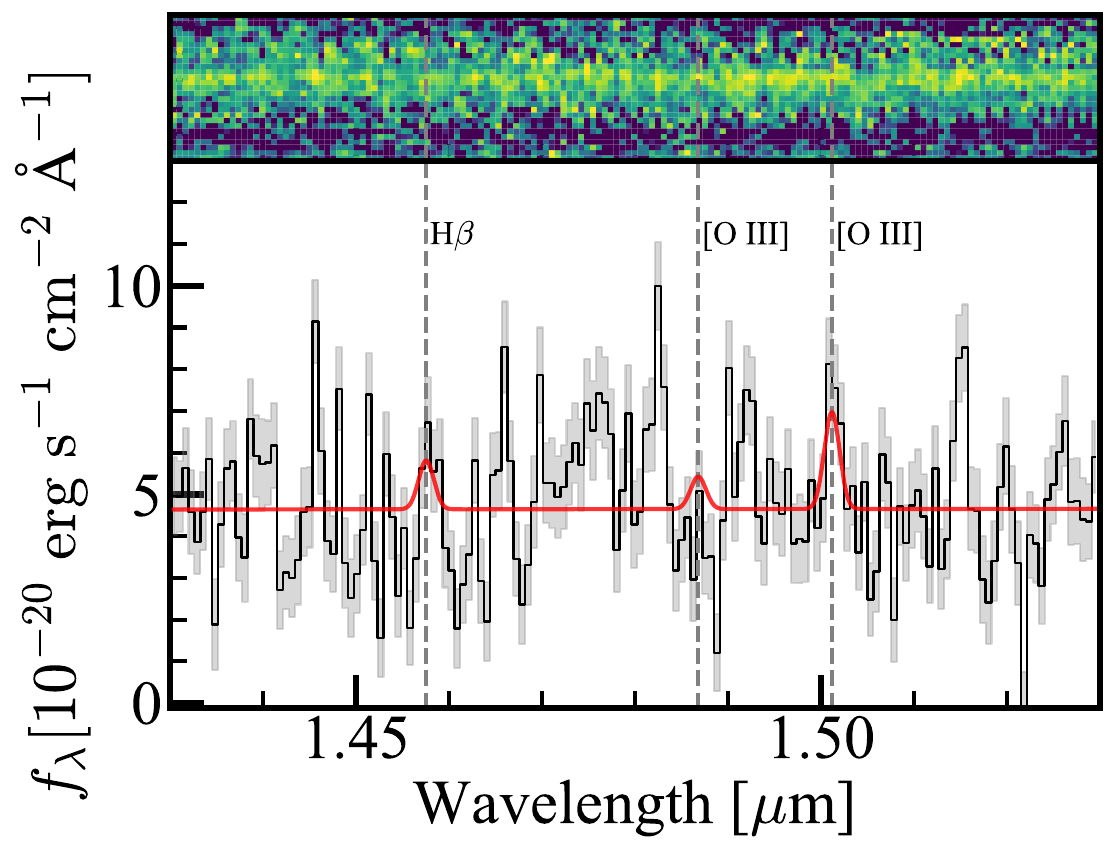}{0.15\textwidth}{}
            \fig{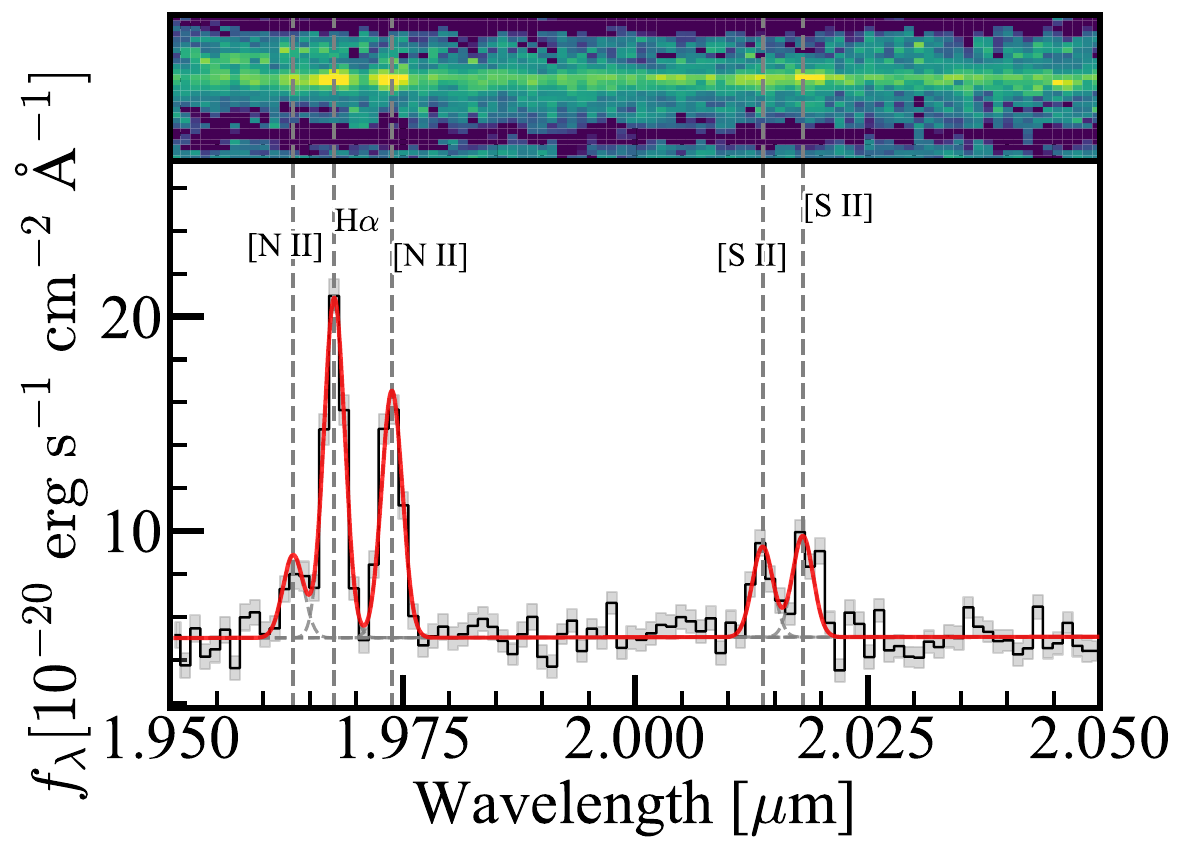}{0.15\textwidth}{}
            \fig{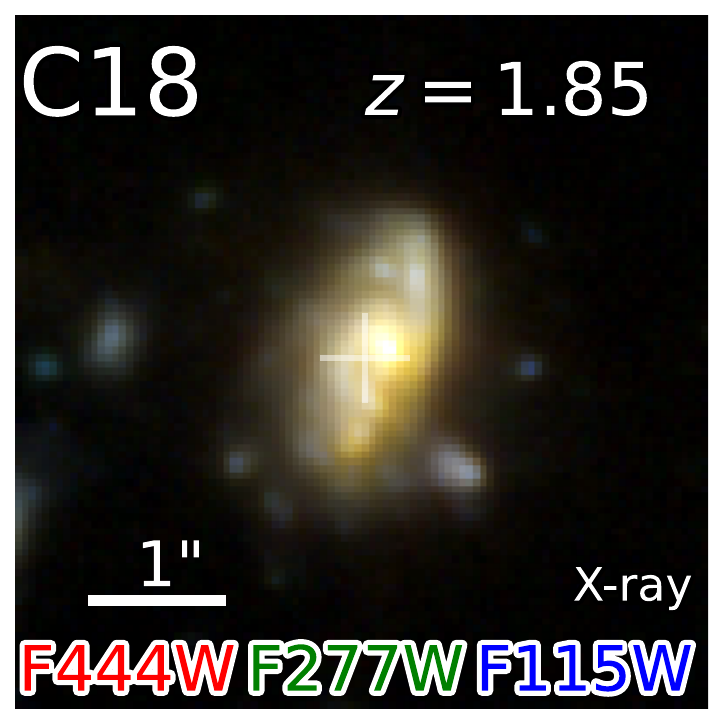}{0.1\textwidth}{}
            \fig{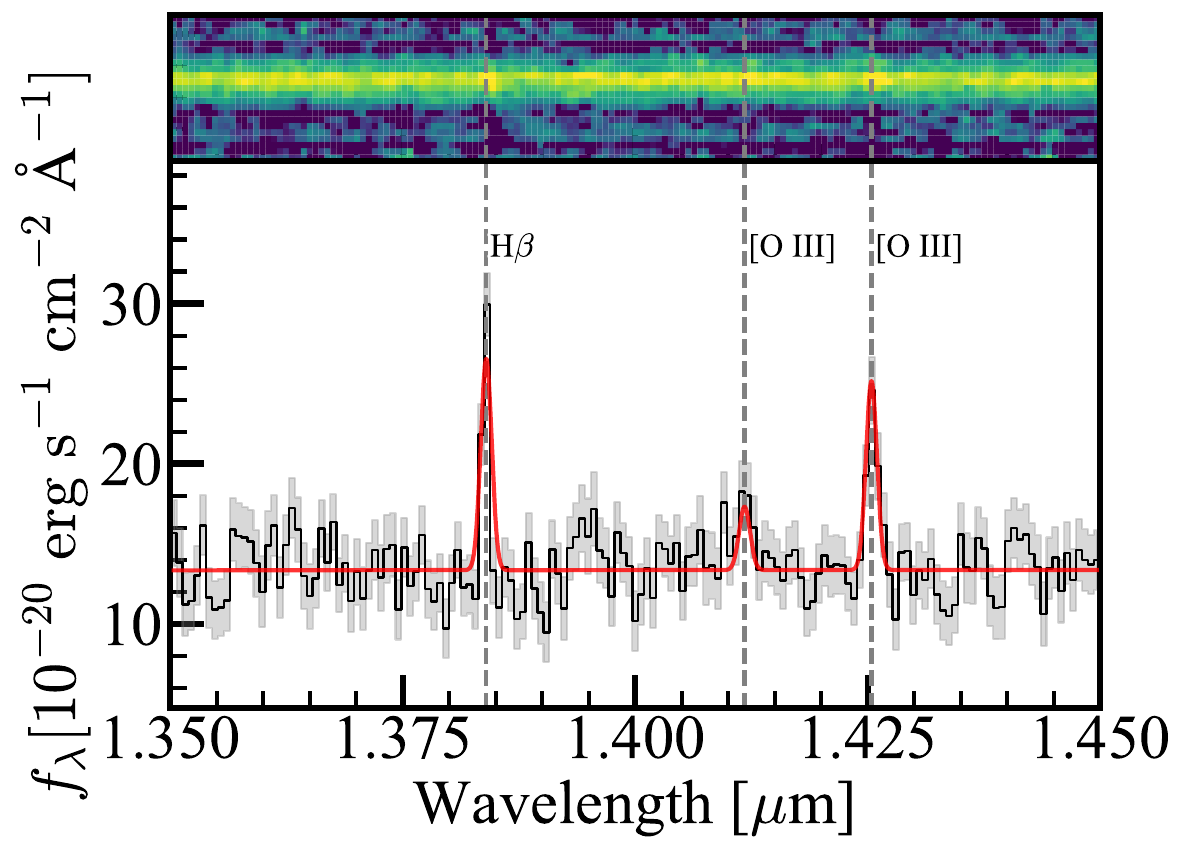}{0.15\textwidth}{}
            \fig{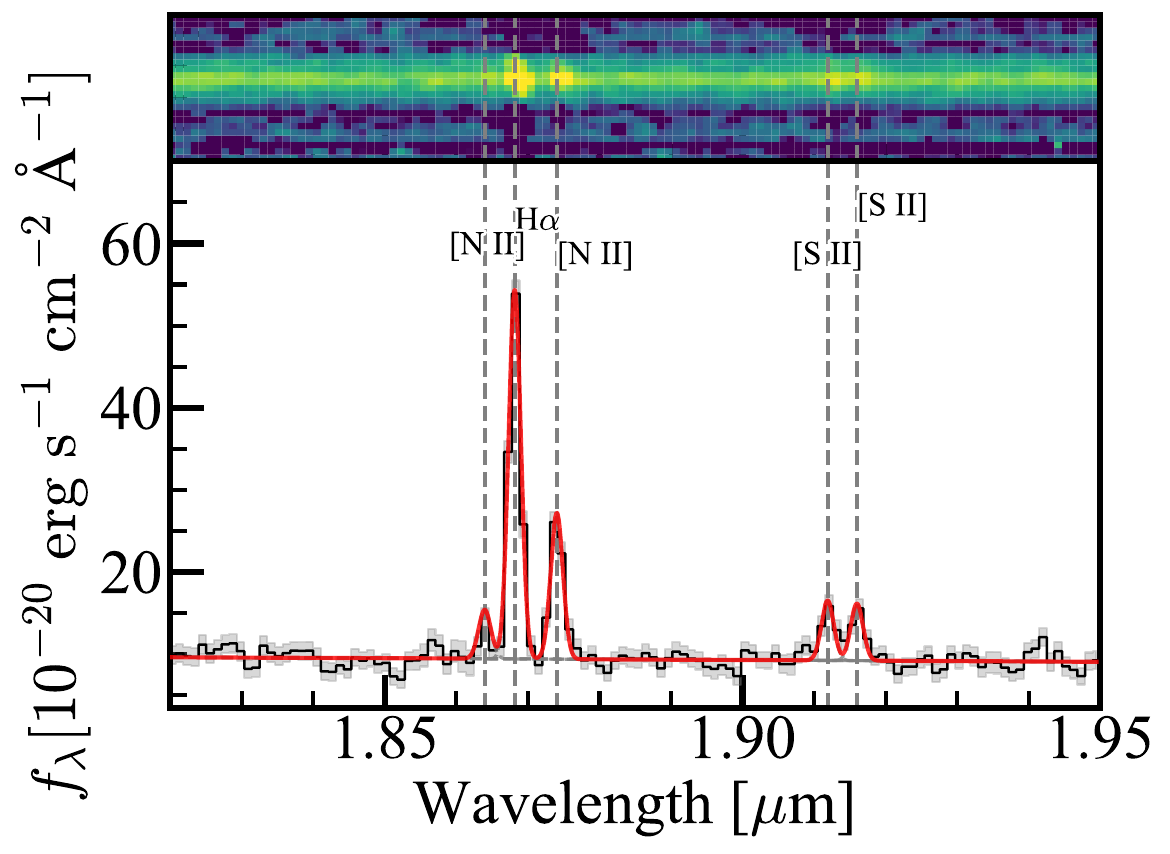}{0.15\textwidth}{}}
  \gridline{\fig{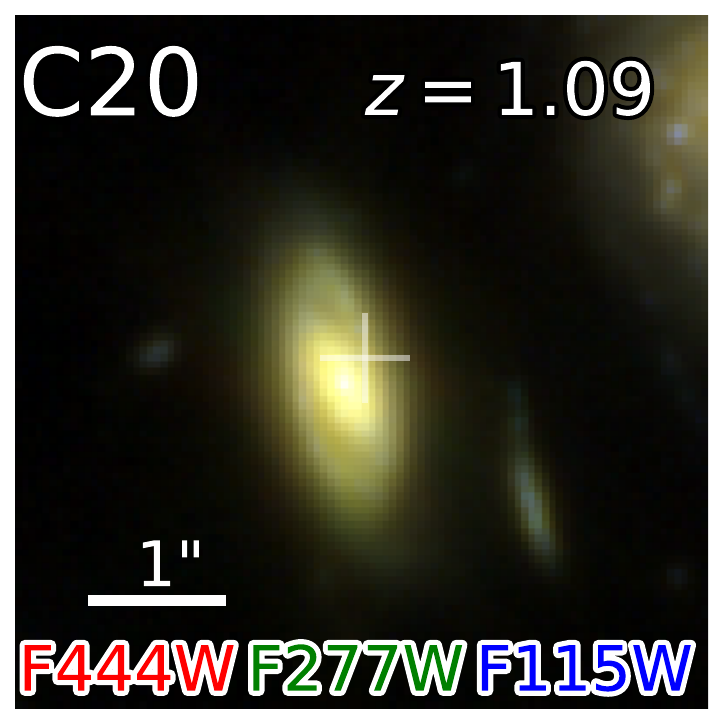}{0.1\textwidth}{}
            \fig{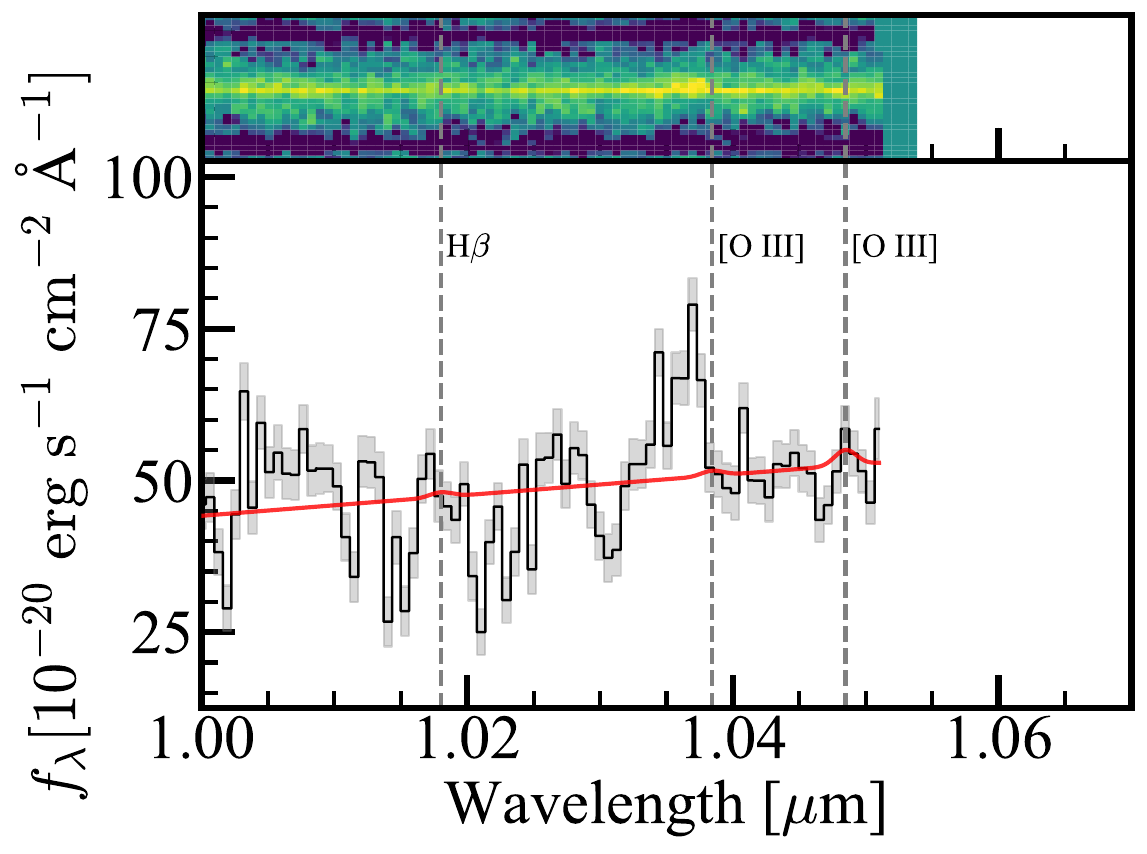}{0.15\textwidth}{}
            \fig{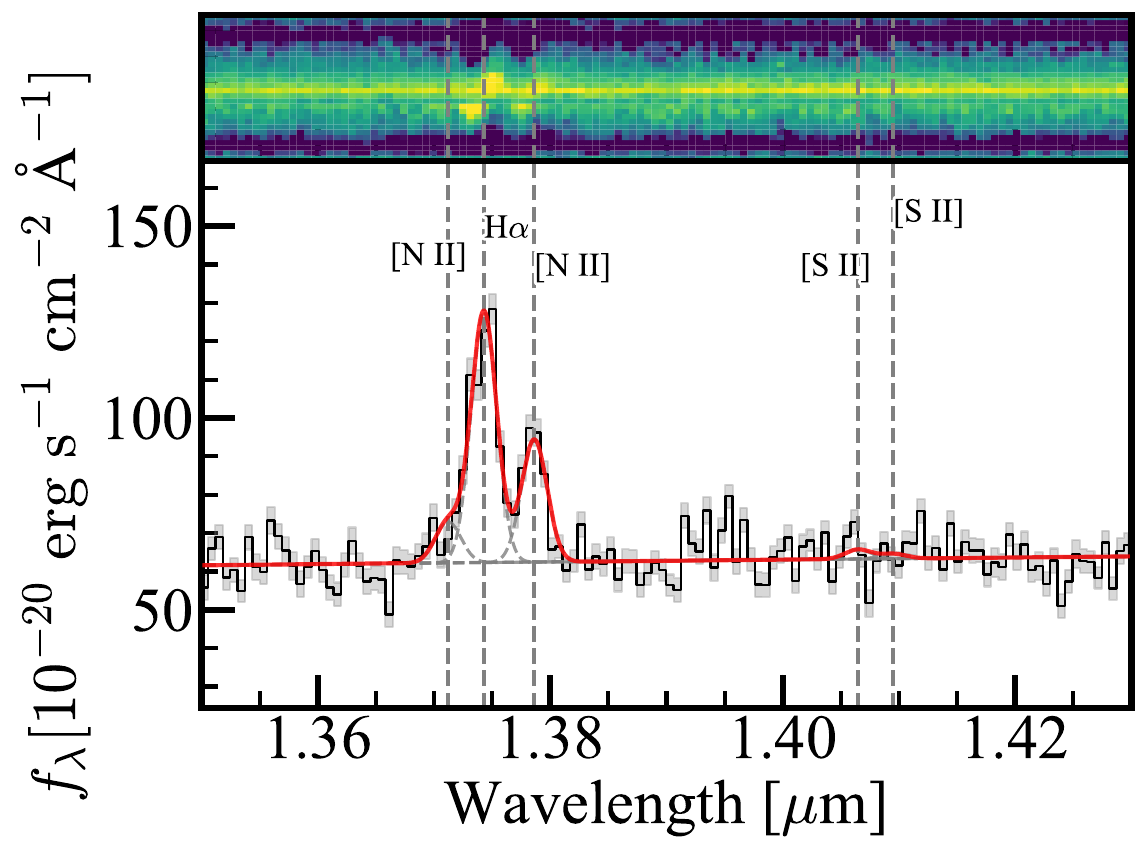}{0.15\textwidth}{}
            \fig{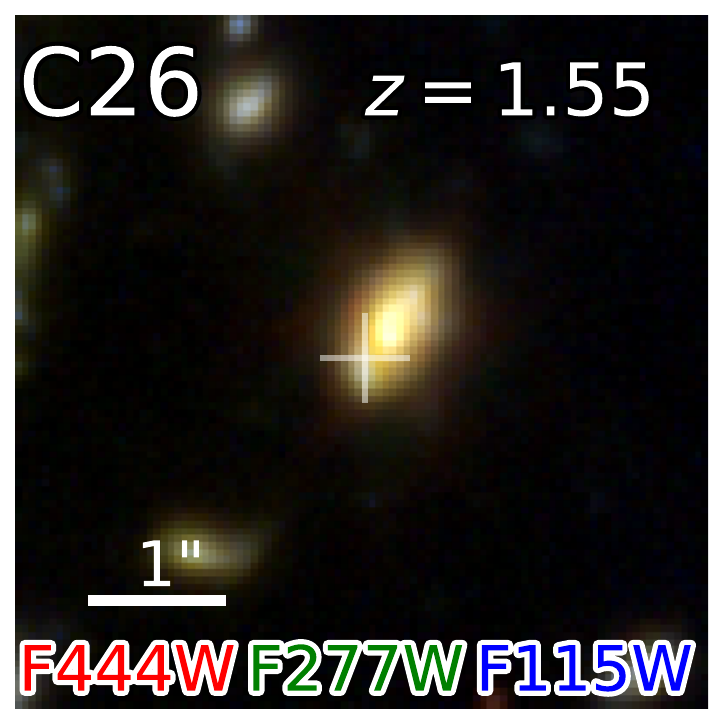}{0.1\textwidth}{}
            \fig{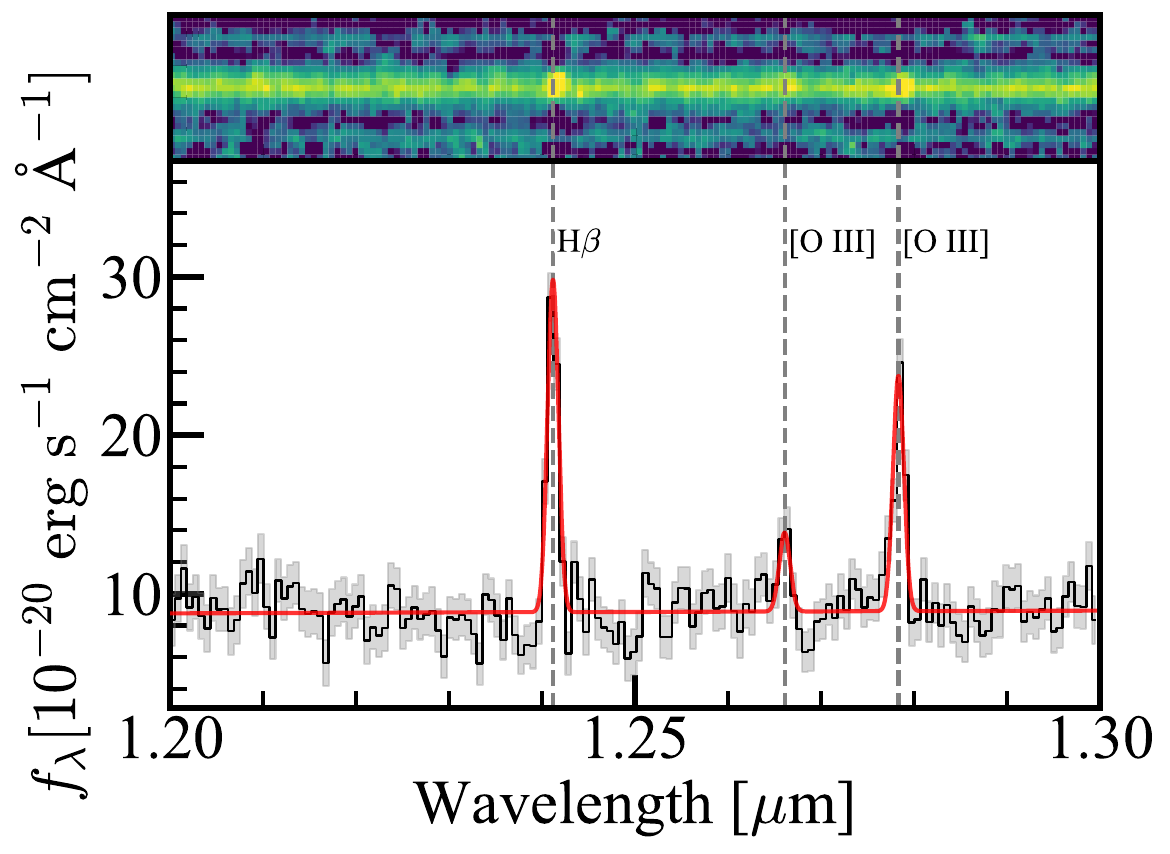}{0.15\textwidth}{}
            \fig{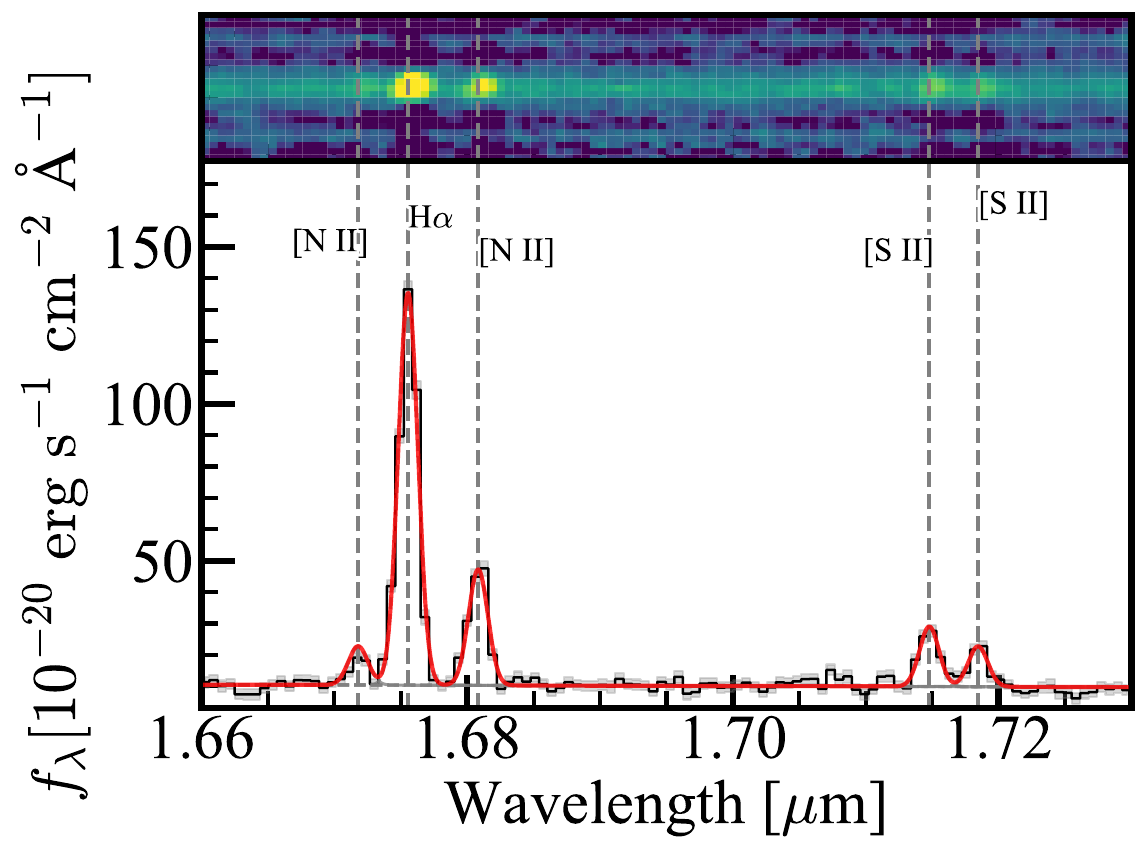}{0.15\textwidth}{}}
  \gridline{
            \fig{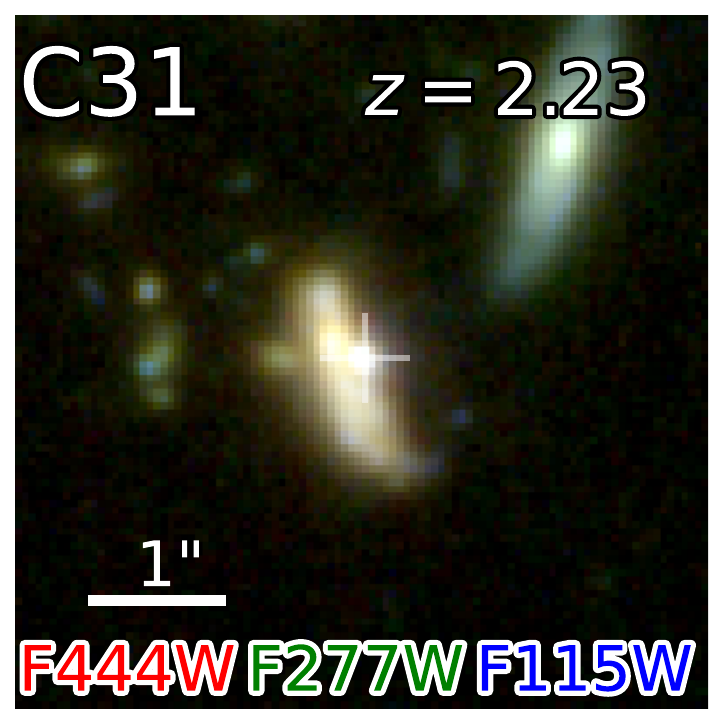}{0.1\textwidth}{}
            \fig{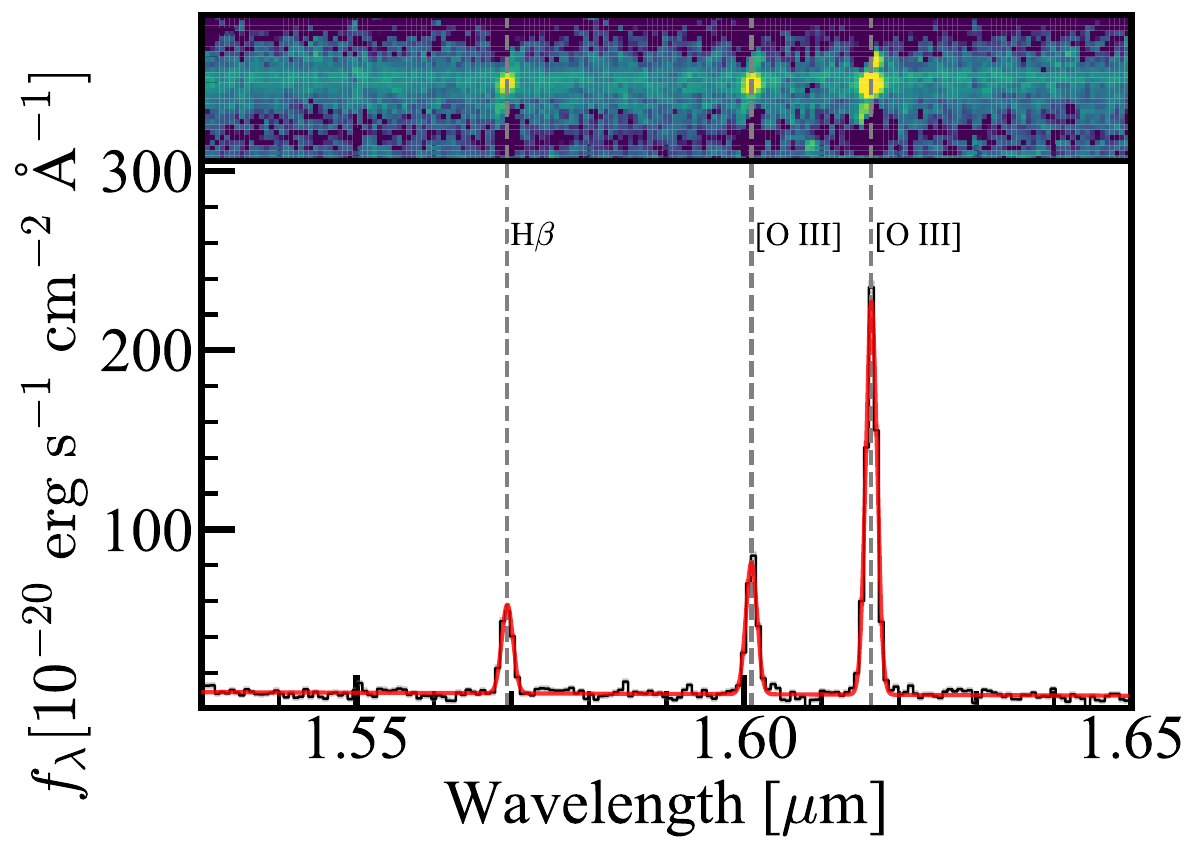}{0.15\textwidth}{}
            \fig{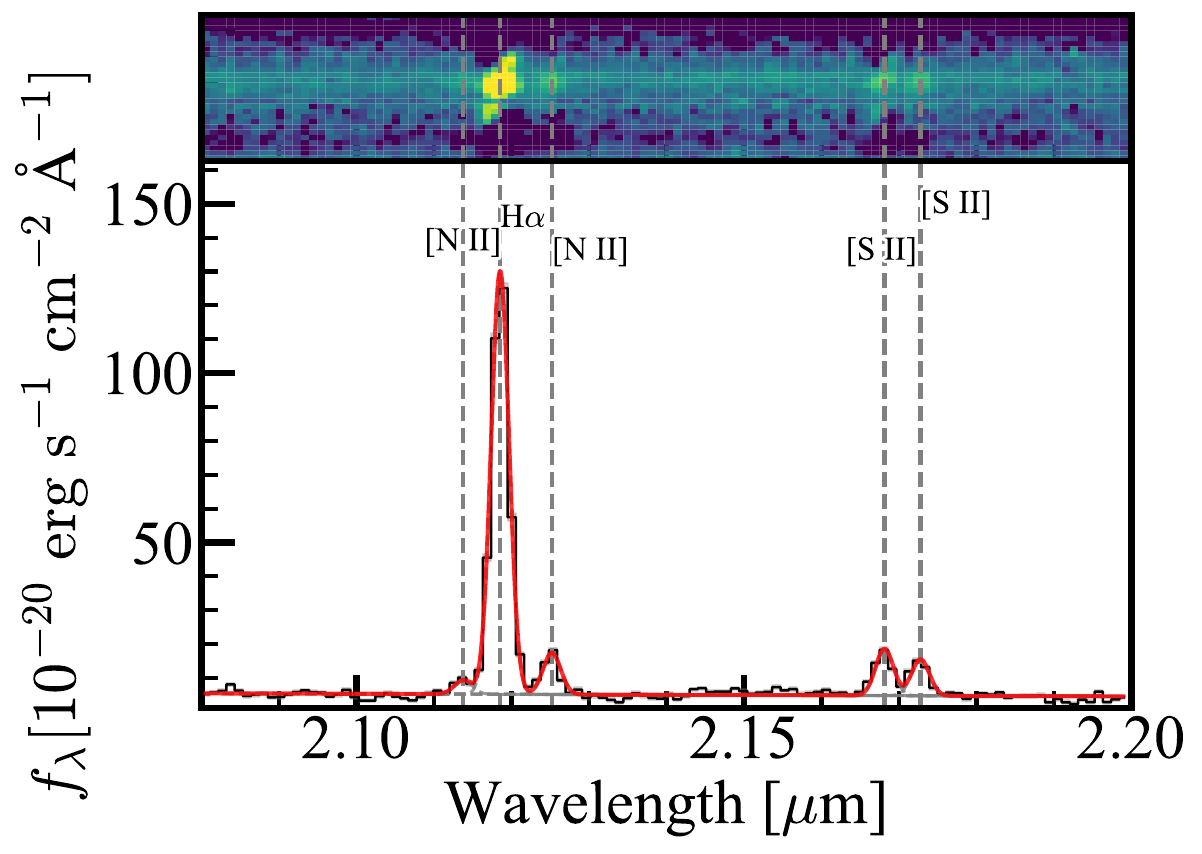}{0.15\textwidth}{}
            \hfill \hfill \hfill \hfill \hfill \hfill
            \hfill \hfill \hfill \hfill \hfill \hfill
            \hfill \hfill \hfill \hfill \hfill \hfill
            \hfill \hfill \hfill \hfill \hfill \hfill
            \hfill \hfill \hfill \hfill \hfill \hfill
            \hfill \hfill \hfill 
            }
  \end{minipage}
}
\caption{Same as in Figure~\ref{fig:snapshots1} but for C08--C31.}
\label{fig:snapshots3}
\end{figure*}
%%%%%%%%%%%%%%%%%%%%%%%%%%%%%%%%%%

Figures~\ref{fig:snapshots1}--\ref{fig:snapshots3} show the NIRCam images and NIRSpec spectra of the sample.  
The left panels of Figure~\ref{fig:snapshots1} show $5\arcsec \times5\arcsec$ cutouts. 
Nearly half of the UDF+ASPECS sources show multiple clumps or tidal features in the rest-frame optical NIRCam images via visual inspection. 
Previous studies suggest the connection between galaxy mergers and SMGs, especially for bright SMGs (e.g., \citealt{hopkins08}). 
\commentred{Consistent with previous studies of dusty star-forming galaxies (e.g., \citealt{gillman24, hodge25}), not all sources show merger-like morphologies.}
% possibly because the UDF+ASPECS sample is fainter than typical bright-SMG samples. 
A subset of sources (e.g., UDF13, C08) shows compact morphology and X-ray detections, suggesting AGN \commentred{activity} (see also Section~\ref{subsec:AGN-SMGs}). 
These morphological features are also reported in other JWST imaging studies (e.g., \citealt{boogaard24} for the ASPECS sources; \citealt{hodge25} for ALESS sources; \citealt{ikeda25} for SMGs at $3.5<z<4.5$; \citealt{gillman24} for ALMA sources in the Cosmic Evolution Survey and UKIDSS Ultra-Deep Survey field; \citealt{mckinney25} and \citealt{ren25} for SCUBA SMGs). 

The middle and right panels of Figure~\ref{fig:snapshots1} show the spectra around H$\beta$+\oiii\ and H$\alpha$+\nii+\sii, respectively. 
All sources show detections ($\mathrm{S/N}>3$) of H$\alpha$ and \nii, while shorter wavelength optical emission lines (H$\beta$ and \oiii) are weak. 
We estimate the dust attenuation $E(B-V)$ from the Balmer decrement, assuming an electron density of $n_e=100~\mathrm{cm^{-3}}$, an electron temperature of $T_e=15,000~\mathrm{K}$, case B recombination \citep{osterbrock06}, and \citet{calzetti00} law. 
We obtain $E(B-V)\sim0.3$--$1.8$ for 12 sources. 
For the remaining four sources, the H$\beta$ line is not detected, and \commentred{we derive} 3$\sigma$ lower limits of $E(B-V)\gtrsim1.0$. 
We list their $E(B-V)$ in Table~\ref{tab:properties}. 
We use these emission-line measurements in the following sections.

\subsection{AGN Signatures} 
\label{subsec:AGN-SMGs} 

It is insightful to search for AGNs in these faint SMGs. 
We identify AGN candidates using three indicators: Chandra X-ray detections, the Baldwin–Phillips–Terlevich (BPT; \citealt{baldwin81}) diagram, and the presence of broad Balmer lines. 

X-ray emission is a key \commentred{signature} of AGN activity. 
As described in Section~\ref{subsec:x-ray}, we cross-matched the UDF+ASPECS sources to the Chandra CDF-S 7~Ms source catalog \citep{luo17}. 
We find seven matches among 16 sources (44\%) within $1\arcsec$. 
We classify these X-ray–matched sources as AGNs (see also \citealt{ueda18}).

\begin{figure}
\plotone{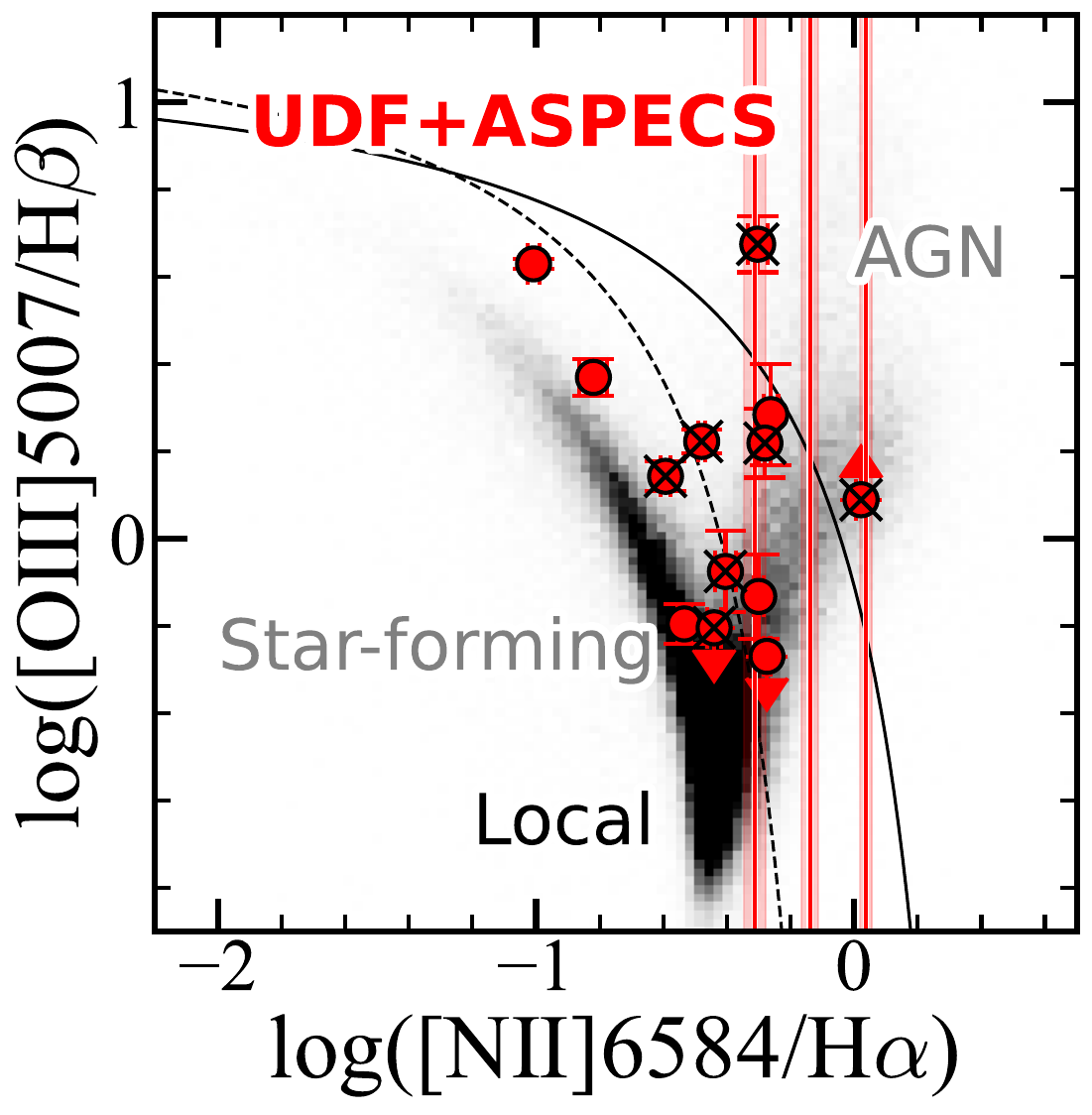}
\caption{
BPT diagram. The UDF+ASPECS sources are plotted as red circles or red lines. 
If neither \oiii\ nor H$\beta$ is detected, we indicate the sources as vertical lines together with red shades indicating $1\sigma$ uncertainties. 
The black crosses indicate the X-ray-detected sources. 
Two demarcation curves between star-forming galaxies and AGNs are shown as dashed \citep{kauffmann03} and solid \citep{kewley01} lines. 
The background gray scales show the distribution of the SDSS galaxies taken from the SDSS DR16 \citep{ahumada20}. 
}
\label{fig:BPT}
\end{figure}

Bringing the rest-frame optical spectra, we also constrain the AGN activity. 
Figure~\ref{fig:BPT} shows the BPT diagram, with the demarcation curves separating star-forming galaxies and AGNs \citep{kewley01, kauffmann03}. 
Sources located above the curves are classified as AGN\commentred{s}. 
In this study, we adopt \commentred{the} \citet{kauffmann03} curve (black dashed curve in Figure~\ref{fig:BPT}) as a demarcation. 
If \oiii\ and H$\beta$ are not detected, we indicate the sources as vertical lines. 
For these sources, we classify a galaxy as an AGN when $\log([\mathrm{N\,II}]\lambda6584/\mathrm{H}\alpha) > -0.2$ (e.g., \citealt{swinbank04}). 
When we calculate the line ratio of the BPT diagram, we correct the dust extinction using $E(B-V)$ obtained from the H$\alpha$/H$\beta$ ratio if both lines are available ($\mathrm{S/N>3}$). 
We do not propagate the uncertainty in $E(B-V)$ into the line-ratio uncertainties. 
\commentred{We do not take into account the line-ratio uncertainties for classification.}

From the BPT diagram, eight of 16 sources (50\%) are classified as AGNs. 
Table~\ref{tab:properties} also lists the BPT-AGN classification for each source. 
\commentred{Several sources lie close to the \citet{kauffmann03} demarcation curve. Their classifications can change within the uncertainties in the emission-line ratios, but the AGN fraction remains high.}
Interestingly, some sources are classified as AGN in the BPT diagram, but have no X-ray counterparts (e.g., UDF2, UDF16). 
This might indicate that the X-rays of these sources are obscured by gas in the dust-sublimation region \citep{mizukoshi24}, and the observed X-ray could be weaker than the CDF-S detection limit. 
In addition, some sources lie between the \citet{kauffmann03} and \citet{kewley01} curves. 
This indicates a composite nature, where star formation alone may not account for the observed excitation. 
Such line ratios can be powered by a mixture of both star formation and AGN activity. 

\commentred{
We also examine the impact of adopting the more conservative \citet{kewley01} demarcation curve for the BPT-based AGN classification (black solid curve in Figure~\ref{fig:BPT}). 
With this criterion, two sources, UDF13 and C08, are classified as AGN (Table~\ref{tab:properties}), and the total AGN fraction, including X-ray-detected AGN, becomes 44\% ($=7/16$). 
Although this is lower than our fiducial estimate, nearly half of the sources are still classified as AGNs. 
Several sources lying between the \citet{kauffmann03} and \citet{kewley01} demarcation curves are detected in X-rays, suggesting that this intermediate region likely includes galaxies with an AGN contribution. 
We therefore regard the \citet{kauffmann03} demarcation curve as our fiducial BPT criterion and the \citet{kewley01} curve as a conservative criterion.
}

We also note that the BPT locus and the demarcation curves may evolve with redshift \citep{kewley13} because they depend on the ionizing radiation field and ISM conditions. 
In this study, we adopt the traditional \citet{kauffmann03} demarcation curve as a criterion \commentred{for AGN classification}, following previous studies (e.g., \citealt{jones24}). 

%%% fig: BPT %%%
%%%%%%%%%%%%%%%%

Broad Balmer emission lines ($\mathrm{FWHM}\gtrsim1000~\mathrm{km~s^{-1}}$) with narrow forbidden lines also suggest that the broad component is powered by AGN activity, especially from the broad line region (e.g., \citealt{maiolino24, greene24}). 
In our sample, C08 ($z=3.71$; Figure~\ref{fig:snapshots3}) shows broad H$\alpha$ lines with $\mathrm{FWHM}\sim6500\,\mathrm{km\,s^{-1}}$ and narrow \oiii\ which is also reported in \citet{juodbalis25} (object ID: JADES-GS-209777, Figure~1 of \citealt{juodbalis25}). 
We therefore classify C08 as a type-1 AGN. 
\citet{juodbalis25} estimate a black hole mass ($M_\mathrm{BH}$) of $\log(M_\mathrm{BH}/M_\odot)=8.9\pm0.3$ from the broad H$\alpha$ assuming the virial relation \citep{reines15}. 
C08 also shows a compact morphology (Figure~\ref{fig:snapshots3}), an X-ray detection (CID: 715, \citealt{luo17}), and a BPT-AGN (the top-right outlier in Figure~\ref{fig:BPT}), further supporting the presence of AGN. 
\commentred{It is worth noting that C08 has an extreme continuum shape with a sharp cutoff around 3000\AA\ \citep[The Hatchling;][see also \citealt{li26}]{ma26}.}
Other sources do not show prominent broad Balmer components ($\mathrm{FWHM}\gtrsim1000\,\mathrm{km\,s^{-1}}$) with narrow \oiii, indicating that other X-ray-selected and/or BPT-AGNs are type-2, and the broad line regions might be obscured. 

UDF13 (X-ray detection, BPT-AGN, and compact) shows a broad H$\alpha$ component with $\mathrm{FWHM}=1400\pm160~\mathrm{km~s^{-1}}$ and $\Delta\mathrm{AIC}=40$. 
However, \oiii\ lines are not detected with sufficient S/N, so we cannot exclude an outflow origin for the broad emission. 
Fitting the H$\alpha$+\nii+\sii\ complex with a model that includes both narrow emission and broad outflow components yields a better fit than a model only with narrow lines. 
Given these ambiguities, we do not further discuss UDF13 as a broad-line AGN. 
We briefly note that UDF13 (JADES ID: 208000) is also identified as a MIRI AGN candidate by \citet{lyu24} and \citet{rieke25}. 

It is worth noting that outflow and galaxy mergers can also broaden the line widths. 
If these mechanisms dominate the line width, we observe broad components in both permitted and forbidden lines. 
However, C08 has the broad component only in the permitted (H$\alpha$ and H$\beta$) lines, suggesting that the broad components are powered by AGN activity.

\begin{figure}
    \centering
    % \includegraphics[width=1.0\linewidth]{figures/fig_AGN_fraction_v1.pdf}
    % \plotone{figures/fig_AGN_fraction_v1.pdf}
    \plotone{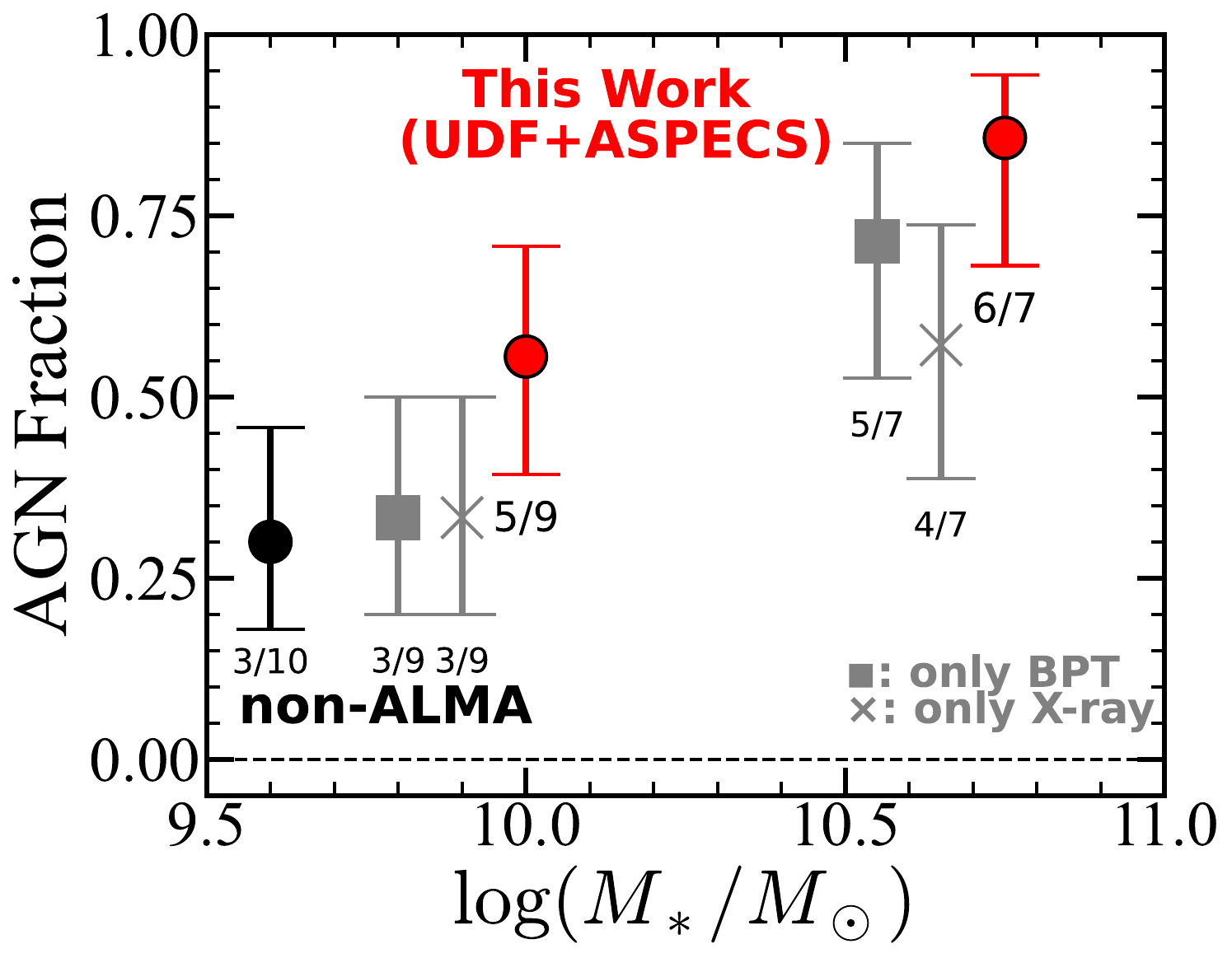}
    \caption{
    Fraction of sources classified as AGNs based on the BPT diagram and/or X-rays as a function of stellar mass. Red circles show the AGN fraction identified via the BPT diagram and/or X-rays, while gray squares and crosses show the fractions based on the BPT diagram only and X-rays only, respectively. 
    Numbers below the plot show the AGN count over the total number of galaxies in each stellar-mass bin.
    Error bars indicate $1\sigma$ binomial uncertainties. 
    The stellar mass bins are $\log(M_*/M_\odot)=9.5$--$10.5$ and $10.5$--$11.0$. 
    For reference, ALMA non-detected sources ($\log(M_*/M_\odot)=9.5$--$10.5$) at $z=1.5$--$3.5$ are also indicated by a black circle.
    For clarity, the points are slightly offset along the x-axis. 
    }
    \label{fig:AGN-fraction}
\end{figure}

In total, 11 out of 16 sources (69\%) are classified as AGNs. 
Figure~\ref{fig:AGN-fraction} summarizes the AGN fraction (identified via the BPT diagram and/or X-rays) as a function of stellar mass. 
In both stellar-mass bins ($\log(M_*/M_\odot)=9.5$--$10.5$ and $10.5$--$11.0$), the AGN fraction exceeds 50\% (56\% and 86\%, respectively). 
The high incidence of AGN signatures in X-rays and/or the BPT diagram is consistent with a scenario in which many of these systems host obscured AGNs (e.g., \citealt{hopkins08}). 
\citet{ueda18} show the Chandra X-ray analysis for the UDF and ALMA twenty-six arcmin$^2$ survey of GOODS-S at one millimeter (ASAGAO; \citealt{hatsukade18}) sources.  
They also report the high AGN fraction for these samples, 38\% for UDF and 67\% for ASAGAO (see also, e.g., \citealt{wang13}). 
We further identify the AGNs that are missed in X-ray surveys and constrain their rest-frame optical properties. 
This result highlights the importance of a multiwavelength approach to AGN diagnostics. 
We note that these extensive surveys have been conducted in other fields. 
For example, \citet{umehata15} reported that $\sim50\%$ of SMGs in the SSA22 protocluster core host X-ray-luminous AGNs based on ALMA and Chandra. 

We consider the detection limit of the broad H$\alpha$ line in the JWST/NIRSpec spectra. 
We can assume a typical stellar mass of the UDF+ASPECS sample to be $\log(M_*/M_\odot)\sim10.5$, and $M_*/M_\mathrm{BH}=10^3$ (e.g., \citealt{kormendy13}). 
This corresponds to the black hole mass of $\log(M_\mathrm{BH}/M_\odot)\sim7.5$.
Assuming a H$\alpha$ broad line width of $\mathrm{FWHM=1500\,km\,s^{-1}}$ and the \citet{reines15} relation, the corresponding H$\alpha$ luminosity is $\sim10^{42}\,\mathrm{erg~s^{-1}}$. 
If there is no dust extinction, we detect the broad H$\alpha$ component, further supporting that the AGNs in the UDF+ASPECS sample are dust-obscured. 

For comparison, we estimate the AGN fraction for ALMA non-detected sources. 
We select sources observed with NIRSpec in the JADES DR4 and SMILES DR2 spectroscopic catalogs and in the ASPECS footprint. 
Following the same procedure as for the UDF+ASPECS sample, we classify AGN candidates using X-ray and BPT, based on the CDF-S 7~Ms catalog and emission-line fluxes reported in JADES DR4 or SMILES DR2 spectroscopic catalogs. 
Since robust stellar masses are not available for all sources in the parent sample, we estimate stellar masses from the F444W absolute magnitude ($M_\mathrm{F444W}$) using empirical calibration. 
F444W probes the rest-frame $\sim1$--$2\,\mu$m at $z\sim2$, and thus primarily traces the stellar continuum. 
We derive a linear relation between $M_\mathrm{F444W}$ and $\log(M_*/M_\odot)$ using the COSMOS-Web photometric catalog (\citealt{casey23, shuntov25}), which provides both photometry and stellar masses from the SED fitting. We select galaxies at $z=1.5$--$3.5$ and $m_\mathrm{F444W}<26$ from the catalog\commentred{, applying the F444W magnitude cut to ensure reliable stellar-mass estimates from sufficiently high-S/N rest-frame optical/near-infrared photometry.}
We then fit a linear model in the $(M_\mathrm{F444W},\,\log(M_*/M_\odot))$ plane. 
This calibration has a scatter of $\sim0.3$ dex. 
We then apply this linear relation to the galaxies at $z=1.5$--$3.5$ to estimate their stellar masses. 
For sources at $z=1.5$--$3.5$ and $\log(M_*/M_\odot)=9.5$--$10.5$, we find an AGN fraction of $\sim30\%$ (3/10; the black circle in Figure~\ref{fig:AGN-fraction}). 
We caution that the inferred AGN fraction for lower-mass galaxies is likely a lower limit. 
Due to the incompleteness of the Chandra X-ray and JWST/NIRSpec observations, we might miss the AGNs for fainter galaxies. 

%%% fig: AGN-fraction %%%%
%%%%%%%%%%%%%%%%%%%%%%%%%%%

\subsection{Relation between [OIII] and X-ray Luminosity}

\begin{figure}
    \centering
    \plotone{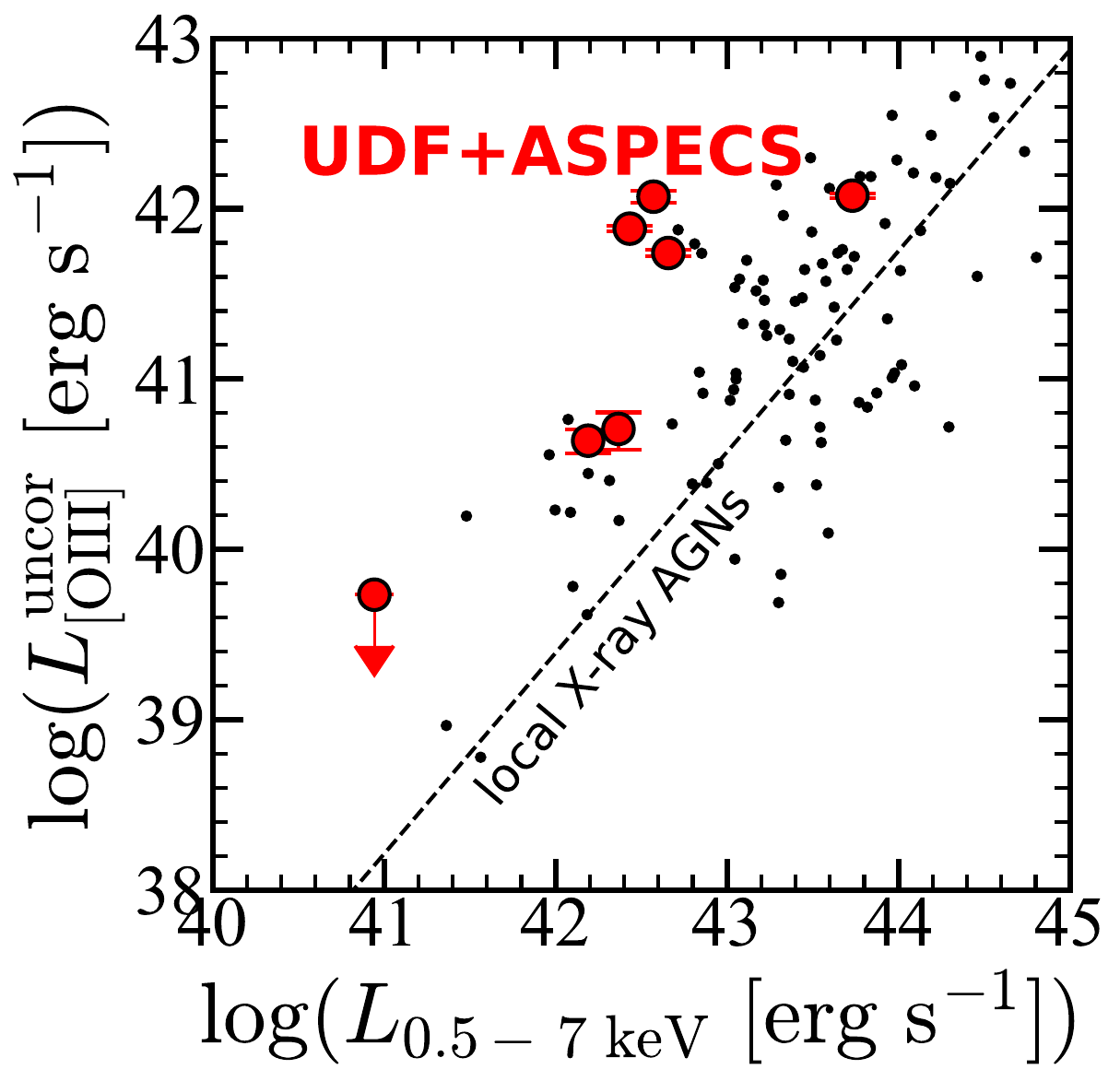}
    \caption{
    Correlation between the intrinsic X-ray luminosity in the 0.5--7~keV band and observed \oiii$\lambda5007$ luminosity. The red circles indicate the X-ray-detected sources in the UDF+ASPECS sample. The black circles and dashed line show the local X-ray AGNs \commentred{in the Swift/BAT 9-month hard X-ray survey} and their best-fit regression lines reported in \citet{ueda15}, respectively. For sources and the best-fit line from \citet{ueda15}, we convert the reported 2--10~keV luminosities to 0.5--7~keV luminosities assuming a photon index of $\Gamma=1.8$.
    }
    \label{fig:OIII-Xray}
\end{figure}

We compare the relation between the observed \oiii$\lambda5007$ luminosity and the intrinsic X-ray luminosity for the X-ray selected AGNs in our UDF+ASPECS sample with that for local X-ray AGNs. 
The \oiii\ emission lines can also trace narrow-line regions, and comparisons with their X-ray luminosities are insightful. 
Figure~\ref{fig:OIII-Xray} shows the relation between the \oiii\ luminosity without dust-extinction correction and the absorption-corrected $0.5$–$7$~keV X-ray luminosity. Relative to the local \oiii--X-ray relation \commentred{(black dashed line)} from \citet{ueda15}, our sources show higher \oiii\ luminosities at a fixed X-ray luminosity by $\gtrsim0.5$~dex. 
For a fair comparison, we plot the local AGNs in \citet{ueda15} after converting the reported 2--10~keV X-ray luminosities to 0.5--7~keV luminosities assuming a photon index of $\Gamma=1.8$. 
\oiii\ luminosities are estimated using slit-loss-corrected spectra provided in JADES DR4 and SMILES DR2.

Several effects could contribute to this offset. 
1) Contamination from star formation. Star formation can contribute to the total \oiii\ flux and could lead to high \oiii\ luminosity. 
2) Different bolometric correction. If these AGNs have larger X-ray bolometric corrections $\kappa_{X}$ (ratio between bolometric luminosity and X-ray luminosity) than the nominal value (e.g., $\kappa_{0.5\text{--}8\,\mathrm{keV}}=5$--$20$; \citealt{rigby09, duras20}), they would become \oiii-bright at a given X-ray luminosity compared to local AGNs (see also discussions in \citealt{ueda18}).
3) Compton-thick case. If the AGNs are Compton-thick, we might observe X-ray emission dominated by scattered components. This could underestimate the X-ray luminosities even after the absorption correction.

%%%%%%%%%%%%%%%%%%%%%%%%%%%%%
%%%%%%%%%%%%%%%%%%%%%%%%%%%%%

\subsection{Gas-phase Metallicity} 
\label{subsec:metallicity-SMGs} 

Bringing the rest-frame optical emission lines, we estimate gas-phase metallicities for the dust continuum sources. 
Since the \commentred{auroral} line \oiii$\lambda4363$ is not detected for those dust continuum sources, we do not use the direct method to estimate metallicity. 
We instead adopt commonly used strong line calibrations. 
In our galaxy sample, all sources have H$\alpha$ and \nii\ lines, and we utilize the N2 index (\nii$\lambda6584$/H$\alpha$) as an indicator of metallicity. 

We use the N2 calibration presented in \citet{sanders21}: 
\begin{equation}
    \log(\mathrm{[\text{N\,\textsc{ii}}]\lambda6584/H\alpha}) = c_0 + c_1x + c_2x^2 + c_3x^3, 
\end{equation}
where $x=12+\log(\mathrm{O/H})-8.69$ and $(c_0, c_1, c_2, c_3) = (-0.606, 1.28, -0.435, -0.485)$. 
This relation is calibrated using star-forming galaxies at $z\sim0$ with the stellar mass range of $\log(M_*/M_\odot)=9.0$--$11.0$. 
We note that the \citet{sanders21} calibration is established using star-forming galaxies and removes AGNs from their sample, using X-ray detections and infrared properties (see Section 2.3 of \citealt{sanders21}). 

To conservatively test the contribution from AGNs to the N2 index measurements, we also use the N2 index calibration presented in \citet{caravalho20}. 
\commentred{This calibration is based on Seyfert 2 AGNs from}
% This is calibrated by the Seyfert 2 AGNs based on the 
Sloan Digital Sky Survey (SDSS; \citealt{york00}): 
\begin{equation}
    Z/Z_\odot = a^{\log(\mathrm{[\text{N\,\textsc{ii}}]\lambda6584/H\alpha})} + b, 
\end{equation}
where $a=4.01$, $b=-0.07$. 
N2 index uncertainties are propagated to the metallicity estimate. 
The systematic calibration uncertainties (0.15~dex for \citealt{sanders21} and 0.10~dex for \citealt{caravalho20}) are also included in quadrature with the metallicity uncertainties, which dominate the total uncertainties. 
The results using \citet{sanders21} and \citet{caravalho20} calibrations are summarized in Table~\ref{tab:properties}. 
Hereafter, if \commentred{a source shows AGN signatures} (X-ray or BPT diagram; Section~\ref{subsec:AGN-SMGs}), we use the metallicity derived from the Seyfert 2 AGN calibration \citep{caravalho20}; otherwise, we use the star-forming galaxy calibration \citep{sanders21}. 

\begin{figure}
\plotone{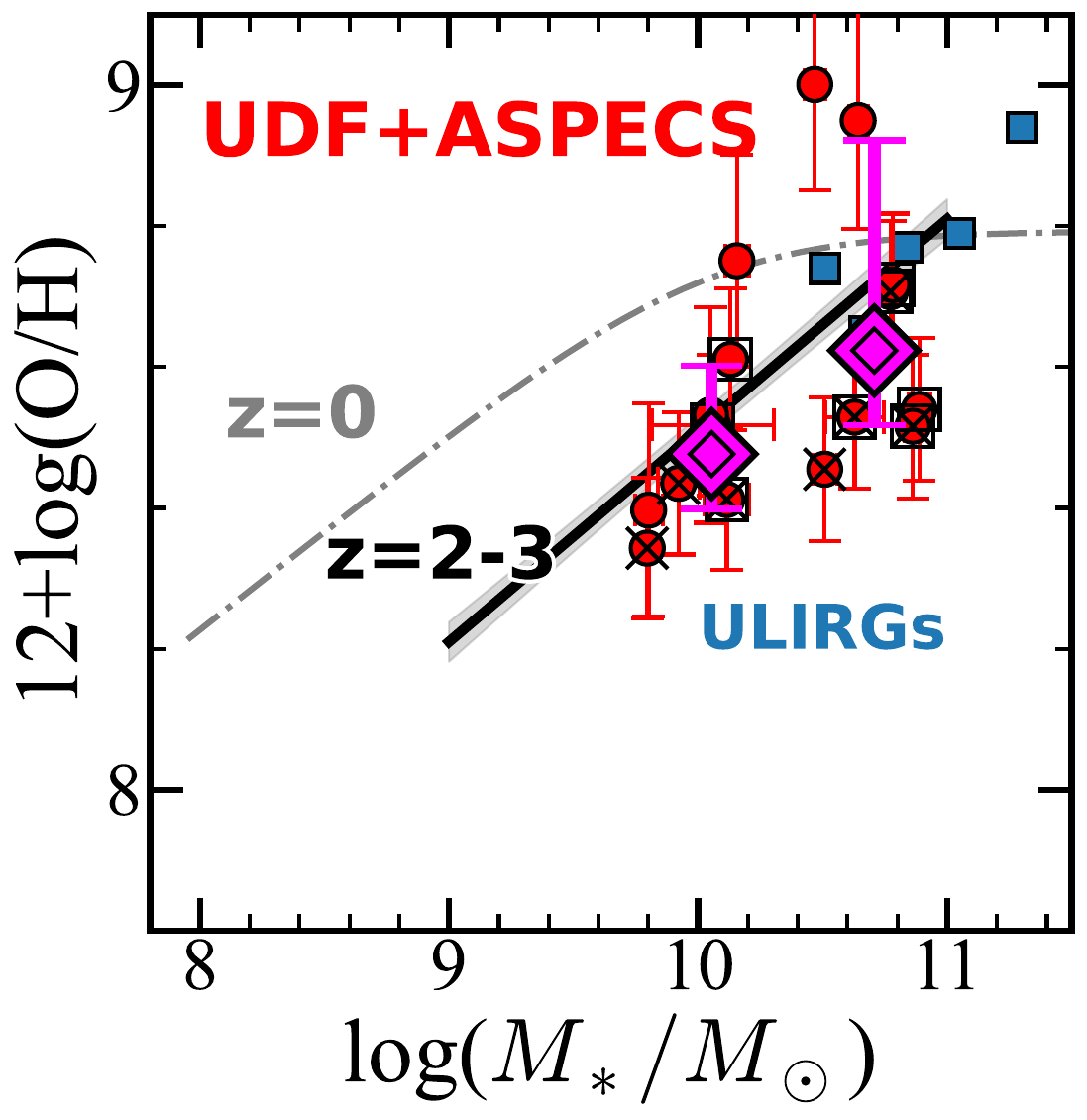}
\caption{
Relation between stellar mass and gas-phase metallicity. 
The red circles show the UDF+ASPECS sample. 
The black crosses and black squares show the X-ray AGN and BPT-AGN, respectively. 
The magenta double diamonds with error bars show the median and the 16th–84th percentiles of the UDF+ASPECS sources in two stellar-mass bins ($M_*<10^{10.4}~M_\odot$ and $M_*>10^{10.4}~M_\odot$).
The black solid line and the gray dot-dashed line show the relations of star-forming galaxies at $z=2$--$3$ \citep{sanders21} and $z=0$ \citep{curti20}, respectively. 
The local ULIRGs are plotted as the blue squares \citep{chartab22}. 
}
\label{fig:MZR}
\end{figure}

Figure~\ref{fig:MZR} shows the relation between gas-phase metallicity and stellar mass. 
The UDF+ASPECS sources (red circles) have metallicities in the range of $12+\log(\mathrm{O/H})=8.3$--$9.0$ (i.e., $0.4$--$2Z_\odot$). 
Compared to the mass metallicity relation at $z=2$--$3$ \citep{sanders21}, the UDF+ASPECS sources have comparable metallicities and broadly follow the mass metallicity relation. 
Since the UDF+ASPECS dust continuum sources have high stellar masses ($\log(M_*/M_\odot)\gtrsim10$), some sources reach nearly solar metallicities. 
\commentred{These nearly solar metallicities are comparable to those observed in local ULIRGs (blue squares in Figure~\ref{fig:MZR}), suggesting similar levels of chemical enrichment.}
\commentred{A similar metallicity range for dusty star-forming galaxies has also been reported in recent JWST/NIRSpec studies by \citet{gillman26}.}
We discuss implications from the metallicity for dusty galaxies in Section~\ref{sec:discussion-SMGs}.

%%% fig: metallicity %%%%%%%%%
%%%%%%%%%%%%%%%%%%%%%%%%%%%%%%

\subsection{Electron Density} 
\label{subsec:electron-density} 

The electron density provides a useful diagnostic of the physical conditions of the ISM (e.g., \citealt{sanders16}).
We derive the electron densities of the dust-continuum sources from the \sii$\lambda\lambda6717,6731$ doublet ratio when both lines are detected with $\mathrm{S/N} > 3$. 
The medium-resolution grating data ($R\sim1000$) are sufficient to resolve the \sii\ doublet. 
We use the PyNeb (version 1.1.28; \citealt{luridiana15}) package \texttt{getTemDen} with the default PyNeb atomic datasets (S$^+$ transition probabilities from \citealt{rynkun19} and S$^+$ collision strengths from \citealt{tayal10}) to convert the observed ratios into electron densities.
In these calculations, we assume an electron temperature of $T_e = 15{,}000$\,K. 
Varying the assumed temperature within a plausible range (e.g., $T_e = 10{,}000$--$20{,}000$\,K) does not affect the general trends discussed in this paper. 
When the observed \sii\ doublet ratio falls beyond the low (high) density limits, we set the $2\sigma$ upper (lower) limits of the electron density. 
The uncertainties in the \sii\ doublet ratio are propagated to the electron density uncertainties. 
The \sii\ doublet ratio and electron densities are summarized in Table~\ref{tab:properties}. 
The median value of the \sii\ electron density of our sample is $n_e([\text{S\,\textsc{ii}}])_\mathrm{med} = 541~\mathrm{cm^{-3}}$. 

%%% fig: electron density %%%
\begin{figure*}
\gridline{\fig{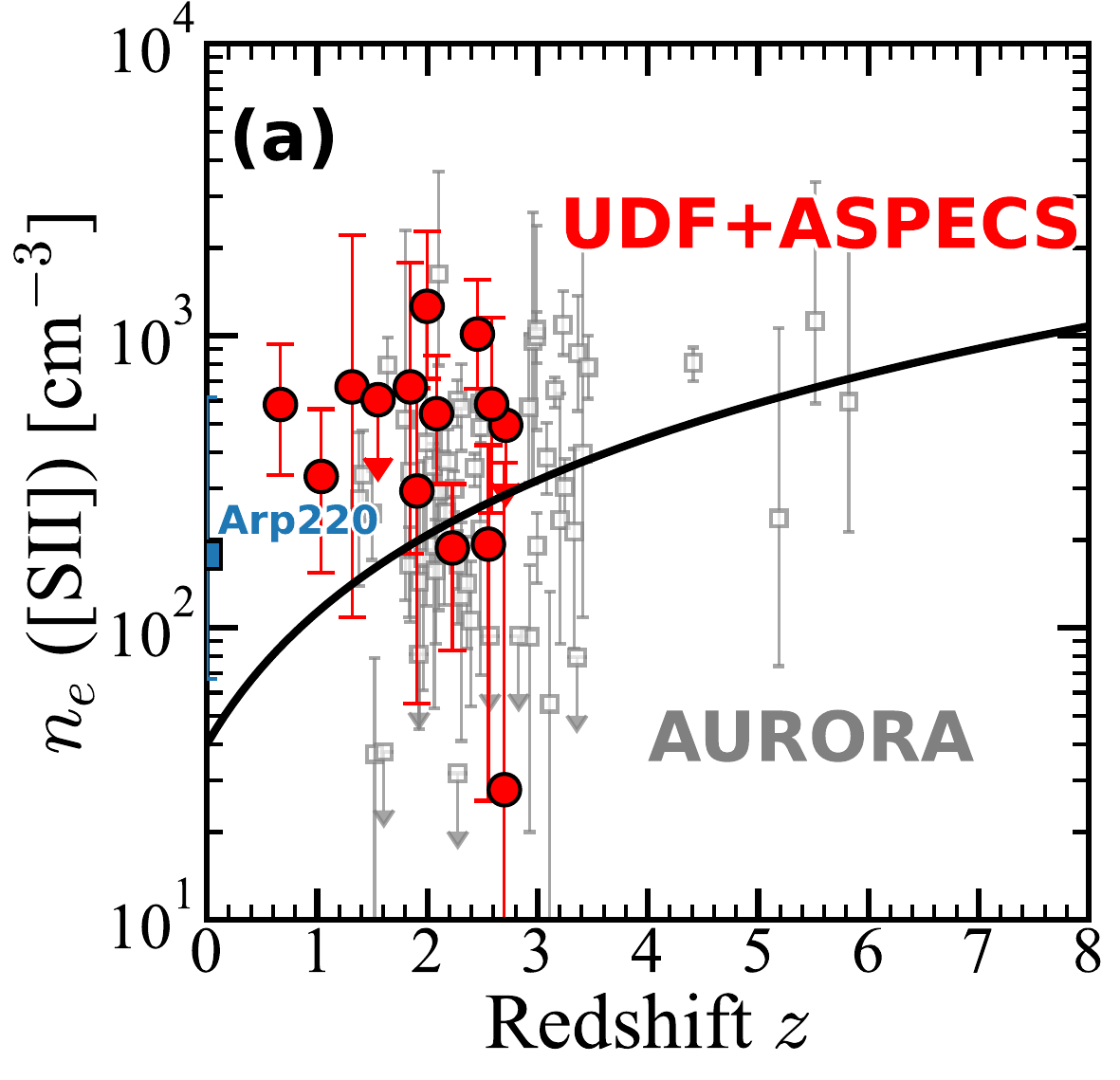}{0.44\textwidth}{}
          \fig{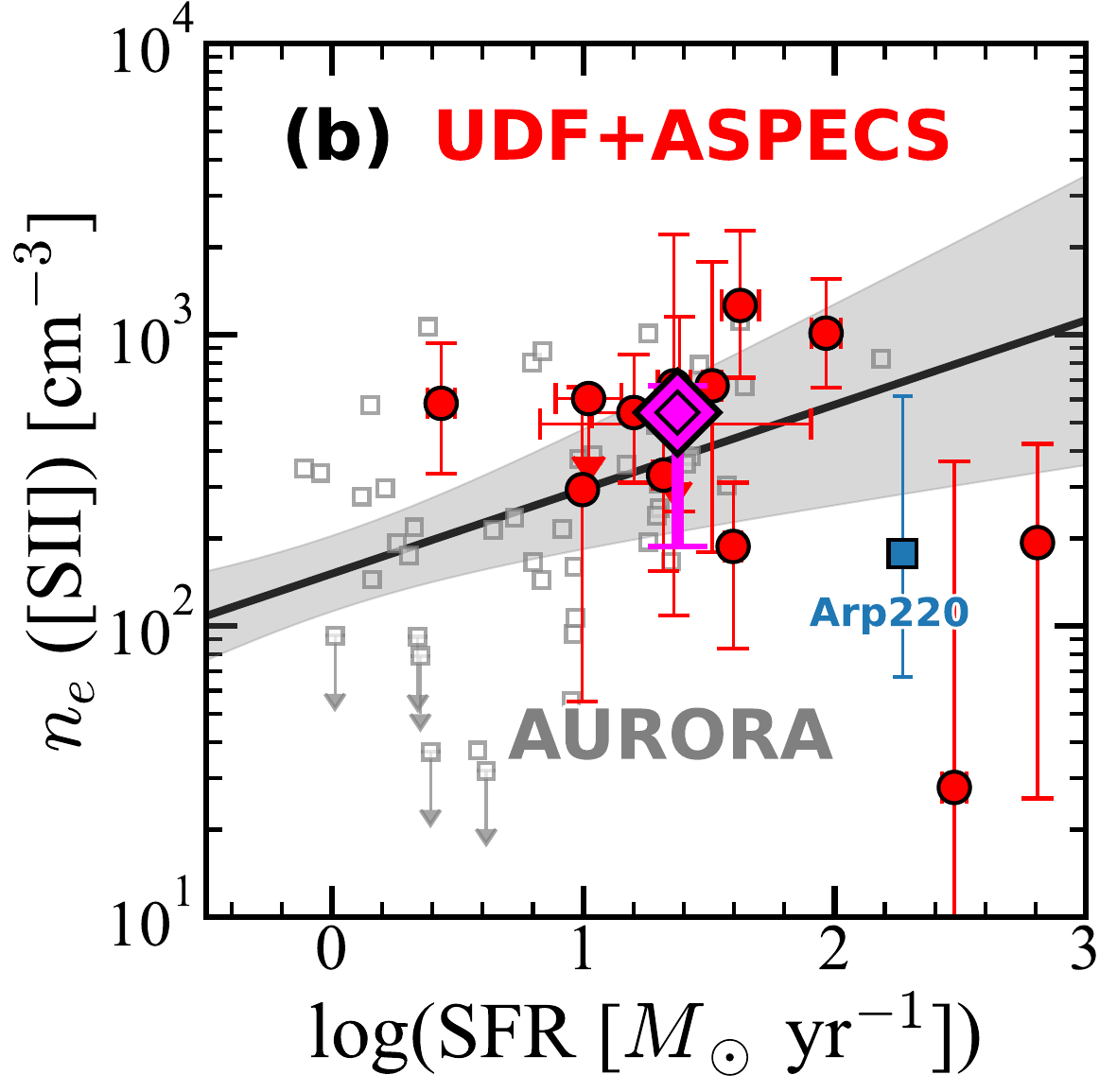}{0.43\textwidth}{}
          }
\gridline{
        \fig{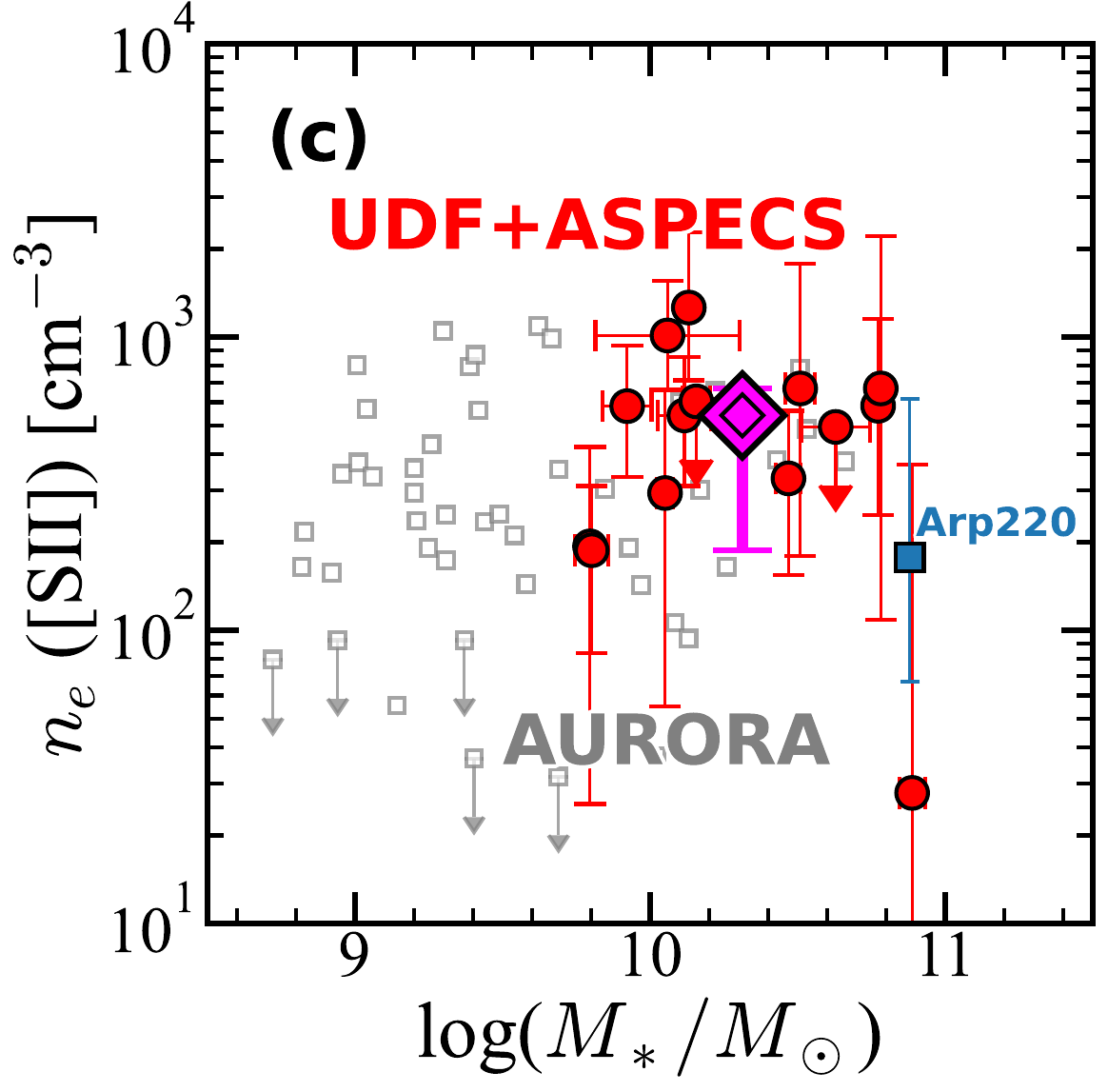}{0.43\textwidth}{}
        \fig{figures/fig_ne_sigmaSFR_v5.pdf}{0.42\textwidth}{}
}
\caption{
\sii\ electron density properties. 
The red circles show the UDF+ASPECS sample. We compare the electron densities to redshift (a), SFR (b), stellar mass (c), and surface SFR density (d). 
The magenta double diamonds with error bars show the median and the 16th–84th percentiles of the UDF+ASPECS sources. 
For comparison, the AURORA sample at $z=1.4$--$5.8$ is plotted as the gray squares \citep{topping25}. 
The redshift evolution curve (black curve; panel (a)) is quoted from \citet{topping25}. 
% The black lines and gray shades in the panels (b) and (d) show the correlations reported in \citet{topping25}. 
The black lines and gray shades in the panels (b) and (d) show the correlations reported in \citet{topping25}. 
\commentred{The black lines and gray shades in the panels (b) and (d) show the correlations and their uncertainties reported in \citet{topping25}, respectively.}
For reference, Arp~220, the nearest ULIRG, is shown as blue squares \citep{perna20, u12, Pereira-Santaella21}. Arp~220 has a surface SFR of $\Sigma_\mathrm{SFR}=2500~M_\odot~\mathrm{yr^{-1}}$ \citep{Pereira-Santaella21}. 
}
% Perna et al. 2020: ne=170, 
% U et al. 2012: logM=10.88, logSFR=2.27
% Pereira-Santaella et al. 2021: Sigma_SFR=2500 Ms/yr
\label{fig:electron-density}
\end{figure*}
%%%%%%%%%%%%%%%%%%%%%%%%%%%%%%

\commentred{
Figure~\ref{fig:electron-density} shows the \sii\ electron densities as a function of redshift, stellar mass, SFR, and SFR surface density ($\Sigma_\mathrm{SFR}^\mathrm{global}$). 
In these figures, we also plot the results taken from the Assembly of Ultradeep Rest-optical Observations Revealing Astrophysics (AURORA; e.g., \citealt{shapley25}) survey for star-forming galaxies at $z=1$--$6$ (gray squares; \citealt{topping25}) and the local ULIRG, Arp 220 (see also Section \ref{subsec:interpretations}). 
We first compare the electron densities of the UDF+ASPECS sample with the redshift evolution presented by \citet{topping25} (black curve in panel (a) of Figure~\ref{fig:electron-density}). 
The UDF+ASPECS sources lie on the high-density side of the relation, but are broadly consistent with the expected redshift evolution within scatter. 

We then compare the electron density with SFR (panel (b) of Figure~\ref{fig:electron-density}). 
The UDF+ASPECS sources have $\log(\mathrm{SFR}~[M_\odot~\mathrm{yr^{-1}}])\sim1$--$3$. 
\citet{topping25} show correlations between electron density and SFR (black line in panel (b)). 
The UDF+ASPECS sample is also broadly consistent with these relations.
}

The relation between the electron density and stellar mass (panel (c) of Figure~\ref{fig:electron-density}) is comparable to that of star-forming galaxies at similar redshifts reported by \citet{topping25}. 
Our galaxy sample is biased toward massive systems ($\log(M_*/M_\odot)\gtrsim10$) due to the selection effects associated with the ALMA dust-continuum detection. 

The panel (d) of Figure~\ref{fig:electron-density} shows the relation between the electron density and surface SFR density. 
To obtain the surface SFR density for the UDF+ASPECS sources, we use the galaxy sizes presented in \citet{vanderwel12}. 
\citet{vanderwel12} conduct global structural parameter measurements for the CANDELS survey using the GALFIT \citep{peng02} Sersic model fits of HST/WFC3 $H_{160}$ images. 
The surface SFR density is calculated as $\Sigma_\mathrm{SFR}^\mathrm{global}=\mathrm{SFR}/(2\pi r_e^2)$. 
Here, $r_e$ is a circularized effective radius, as $r_e=r\sqrt{b/a}$, where $r$ is a half-light radius along the major axis, and $b/a$ is an axis ratio from the Sersic model fitting. 
The SFR values are obtained from the SED fitting (Section~\ref{subsec:SED-fitting}). 
The UDF+ASPECS sources have surface SFR densities of $\log(\Sigma_\mathrm{SFR}^\mathrm{global}~[M_\odot~\mathrm{yr^{-1}~kpc^{-2}}])\sim-1$--$1$ (see also \citealt{rujopakarn16}). 
Relative to the tentative $\Sigma_\mathrm{SFR}$–$n_e$ correlation reported by \citet{topping25} (black solid line; panel (d) of Figure~\ref{fig:electron-density}), the UDF+ASPECS sources show electron densities that are consistent with the relation. 
\commentred{A similar trend has also been reported by \citet{gillman26} for dusty star-forming galaxies with JWST/NIRSpec studies, including some sources in the UDF+ASPECS sample.}

\subsection{PAH Emission} 
\label{subsec:PAH}

PAH features (e.g., 3.3\,$\mu$m, 6.2\,$\mu$m, and 7.7\,$\mu$m; e.g., \citealt{tielens08}) fall within the JWST/MIRI wavelength coverage for the UDF+ASPECS sample and could provide an additional probe of dust conditions. 
As shown in Figure~\ref{fig:SED-expample}, MIRI photometry successfully traces the PAH emission lines ($\lambda_\mathrm{obs}\sim10$--$20\,\mu\mathrm{m}$). 
In the UDF+ASPECS sample, sources classified as AGN (e.g., UDF3 and C08; Figure~\ref{fig:SED-expample}) show weaker PAH features in their SEDs than those of the non-AGN sources (e.g., C13; Figure~\ref{fig:SED-expample}). 
This qualitative trend is consistent with scenarios in which intense AGN radiation fields suppress PAH emission (e.g., \citealt{genzel98}). 
A quantitative analysis of the PAH strengths of the ALMA sources is beyond the scope of this paper.
\commentred{MIRI spectroscopic observations, including slit, integral field unit, and wide-field slitless spectroscopy, will enable a more detailed characterization of the PAH emission in these ALMA sources} (see also, e.g.,  \citealt{spilker23, shivaei24, rieke25, kendrew26}). 

These studies can be extended to higher-redshift dusty galaxies with the proposed PRobe Infrared Mission for Astrophysics (PRIMA; \citealt{glenn25}). 
Its Far-Infrared Enhanced Survey Spectrometer (FIRESS; \citealt{bradford25}) will provide spectroscopic data over $24$--$235\,\mu\mathrm{m}$, allowing us to trace PAH emission and other mid-/far-IR diagnostics over a wide redshift range ($z \gtrsim 1$) with spectroscopy.

\subsection{Dust Properties} 
\label{subsec:dust-SMGs} 

Comparing the dust masses of these faint SMGs with other physical properties, especially the metallicities newly constrained by JWST spectroscopy, provides insight into their nature. 
We estimate dust masses of the UDF sources reported in \citet{dunlop17} using the following modified blackbody: 
\begin{equation}
    M_\mathrm{dust} = \frac{D_\mathrm{L}^2 S_{\nu, \mathrm{obs}}}{(1+z) \kappa_\mathrm{d}(\nu_\mathrm{rest}) [B_\nu(T_\mathrm{dust})-B_\nu(T_\mathrm{CMB}(z))]}, 
\end{equation}
where $D_\mathrm{L}$ is the luminosity distance, $S_{\nu, \mathrm{obs}}$ is the observed flux \commentred{density}, $\kappa_\mathrm{d}(\nu_\mathrm{rest})$ is the rest-frame dust mass absorption coefficient, and $B_\nu(T_{\mathrm{dust}})$ is the Planck function at a rest-frame frequency $\nu_\mathrm{rest}$ and dust temperature $T_\mathrm{dust}$. 
In this equation, $B_\nu(T_\mathrm{CMB}(z))$ is used to correct for the cosmic microwave background (CMB) effects (e.g., \citealt{ota14}). 
We adopt the emissivity index of $\beta_\mathrm{IR}=1.5$, and assume $\kappa_\mathrm{d}(\nu_\mathrm{rest})=0.77(850~\mathrm{\micron}/\lambda_\mathrm{rest})^{\beta_\mathrm{IR}}~\mathrm{cm^2~g^{-1}}$ \citep{dunne00}, where $\lambda_\mathrm{rest}$ is the rest-frame wavelength in units of $\mu$m. 
We use $T_\mathrm{dust}=40\,\mathrm{K}$, which is typical for high-redshift galaxies (e.g., \citealt{aravena20, bouwens20}). 
In this calculation, we use the 1.3-mm fluxes reported in \citet{dunlop17}. 
The dust masses of the ASPECS sources are derived by \citet{aravena20}
\footnote{\citet{aravena20} provide SED-based molecular gas masses ($M_{\rm mol, SED}$) by assuming $M_{\rm mol, SED}=200\,M_{\rm dust}$. We convert them to dust masses via $M_{\rm dust}=M_{\rm mol, SED}/200$ and adopt those values.}, and we utilize them in the following analysis. 

\begin{figure}
% \plotone{figures/fig_DTS_SFR_v2.pdf}
\plotone{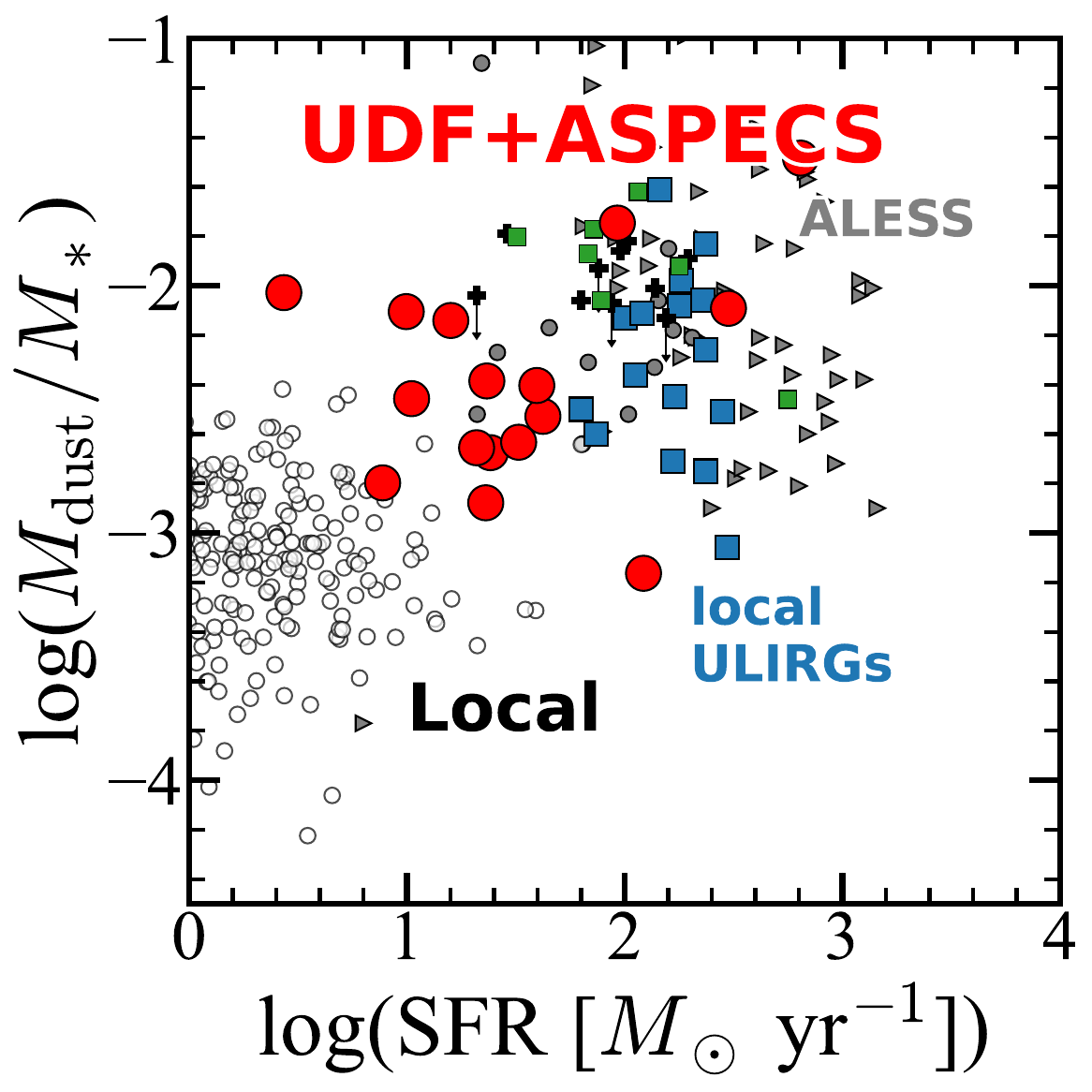}
\caption{
Relation between dust-to-stellar mass ratio (DTS) and SFR. 
The red circles show the UDF+ASPECS sample. 
The local galaxies \citep[Dustpedia, black open circle;][]{pavesi19}, local ULIRGs \citep[blue circles;][]{dacunha10}, ALESS \citep[\commentred{gray} triangle;][]{dacunha15}, \commentred{MOSDEF-ALMA \citep[black plus symbols;][]{sanders23-co, popping23}, ALPINE-CRISTAL \citep[green squares;][]{bethermin20,mitsuhashi24,faisst26}, and} REBELS \citep[gray circles;][]{algera25, rowland25} are also shown.
}
\label{fig:DTS-SFR}
\end{figure}

Figure~\ref{fig:DTS-SFR} shows the dust-to-stellar mass ratio (DTS) and SFR relation. 
The DTS of UDF+ASPECS sample ranges from $\log(\mathrm{DTS})\sim-3$ to $-2$. 
These values are comparable to the local ULIRGs (blue squares) and the ALESS sample (\commentred{gray} triangles), indicating that the UDF+ASPECS sources have dust contents similar to those of local ULIRGs and ALESS SMGs. 
Their high DTS compared to the local galaxies ($\log(\mathrm{DTS})\sim-3$; black open circles) might be due to their higher SFRs. 

\begin{figure*}
% \gridline{\fig{figures/fig_DTG_Z_v2.pdf}{0.45\textwidth}{}
%           \fig{figures/fig_DTS_Z_v2.pdf}{0.45\textwidth}{}
%           }
\gridline{\fig{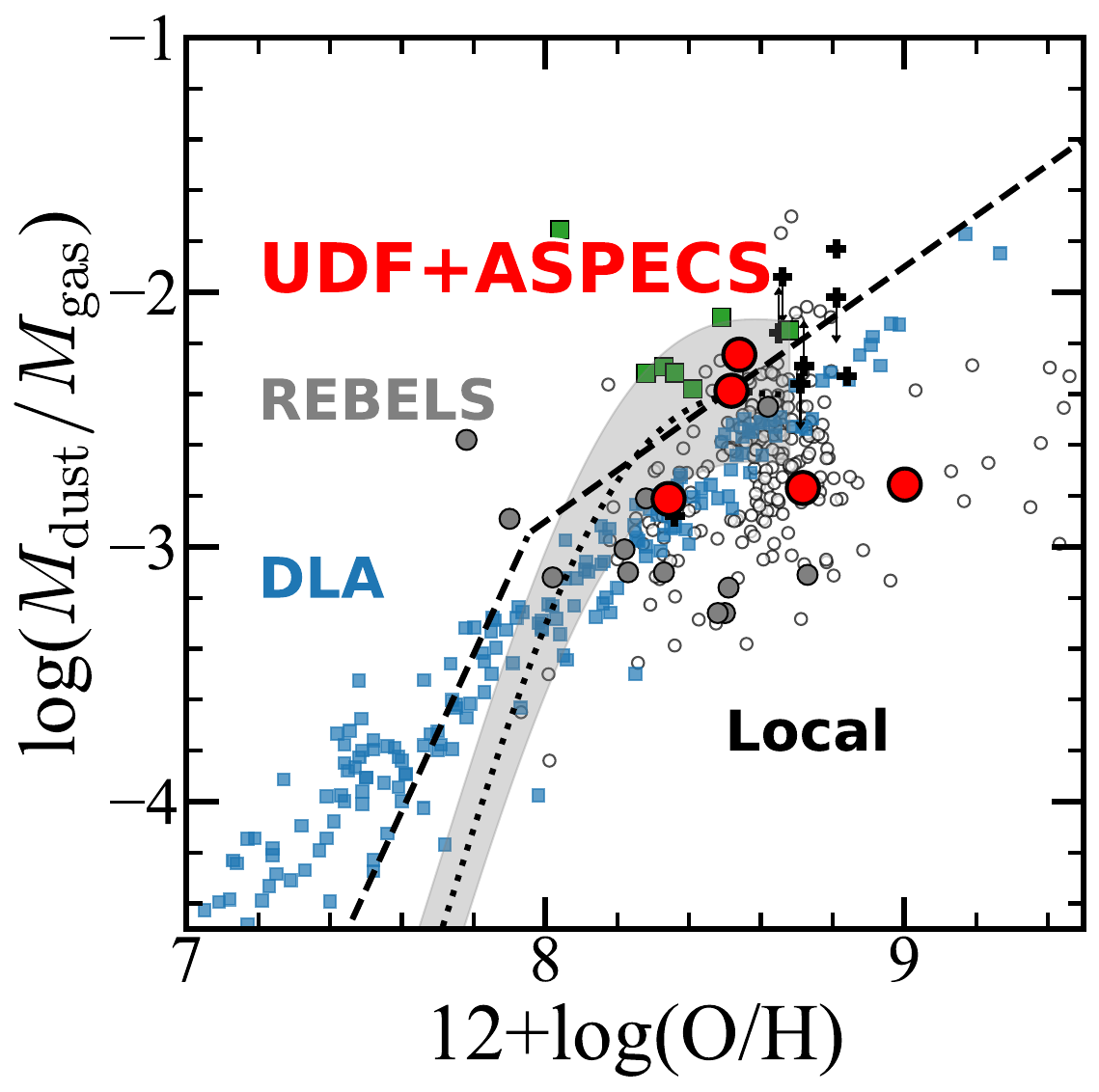}{0.45\textwidth}{}
          \fig{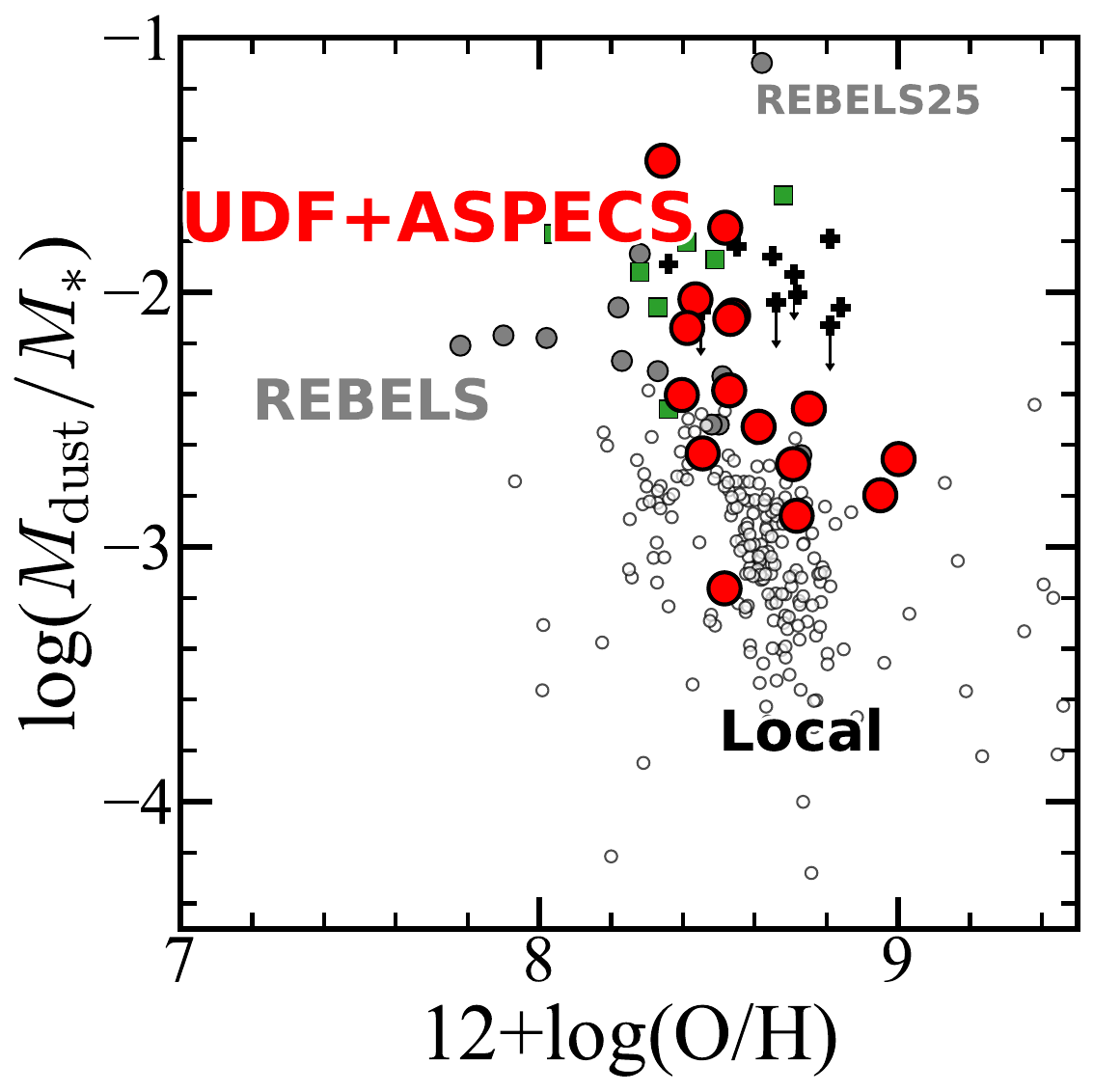}{0.45\textwidth}{}
          }
\caption{
The relation between DTG (DTS) and metallicity is shown in the left (right). The red circles show the UDF+ASPECS sample. For the DTG, we only plot the sources with molecular gas mass estimated from the CO lines \citep{aravena20}. 
The white circles and blue squares show the local galaxies (Dustpedia; \citealt{devis19, casasola20}) and DLA \citep{peroux20}, respectively. 
\commentred{We also show the MOSDEF-ALMA \citep[black plus symbols;][]{sanders23-co, popping23}, ALPINE-CRISTAL \citep[green squares;][]{bethermin20,mitsuhashi24,faisst26}, and the REBELS \citep[gray circle;][]{rowland25, algera25} from the literature values.}
The black dashed and dotted lines show local relations \citep{remy-ruyer14, galliano21}. \commentred{The gray shaded region indicates the uncertainty associated with the local relation shown by the dotted line.}
}
\label{fig:DTG}
\end{figure*}

The left panel of Figure~\ref{fig:DTG} shows the relation between dust-to-gas mass ratio (DTG) and metallicity. 
We compile \commentred{molecular} gas masses for galaxies in the UDF+ASPECS sample with CO emission-line detections from the literature (see Table~\ref{tab:FIR}). 
The UDF+ASPECS sources (red circles) show comparable or smaller DTG compared to the local relation (black dashed and dotted lines; \citealt{remy-ruyer14, galliano21}) and the damped Ly$\alpha$ (DLA) sample (blue square; \citealt{peroux20}). 
\commentred{the ALPINE-CRISTAL, REBELS, and MOSDEF-ALMA samples are also on the local relation, indicating similar properties regarding DTG and metallicity.} 
The right panel of Figure~\ref{fig:DTG} presents the DTS and metallicity relation (see also, e.g., \citealt{calura17}). 
The UDF+ASPECS sources have higher DTS than local galaxies (Dustpedia; \citealt{devis19}) while being comparable to the DTS of other samples. 

%%% fig: dust properties 1 %%%
%%%%%%%%%%%%%%%%%%%%%%%%%%%%%%

%%% fig: dust properties 2 %%%
%%%%%%%%%%%%%%%%%%%%%%%%%%%%%%

\section{Discussion} 
\label{sec:discussion-SMGs}

\subsection{Comparisons with Other Galaxies} 
\label{subsec:conmparison}

\commentred{
Comparisons with other galaxy populations help characterize the faint UDF+ASPECS SMGs. Their positions in the SFR--stellar mass and mass--metallicity planes overlap those of massive star-forming galaxies at $z\sim2$--$3$ (Figures~\ref{fig:LIR-Mstar} and~\ref{fig:MZR}). Some sources reach nearly solar metallicity, comparable to local ULIRGs \citep{chartab22}. ALESS galaxies generally have brighter dust continuum emission, larger inferred dust masses, and higher SFRs than UDF+ASPECS.

MOSDEF-ALMA provides a UV/optically selected comparison at similar redshifts, with optical metallicities and ALMA dust and molecular gas constraints \citep{sanders23-co,popping23}. Its stellar masses and dust mass ratios partly overlap those of UDF+ASPECS, while our sample extends to lower SFRs and dust-to-stellar mass ratios (Figures~\ref{fig:LIR-Mstar}, \ref{fig:DTS-SFR}, and~\ref{fig:DTG}). The ALPINE-CRISTAL-JWST and REBELS-IFU extend the comparison to higher redshifts through their combined JWST metallicity and ALMA dust/gas measurements \citep{faisst26, faisst26b, rowland25, algera25}.
We also note that complementary comparisons are provided by Herschel-selected main-sequence galaxies \citep{valentino20} and strongly lensed star-forming galaxies \citep{Dessauges-Zavadsky15}, whose dust mass ratios ($\log(M_{\mathrm{dust}}/M_\star)\sim-2$ to $-3$) also overlap those of our sample (Figure~\ref{fig:DTS-SFR}). 

Together, these comparisons suggest that faint ALMA-selected SMGs span a broad range of star formation activity while having dust contents, relative to their stellar and molecular gas masses, comparable to other star-forming populations. This connects their dusty nature to the global properties of a broader population of chemically enriched galaxies.
}

\subsection{Critical Metallicity} 
\label{subsubsec:critical-metallicity}

%%% fig: Mdust vs. 12+log(O/H) %%%
\begin{figure}
\plotone{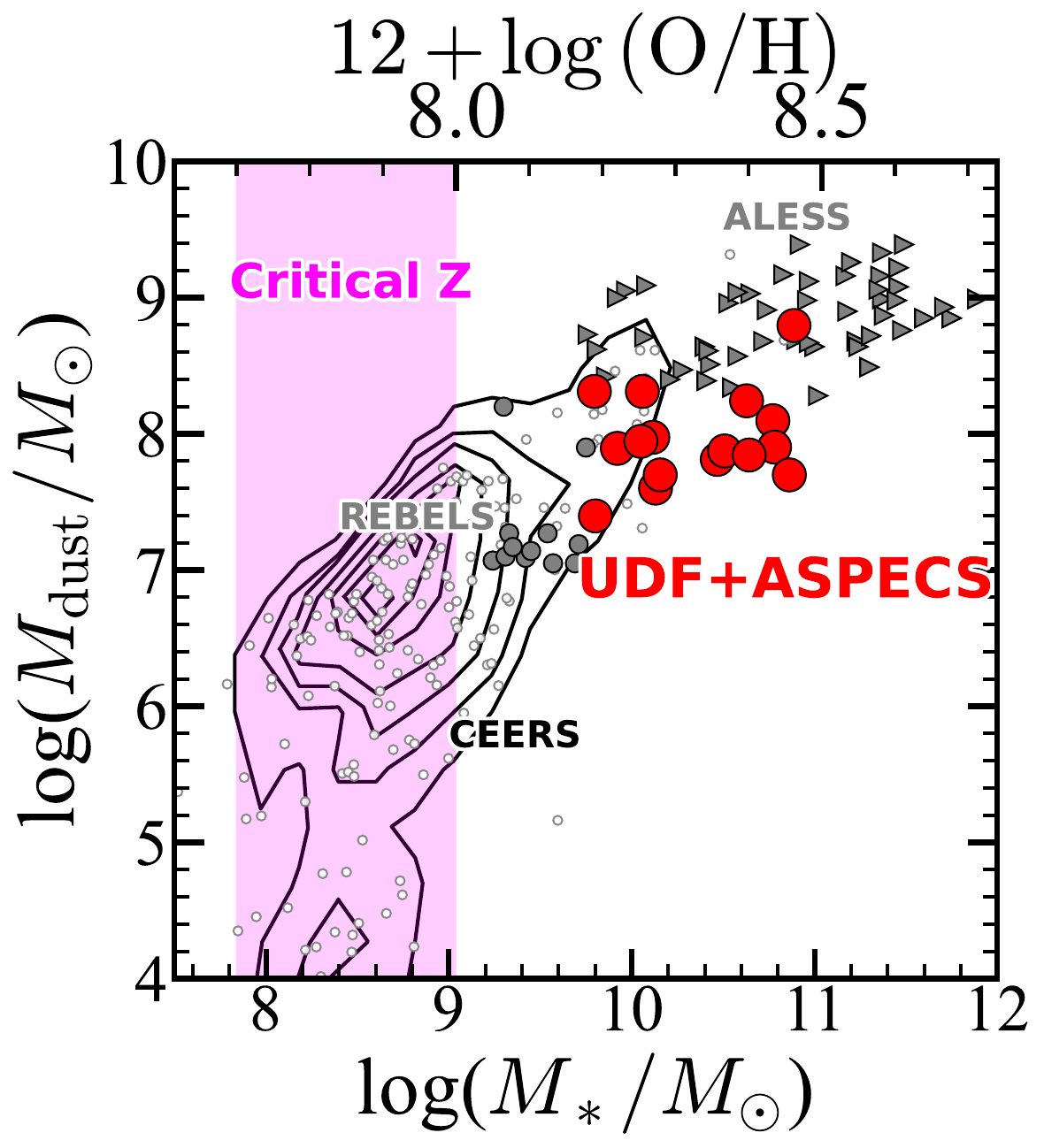}
\caption{
Relation between dust mass and stellar mass. 
The red circles show the UDF+ASPECS sample. 
The gray triangles and circles show the ALESS and REBELS samples, respectively. 
The background white circles show the JWST Cosmic Evolution Early Release Science (CEERS) survey galaxies at $4<z<11$ reported in \citet{burgarella25}, and the black contours show their distribution. 
The dust mass of the CEERS galaxies is estimated from rest-frame UV-NIR SED fitting.
The magenta shaded band indicates the model-predicted critical metallicity ($0.1$--$0.2Z_\odot$; \citealt{asano13, burgarella25}). 
The ticks at the top represent metallicity, converted from stellar mass, using the mass–metallicity relation at $z = 4$--$10$ as presented in \citet{nakajima23}. 
\label{fig:Md-Ms-SMGs}}
\end{figure}
%%%%%%%%%%%%%%%%%%%%%%%%%%%%%%

With JWST/NIRSpec spectroscopy, we can now directly measure gas-phase metallicities for the faint UDF+ASPECS sample (Section~\ref{subsec:metallicity-SMGs}), enabling a direct comparison between dust mass and chemical enrichment in this population. 
Figure~\ref{fig:Md-Ms-SMGs} shows the relation between dust mass and gas-phase metallicity (see also \citealt{kiyota25}). 
The UDF+ASPECS sources have metallicities above the model-predicted critical metallicity ($Z_{\rm crit}\sim0.1$--$0.2\,Z_\odot$; e.g., \citealt{asano13, inoue11, zhukovska14, burgarella25}). 
Above $Z_{\rm crit}$, 
% growth of ISM grain by accretion becomes efficient and can dominate the dust mass build-up. 
\commentred{grain growth in the ISM through accretion becomes efficient}
and can dominate the dust mass build-up. 
The fact that our SMGs lie above $Z_{\rm crit}$ therefore supports a scenario in which their dust masses are at least partly sustained by ISM grain growth, which naturally explains the dusty nature of the SMGs. 

We also find that the UDF+ASPECS sample largely follows the typical DTG-metallicity relation (Figure~\ref{fig:DTG}), with DTG values consistent with those of other galaxies at fixed metallicity. 
This result suggests their dust content is broadly consistent with their chemical enrichment. 
Consequently, the large dust masses of the UDF+ASPECS sample likely also reflect large gas reservoirs (see also \citealt{algera25} for the REBELS sample), together with above the critical metallicities.

\subsection{Interpretations} 
\label{subsec:interpretations}

%%% fig: SMGs interpretation figure %%%
\begin{figure*}
    \centering
    \includegraphics[width=0.7\linewidth]{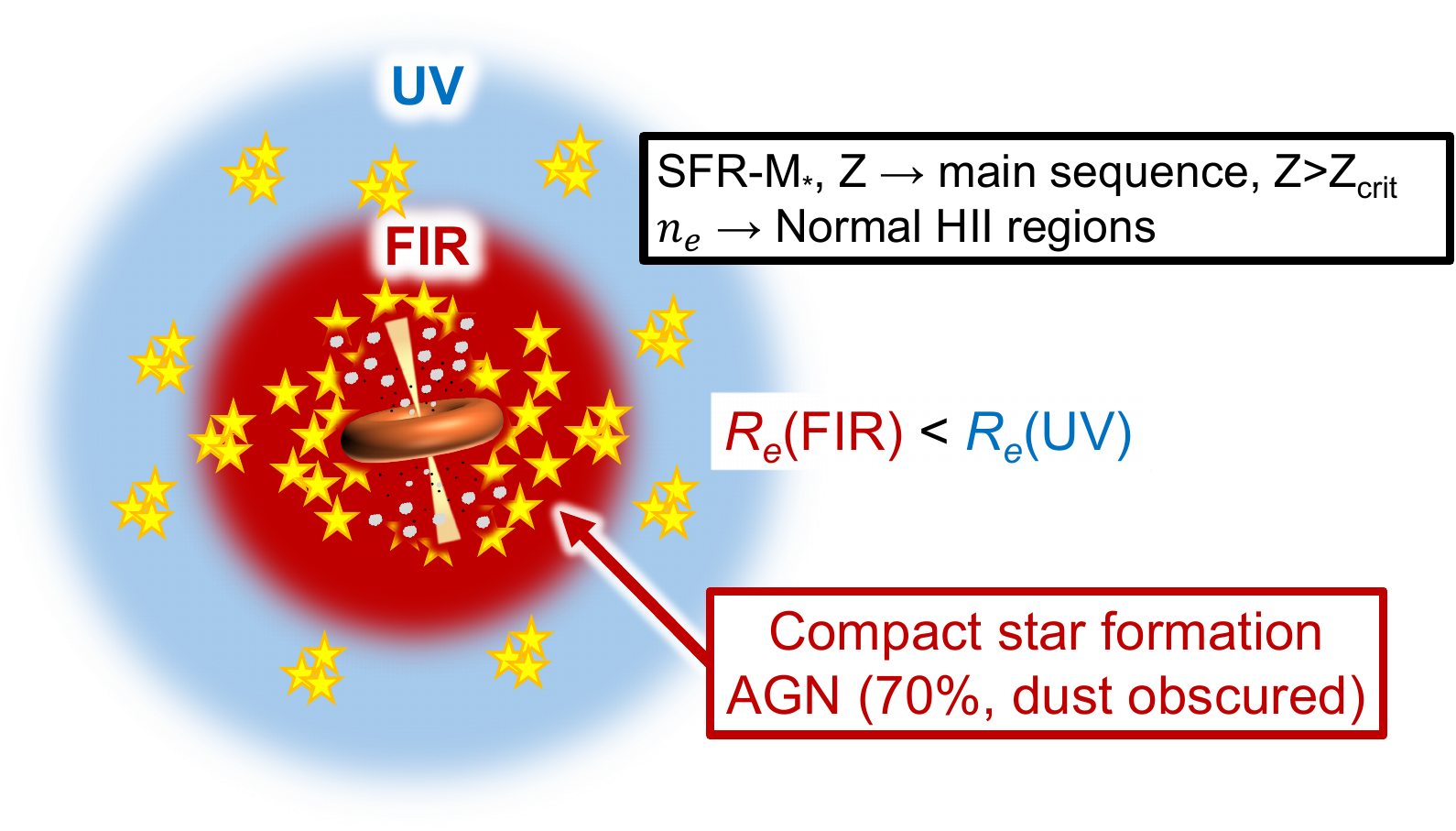}
    \caption{
    Schematic illustration of a possible physical picture for \commentred{faint SMGs}, motivated by the UDF+ASPECS sample and literature. 
    Previous work has shown that FIR size is more compact than UV size ($R_e(\mathrm{FIR})<R_e(\mathrm{UV})$) and that faint SMGs typically lie on the star-forming main sequence. In this study, the ISM properties are broadly consistent with those of other star-forming galaxies (metallicity and \sii\ electron density). We also find that $\sim70\%$ of sources host AGNs, and most of them are obscured (type-2). 
    In this scenario, compact central star formation and AGN activity produce a compact FIR component, yielding $R_e(\mathrm{FIR})<R_e(\mathrm{UV})$ (FIR: red; UV/optical: blue). Because the \sii\ doublet has a relatively low critical density ($\commentred{n_\mathrm{crit}}\sim10^3~\mathrm{cm^{-3}}$), \sii\ primarily traces the bulk star-forming ISM rather than the highest-density gas near the nucleus. The metallicity exceeds the critical value ($Z_\mathrm{crit}$), placing these galaxies in a regime where efficient dust growth can proceed. 
    }
    \label{fig:smgs-manga}
\end{figure*}
%%%%%%%%%%%%%%%%%%%%%%%%%%%%%%%%%%%%%%%

Here, we discuss possible physical interpretations of faint SMGs.
Figure~\ref{fig:smgs-manga} summarizes the results of our analysis and previous studies, and then presents a possible interpretation as a schematic illustration. 
As discussed in Section~\ref{subsec:conmparison}, the UDF+ASPECS sources largely follow the star-forming main sequence (Figure~\ref{fig:LIR-Mstar}) and mass-metallicity relation (Figure~\ref{fig:MZR}). 
In addition, their electron densities are comparable to those of other star-forming galaxies with similar stellar mass and SFR (Figure~\ref{fig:electron-density}). 
These results suggest that the global ISM conditions are not exceptionally extreme, and the UDF+ASPECS sources represent typical massive star-forming galaxies with \commentred{an} AGN fraction of $\sim70\%$ (Section~\ref{subsec:AGN-SMGs} and Figure~\ref{fig:AGN-fraction}). 

We caution, however, that the \sii\ electron density results do not rule out compact starbursts. 
The \sii\ diagnostic saturates above its critical density ($n_\mathrm{crit}\sim10^{3.5}$\,cm$^{-3}$) and is therefore insensitive to much higher densities. 
If nuclear star formation exists, it could still contribute to the observed submillimeter continuum. 
One example is Arp~220 (e.g., \citealt{arp66}), the nearest ULIRG, which shows the \sii\ electron density of $n_e([\text{S\,\textsc{ii}}]) = 170~\mathrm{cm^{-3}}$ (\commentred{blue squares} in Figure~\ref{fig:electron-density}; e.g., \citealt{perna20}). 
Its electron density from \feii\,$1.644\,\mu\mathrm{m}, 1.677\,\mu\mathrm{m}$ is $n_e([\text{Fe\,\textsc{ii}}])\sim5000~\mathrm{cm^{-3}}$ especially around the nuclear outflow region \citep{ulivi25}. 
The critical density of [Fe\,{\sc ii}] is higher ($n_\mathrm{crit}\sim10^{4}$\,cm$^{-3}$) than that of \sii, and \feii\ can trace the higher density gas \citep{ulivi25}. 

The FIR sizes of ALMA-selected galaxies are often more compact than their rest-frame UV/optical sizes, supporting compact star formation \citep[e.g.,][]{fujimoto17, rujopakarn16, elbaz18, kaasinen20, gomez-guijarro22}. 
For the ALMA UDF sources, we compile literature measurements from \citet{elbaz18} and find the median effective radius of $R_{e}({\rm FIR})=1.77$~kpc and $R_{e}({\rm UV})=3.15$~kpc (HST/WFC3 $H$ band). 
Consistent with this picture, \citet{boogaard24} find that the sizes of the ASPECS sources measured in JWST/MIRI F560W, which corresponds to their NIR wavelength, are smaller than those in HST F160W ($R_e(\mathrm{F560W})/R_e(\mathrm{F160W})\sim0.7$). 
These multi-wavelength size offsets are naturally explained by centrally concentrated, dust-obscured star formation (and/or stronger dust attenuation toward the inner regions), which likely governs the observed FIR emission (see Figure~\ref{fig:smgs-manga}). 

For individual sources in the UDF+ASPECS sample, UDF2 and UDF7 are analyzed in \citet{rujopakarn19} based on the high-resolution (30~mas, $200~\mathrm{pc}$) ALMA Band~7 data. 
They report compact dust emission on $\sim1$--$3\,\mathrm{kpc}$ scale and surrounding UV-emitting clumps with no dust detections. 
These results further support a picture of compact FIR dust emission (see also \citealt{rujopakarn23}). 

Regarding AGN activity, previous studies \citep{hopkins08, toft14} suggest that dusty galaxies can be progenitors of AGNs and QSOs based on their number density and gas depletion timescales. 
The high fraction of AGNs (Figure~\ref{fig:AGN-fraction}), even in the faint SMGs, is consistent with these scenarios and provides direct evidence of the co-evolution of the stellar mass growth and supermassive black hole growth (see also \citealt{fujimoto22}). 

We note that AGN heating may also contribute to dust emission, and could potentially affect even the FIR continuum (e.g., \citealt{nandra07, symeonidis16, symeonidis17, mckinney21, tsukui23}). 
If AGN UV radiation escapes to the ISM scales, it can contribute to dust heating and may also be observed as the cold dust emission ($T_\mathrm{dust}\sim40\,\mathrm{K}$). 
The dust continuum emission peaks of the faint SMGs are sometimes located near the central regions of the sources (Figures~\ref{fig:snapshots1}--\ref{fig:snapshots3}), raising the possibility that some of the dust emission is associated with central AGN activity.
However, other discussions also argue for 
\commentred{a smaller AGN contribution to the FIR emission than that from star formation, based on studies of bright QSOs}
(e.g., \citealt{silverman25}). 
The current ALMA dust continuum data of the UDF and ASPECS have limited spatial resolution ($\sim1\arcsec$). 
It remains challenging to determine the relative contributions of AGN activity and compact, nuclear star formation to the FIR dust emission. 
Future high-resolution ALMA dust continuum observations \commentred{will} be useful for determining the origins of the FIR emission. 
% (e.g., the ALMA Cycle 12 large program, HIDING in the HUDF; PI: Leindert Boogaard). 

%%% Detection Fraction %%%%%%%
\begin{figure}
    \centering
    \plotone{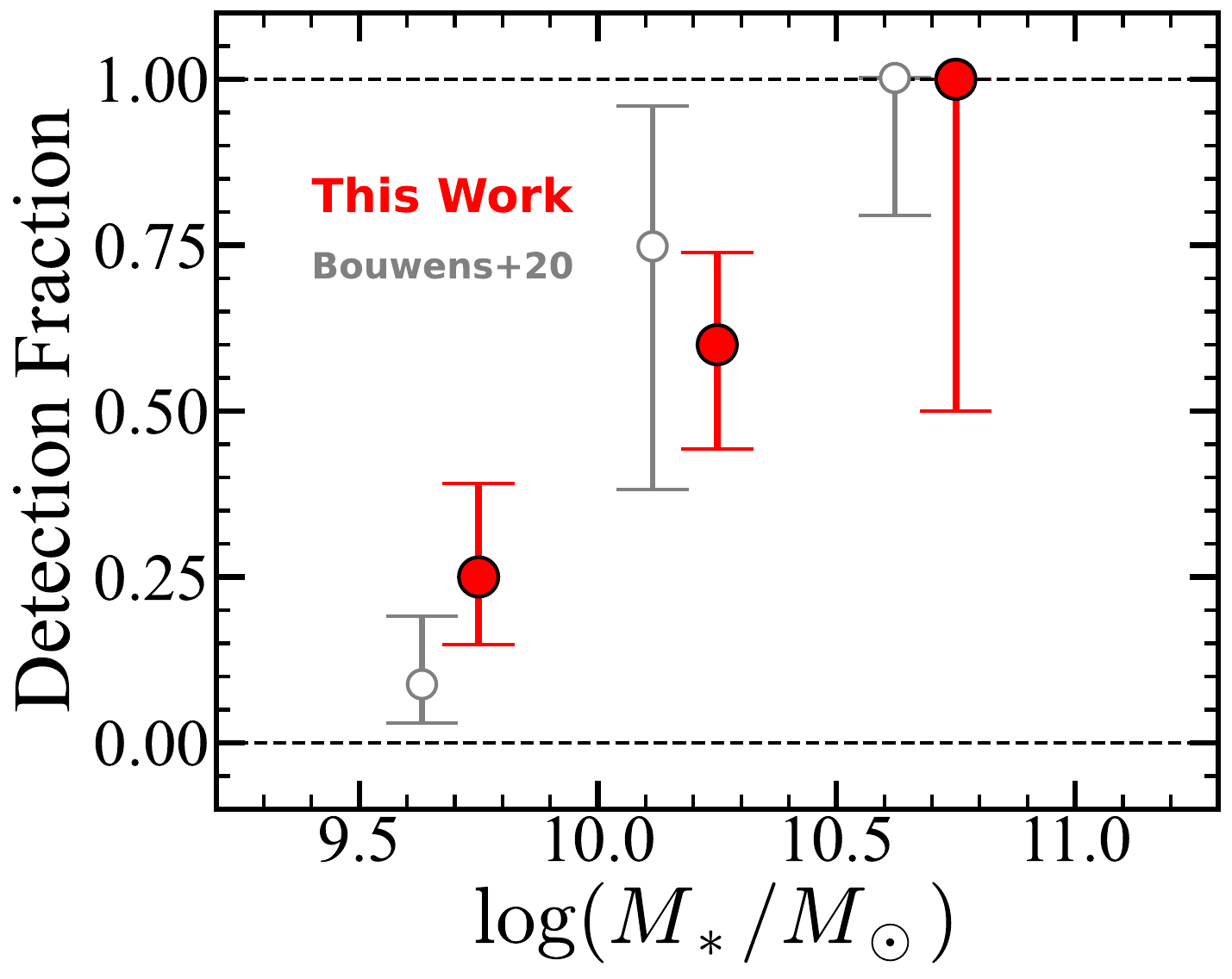}
    \caption{
    Relation between the fraction of galaxies at $z=1.5$--$3.5$ that are detected in the UDF or ASPECS programs and stellar mass. 
    The error bars show the binomial proportion confidence intervals.  
    We only use galaxies that have $z_\mathrm{spec}$ from JADES and SMILES. 
    The stellar mass bins are $\log(M_*/M_\odot)=9.5$--$10.0$, $10.0$--$10.5$, and $10.5$--$11.0$. 
    The open circles show the fraction calculated in \citet{bouwens20} based on HST photo-$z$ galaxies and the ASPECS-detected sources where the ASPECS $1\sigma$ continuum sensitivity is the highest ($<20~\mathrm{\mu Jy~beam^{-1}}$). 
    } 
    \label{fig:ALMA-detection-fraction}
\end{figure}
%%%%%%%%%%%%%%%%%%%%%%%%%%%%%%

To discuss the physical properties of ALMA-detected galaxies, it is useful to examine the ALMA detection fraction as a function of stellar mass. Figure~\ref{fig:ALMA-detection-fraction} shows the fraction of galaxies at $z=1.5$--$3.5$ detected in the UDF and/or ASPECS programs, which have spectroscopic redshifts from JADES or SMILES. 
We define ALMA-detected sources as those reported in Table 2 of \citet{dunlop17} or Table 1 of \citet{aravena20}. 
We construct the parent sample from the JADES and SMILES spectroscopic catalogs by selecting sources within the ASPECS footprint where the ALMA primary-beam response (PB) satisfies $\mathrm{PB}>0.2$ (e.g., \citealt{gonzalezlopez20}). 
We then restrict the sample to $z=1.5$--$3.5$ using the spectroscopic redshifts provided in the JADES or SMILES catalog. 
The stellar masses of all sources in this analysis are estimated with the empirical calibration between $M_\mathrm{F444W}$ and $\log(M_*/M_\odot)$ as described in Section~\ref{subsec:AGN-SMGs}. 

Figure~\ref{fig:ALMA-detection-fraction} indicates that the detection fraction is $\sim60\%$ and $\sim100\%$ at $\log(M_*/M_\odot)=10.0$--$10.5$ and $\log(M_*/M_\odot)>10.5$, respectively. 
The detection fraction decreases toward lower stellar masses, likely reflecting the limited depth of the current ALMA data for low-mass systems. 
\citet{bouwens20} report a consistent trend using HST photometry as a parent sample (open circles in Figure~\ref{fig:ALMA-detection-fraction}). 
These results indicate that ALMA sources provide an efficient census of the most massive star-forming galaxies at $z\sim2$--$3$. 
Taken together, we explore representative massive star-forming galaxies and their rest-frame optical spectroscopic properties, using dust continuum emission as a tracer.

\section{Summary and Conclusions} \label{sec:conclusions}

We present a JWST/NIRSpec rest-frame optical spectroscopic census of 16 ALMA 1-mm flux-limited sources in the HUDF. Our parent sample is taken from the ALMA UDF and ASPECS deep surveys ($S_{1\,\mathrm{mm}}\gtrsim 0.1\,\mathrm{mJy}$). 
\commentred{This sample complements other high-redshift ALMA surveys, such as the ALPINE and REBELS, by probing galaxies that are fainter in the UV than their target galaxies.}
From these sources, we select those with medium-resolution NIRSpec spectra available from the public JWST surveys, JADES, and SMILES. The resulting sample covers $z_\mathrm{spec}=1$--$4$ and $\log(L_\mathrm{IR}/L_\odot)=11$--$13$. Our findings and discussions are summarized as follows: 

\begin{enumerate}
    \item 
    We explore the rest-frame optical emission lines of H$\beta$, \oiii$\lambda\lambda4959, 5007$, H$\alpha$, \nii$\lambda\lambda6548, 6584$, \sii$\lambda\lambda6717, 6731$. 
    We obtain a nebular attenuation of $E(B-V)\sim0.3$--$1.8$ for 12 sources. 
    For the remaining four sources, the H$\beta$ line is not detected, which implies $E(B-V)>1.0$. 
    \item 
    Based on the Chandra X-ray data and the BPT diagram, nearly 70\% (11/16) of the UDF+ASPECS sources are classified as AGNs. 
    We find that only one source shows prominent broad Balmer lines (H$\alpha$ and H$\beta$, $\mathrm{FWHM_{H\alpha}\sim6500~km~s^{-1}}$) with narrow \oiii\ lines, indicating a type-1 AGN. 
    These results suggest that the faint SMGs preferentially host obscured AGNs. 
    \item 
    The \sii\ electron densities are $n_e\sim10^2$--$10^3~\mathrm{cm^{-3}}$, comparable to those of other star-forming galaxies with similar stellar mass and SFR. 
    The metallicities estimated from the N2 index are moderately high ($12+\log(\mathrm{O/H})=8.3$--$9.0$; i.e., $Z=0.4$--$2Z_\odot$), suggesting that these faint SMGs are above the critical metallicity for efficient dust growth. 
    \item 
    The UDF+ASPECS sample is mainly on the SFR–$M_*$, mass–metallicity, and dust mass–other properties (e.g., stellar mass, molecular gas mass) relation, indicating that 
    these sources are representative of massive star-forming galaxies at similar redshifts.
    Combining the previous reports on the galaxy morphologies ($R_e(\mathrm{FIR}) < R_e(\mathrm{UV})$), these sources show compact star formation, obscured (type-2) AGNs, and the bulk ISM conditions are comparable to those of the other star-forming galaxies. 
\end{enumerate}

\begin{acknowledgments} 
\label{ack}

\commentred{The authors thank the anonymous referee for the careful reading and valuable comments that improved our manuscript.}
The authors thank \commentred{Hiddo S. B. Algera, Caitlin M. Casey}, Masatoshi Imanishi, \commentred{Akio Inoue}, Nozomu Kawakatsu, \commentred{Sarah Kendrew, Yuichi Matsuda}, Yurina Nakazato, Yoshiki Toba, \commentred{and Peixin Zhu} for the valuable comments and discussions. 
T.K. thanks Ryosuke Uematsu for technical support with the SED analysis. 

This work is based in part on observations made with the NASA/ESA/CSA James Webb Space Telescope. The data were obtained from the Mikulski Archive for Space Telescopes at the Space Telescope Science Institute, which is operated by the Association of Universities for Research in Astronomy, Inc., under NASA contract NAS 5-03127 for JWST. These observations are associated with programs GTO-1180, GTO-1181, GTO-1210, GTO-1286, GTO-1287, GO-3215 (JADES), GO-1963 (JEMS), GTO-1207 (SMILES). 
The authors acknowledge JADES (PIs: Daniel J. Eisenstein, Nora Luetzgendorf, and Kate Isaak), JEMS (PI: Christina C. Williams), and SMILES (PI: George Rieke) for developing their observing programs with a zero-exclusive-access period. 
\commentred{
All the JWST data used in this paper can be found in MAST in the JADES \dataset[10.17909/8tdj-8n28]{http://dx.doi.org/10.17909/8tdj-8n28}, JEMS \dataset[10.17909/fsc4-dt61]{http://dx.doi.org/10.17909/fsc4-dt61}, and SMILES \dataset[10.17909/et3f-zd57]{http://dx.doi.org/10.17909/et3f-zd57} High Level Science Product repositories. 
}
Data analysis was in part carried out on the Multi-wavelength Data Analysis System operated by the Astronomy Data Center (ADC), National Astronomical Observatory of Japan. 

This publication is based upon work supported by the World Premier International Research Center Initiative (WPI Initiative), MEXT, Japan, KAKENHI (21H04489, 22H04939, 23K20035, 24H00004, 25H00674, \commentred{26KJ1232}) through the Japan Society for the Promotion of Science, and JST FOREST Program (JP-MJFR202Z). 
This work was supported by the joint research program of the Institute for Cosmic Ray Research (ICRR), University of Tokyo. 
\commentred{T.K. was partially supported by the SOKENDAI Student Dispatch Program (2025, 2026), The Graduate University for Advanced Studies, SOKENDAI. T.K. was partially supported by the Overseas Travel Fund for Students (2026) of the Astronomical Science Program, The Graduate University for Advanced Studies, SOKENDAI.}
S.F. acknowledges support from the Dunlap Institute, funded through an endowment established by the David Dunlap family and the University of Toronto. 
Y.U. acknowledges support from the Kyoto University Foundation. 
M.N. is supported by JST, the establishment of university fellowships towards the creation of science technology innovation, Grant Number JPMJFS2136.
The English writing in this paper has been improved with the help of ChatGPT and Grammarly, while the software does not generate sentences from scratch. 

\end{acknowledgments}

\vspace{5mm}
\facilities{JWST (NIRSpec, NIRCam, MIRI)}
\software{Astropy \citep{astropy:2013, astropy:2018, astropy:2022},  
          CIGALE \citep{burgarella05, noll09, boquien19, yang20, yang22},
          Matplotlib \citep{hunter07},
          NumPy \citep{harris20}, 
          pandas \citep{mckinney2010data}
          PyNeb \citep{luridiana15},
          SciPy \citep{virtanen20}, 
          }

\appendix

\restartappendixnumbering
\section{CIGALE Parameters} 
\label{appendix:sed-model}

In Table~\ref{tab:param}, we summarize the parameters used for the SED fitting with CIGALE (Section~\ref{subsec:SED-fitting}). 

\begin{deluxetable}{lcc}[!h]
    \tablecaption{CIGALE Modules and Parameters \label{tab:param}}
    \tablewidth{0pt}
    \tablehead{
    \colhead{Parameter} & \colhead{Symbol} & \colhead{Value}
    }
    \startdata
    \multicolumn{3}{c}{Delayed SFH and recent burst (\texttt{shfdelayedbq}; \citealt{ciesla17})}\\
    \hline
    e-folding time of the main stellar population & $\log(\tau_{\mathrm{main}})$ (Myr) & [$2.0$, $4.1$] (step 0.3) \\ 
   Age of the main stellar population & $\log(\mathrm{age}_{\mathrm{main}})$ (Myr) & [$2.0$, $4.1$] (step 0.3) \\
    Age of the burst and quench episode & age$_{\mathrm{bq}}$   (Myr) & 10, 100 \\
    Ratio of the SFR after/before age$_{\mathrm{bq}}$ & $R_\mathrm{SFR}$ & 1, 10, 100, 1000 \\
    \hline
    \multicolumn{3}{c}{SSP (\texttt{bc03}; \citealt{bruzual03})}\\
    \hline
    Initial mass function  & \nodata & \citet{chabrier03} \\
    Metallicity            & $Z$ & 0.008, 0.02 \\
    \hline
    \multicolumn{3}{c}{Attenuation law (\texttt{dustatt\_modified\_starburst}; \citealt{calzetti00, leitherer02})}\\
    \hline
    Color excess of the nebular lines light & $E(B-V)\mathrm{_{lines}}$ & 0.1, [$0.3$, $2.4$] (step 0.3) \\
    Factor for the stellar continuum attenuation & $E(B-V)\mathrm{_{factor}}$ & 0.44 \\
    Power law slope for the attenuation curve & $\delta$ & [$-0.8$, $0.4$] (step 0.4) \\
    \hline
    \multicolumn{3}{c}{Dust emission (\texttt{themis}; \citealt{Jones17})}\\
    \hline
    Mass fraction of hydrocarbon solids & $q_\mathrm{hac}$  & $0.02, 0.06, 0.10, 0.14, 0.20, 0.28, 0.36$ \\
    Minimum radiation field & $U_\mathrm{min}$  & 0.6, 1.5, 4.0, 10, 25, 50 \\
    Power law slope of $dU/dM \propto U^\alpha$ & $\alpha$ & 2.0 \\ %, 2.5 \\
    Fraction illuminated from $U_\mathrm{min}$ to $U_\mathrm{max}$ & $\gamma$ & 0.02 \\
    \hline
    \multicolumn{3}{c}{AGN (\texttt{skirtor2016}; \citealt{stalevski12, stalevski16})}\\
    \hline
    Average edge-on optical depth at 9.7~$\micron$ & $\tau_{9.7}$ & 5, 9 \\
    Inclination angle & $i$ $(\degr)$ & 30, 70 \\
    AGN fraction & $f_{\mathrm{AGN}}$ & 0.0, 0.05, [0.1, 0.9] (step 0.2) \\
    Temperature of polar dust & $T$ (K) & 150 \\
    \enddata
    \tablecomments{Other parameters are the same as the CIGALE default values.}
\end{deluxetable}

% \clearpage
\onecolumngrid
\restartappendixnumbering
\commentred{
\section{SED Fitting Results} 
\label{appendix:sed-results}

Figure~\ref{fig:all_sed} shows the multiwavelength photometry and the best-fit SEDs estimated with CIGALE. 

\begin{figure}[!h]
    \centering
       \includegraphics[keepaspectratio, scale=0.34]{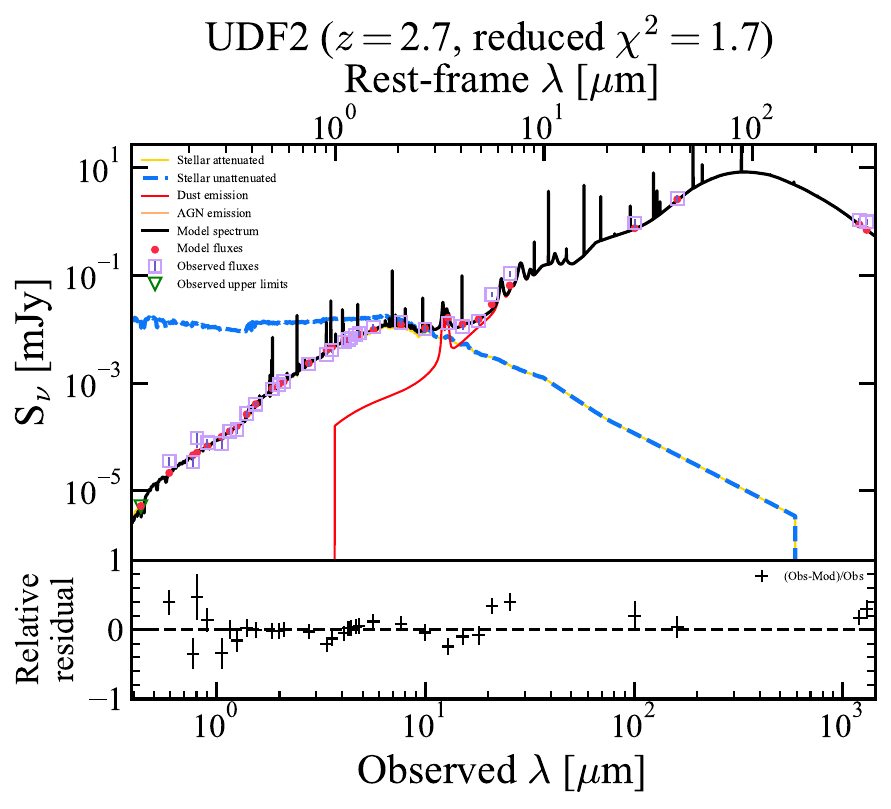}
       \includegraphics[keepaspectratio, scale=0.34]{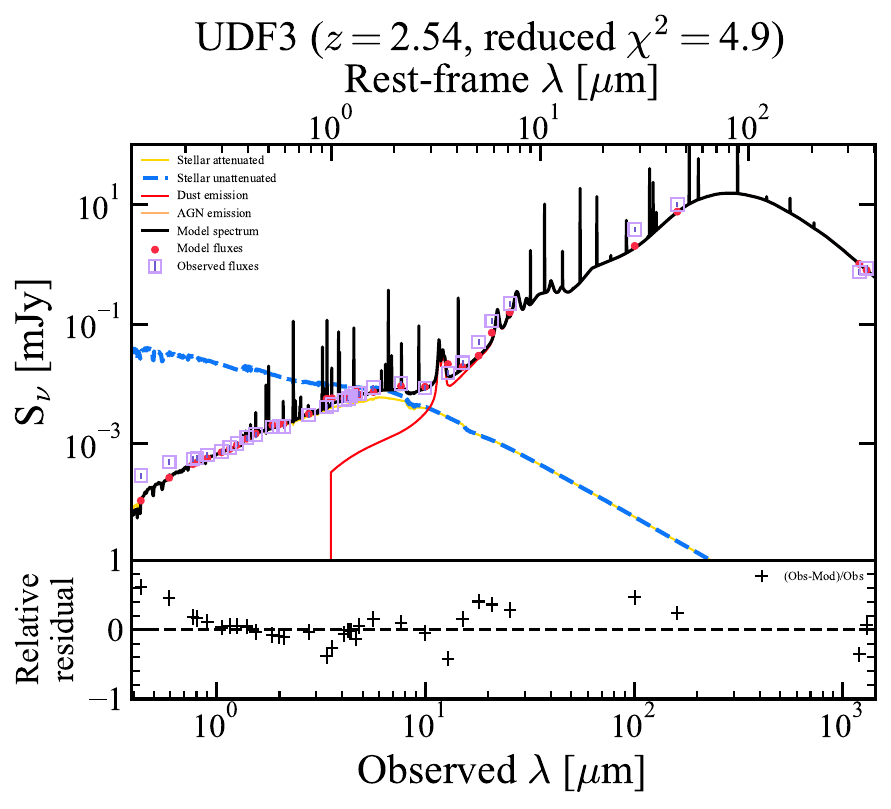}
       \includegraphics[keepaspectratio, scale=0.34]{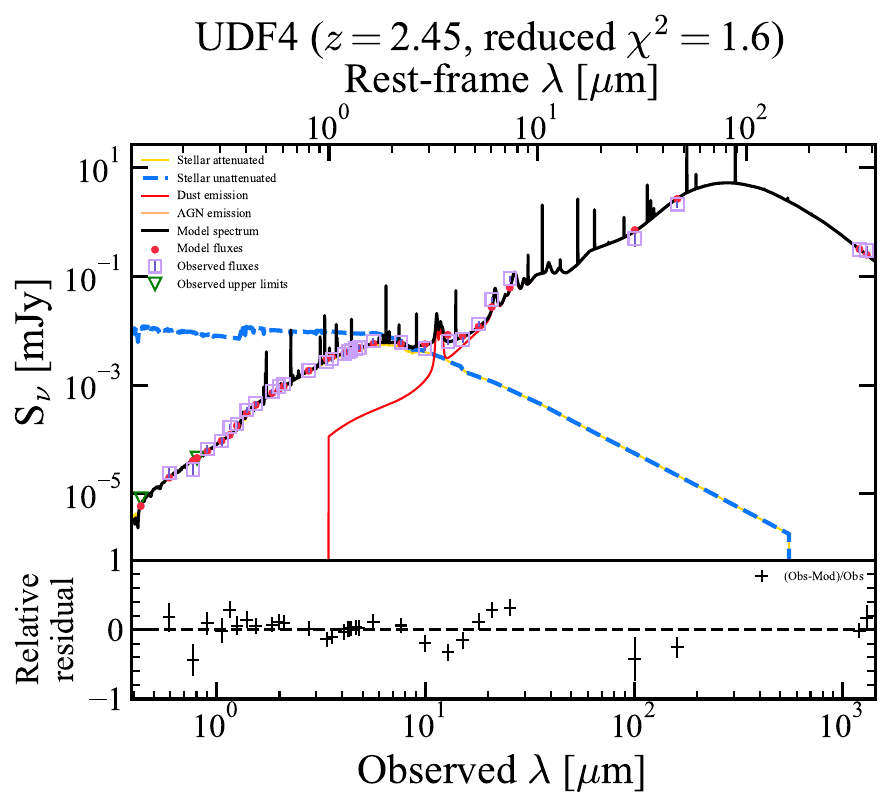}
       \includegraphics[keepaspectratio, scale=0.34]{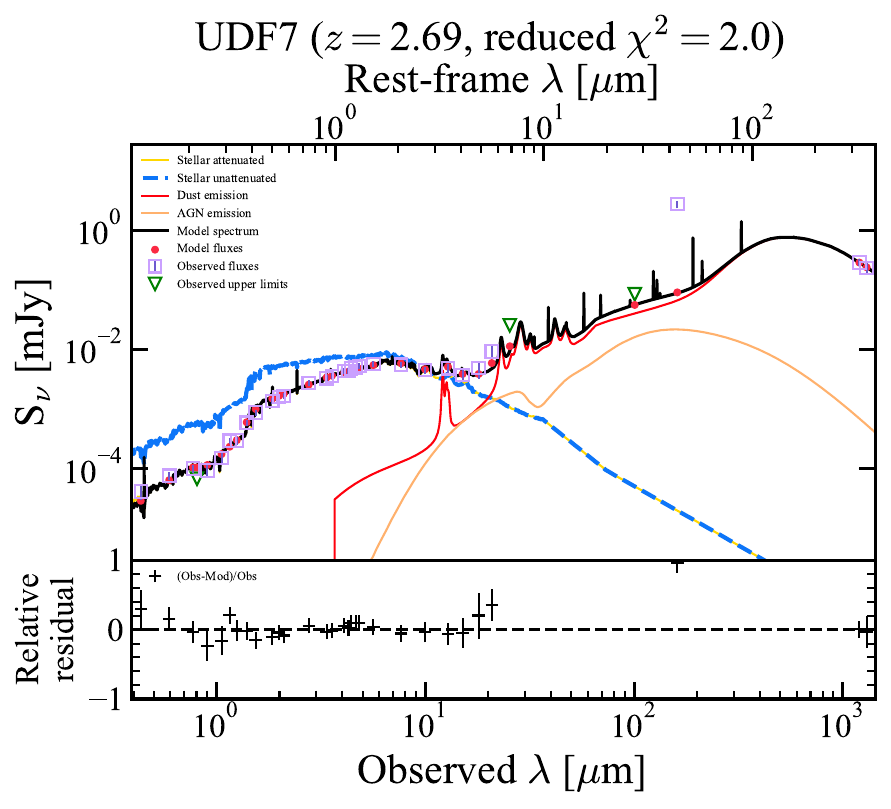}
       \includegraphics[keepaspectratio, scale=0.34]{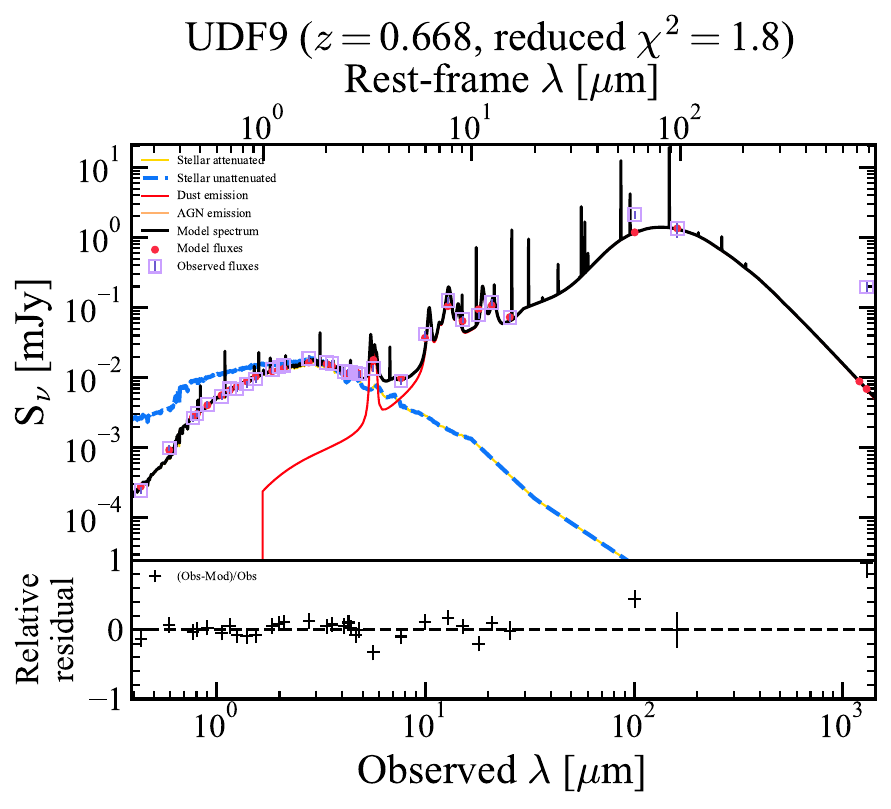}
       \includegraphics[keepaspectratio, scale=0.34]{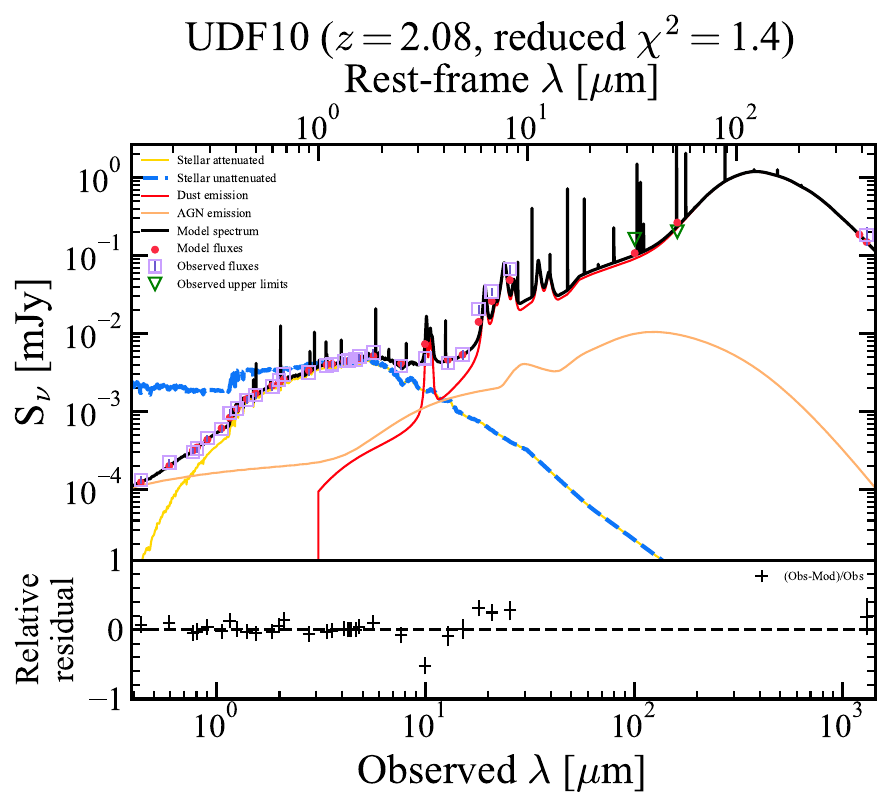}
       \includegraphics[keepaspectratio, scale=0.34]{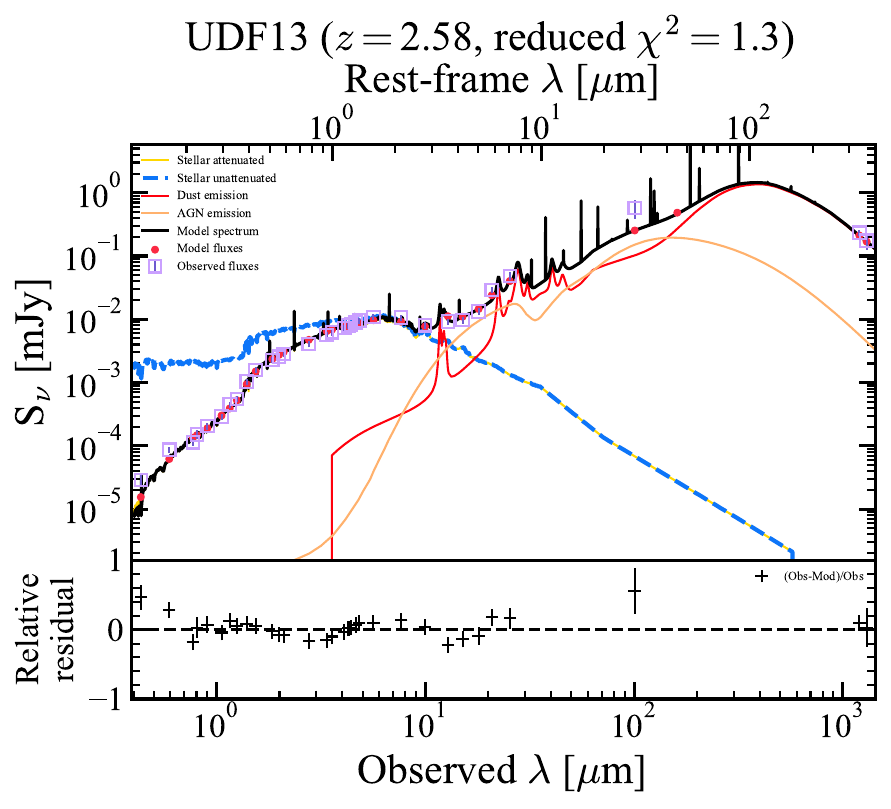}
       \includegraphics[keepaspectratio, scale=0.34]{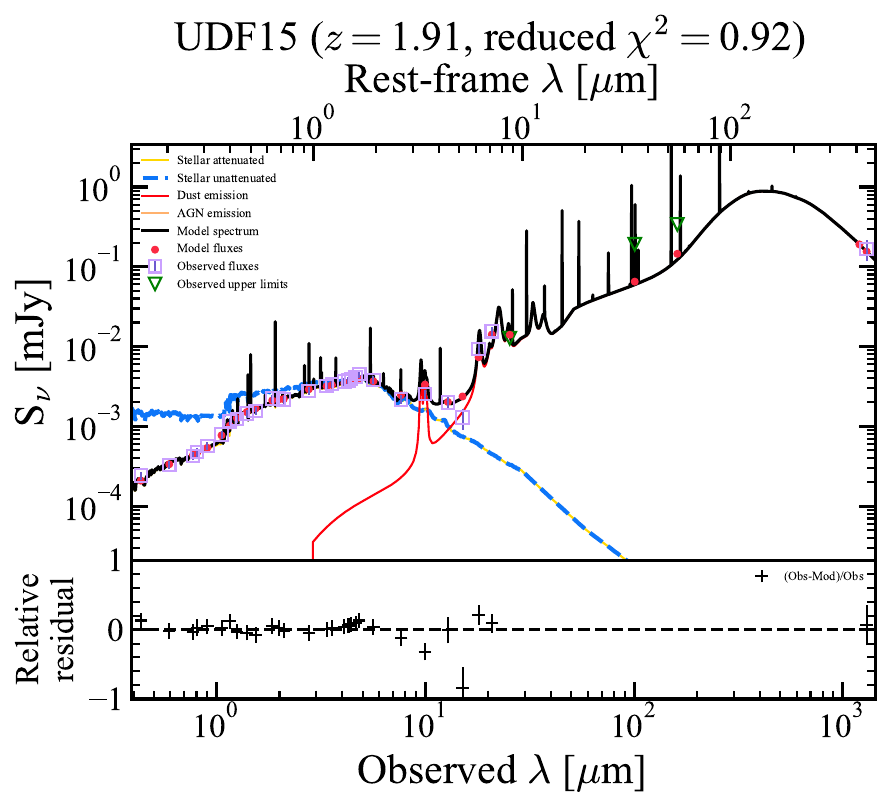}
       \includegraphics[keepaspectratio, scale=0.34]{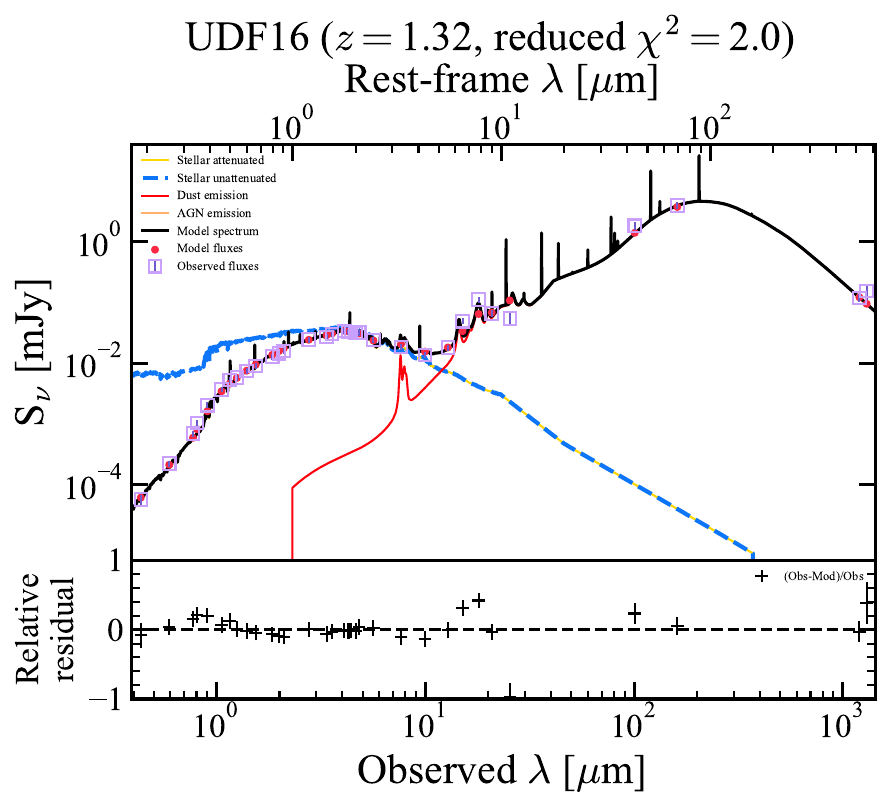}
       \includegraphics[keepaspectratio, scale=0.34]{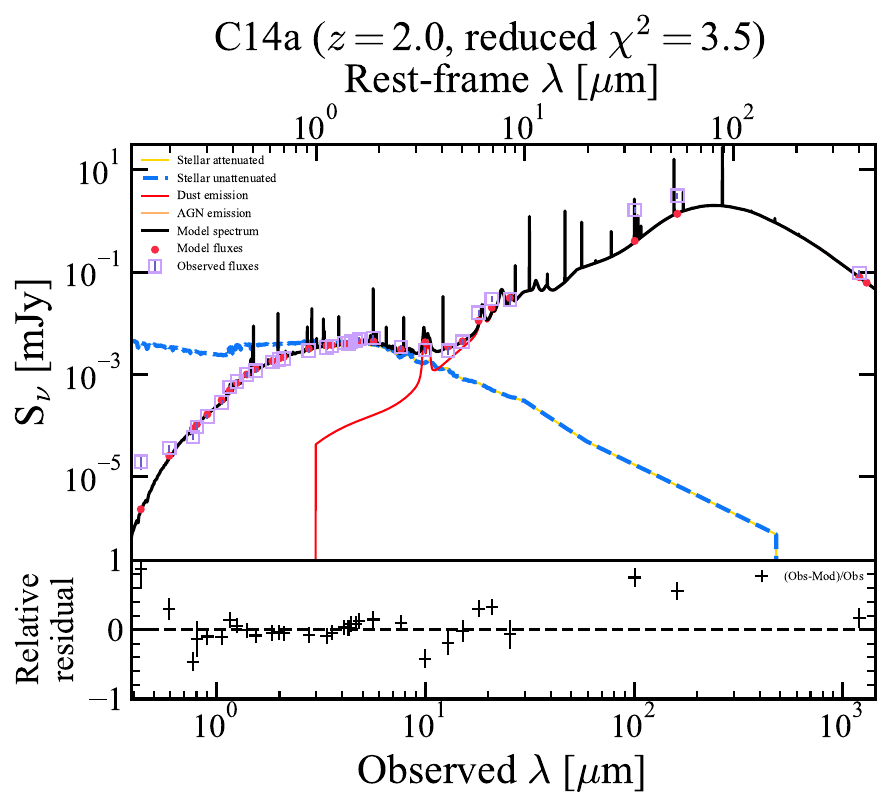}
       \includegraphics[keepaspectratio, scale=0.34]{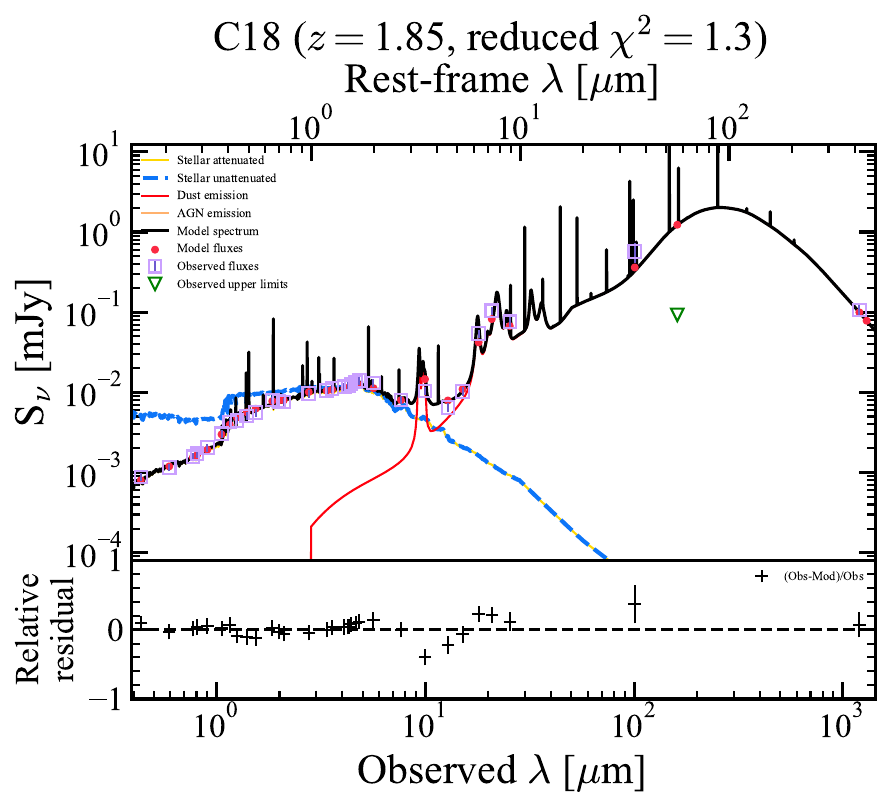}
       \includegraphics[keepaspectratio, scale=0.34]{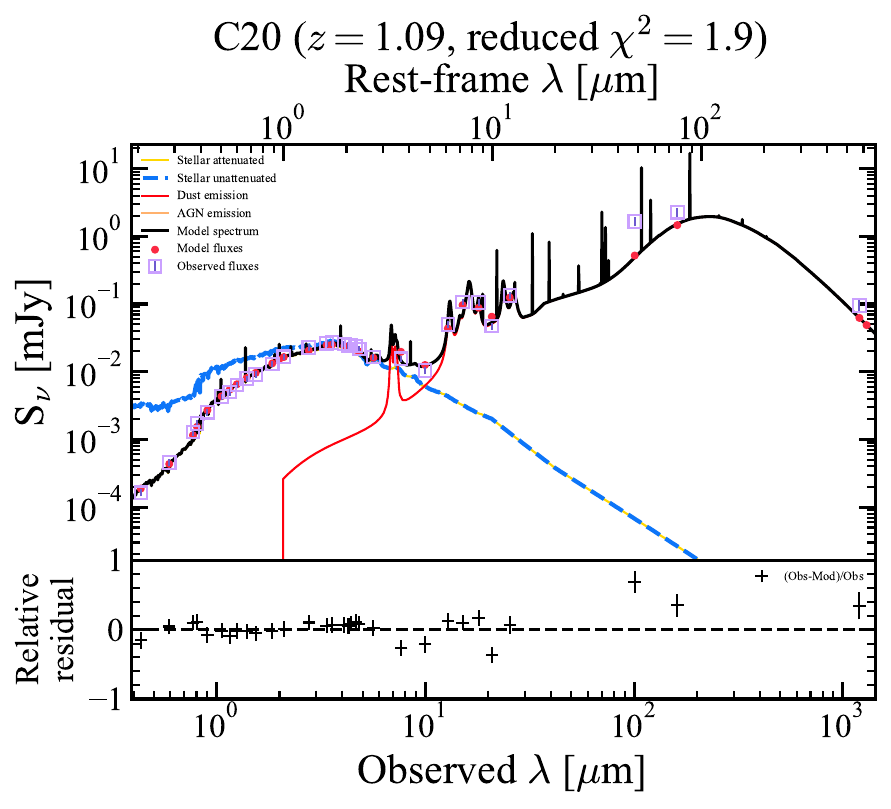}
       \includegraphics[keepaspectratio, scale=0.34]{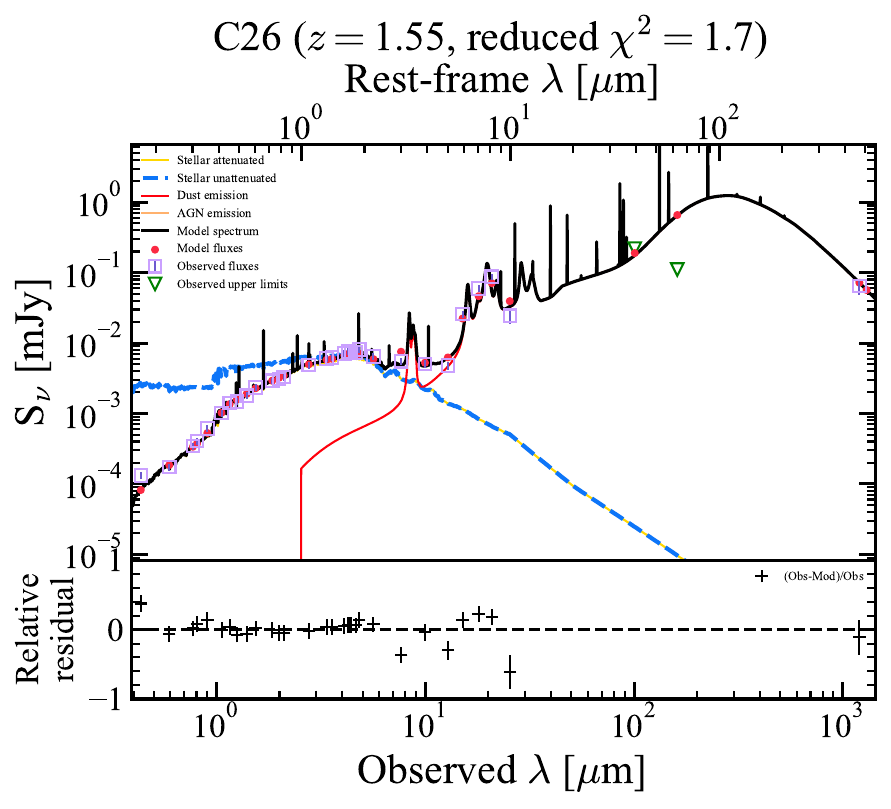}
       \includegraphics[keepaspectratio, scale=0.34]{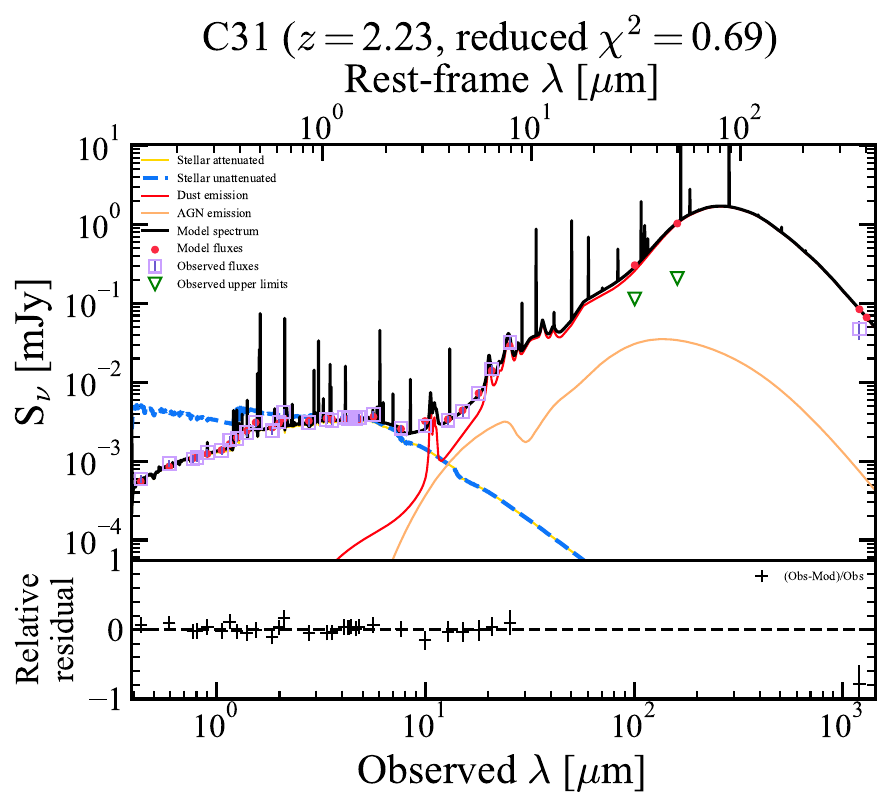}
        \caption{
        \commentred{
        SEDs of all sources, except for C08 and C13 shown in Figure~\ref{fig:SED-expample}. 
        The symbols are the same as in Figure~\ref{fig:SED-expample}.
        }
        }
\label{fig:all_sed}
\end{figure}

}

\clearpage
\bibliography{references}{}
\bibliographystyle{aasjournalv7}

\end{document}